\documentclass[12pt]{article}
\usepackage{theorem}
\usepackage{amsmath,amssymb}
\usepackage{graphicx,makeidx}
\usepackage{caption,subcaption}
\usepackage[mathscr]{eucal}

\usepackage{titlesec}
\titleformat{\subsection}[runin]
 {\normalfont\normalsize\bfseries}{\thesubsection}{1em}{\!\!}
\titleformat{\subsubsection}[runin]
 {\normalfont\normalsize\bfseries}{\thesubsubsection}{1em}{\!\!}
{\theorembodyfont{\upshape} }
\newcommand{\boma}[1]{{\mbox{\boldmath $#1$} }}

\begin{document}
\def\ov{\overline}
\def\Ro{\mathfrak{R}}
\def\ns{{p }}
\def\nn{p}
\def\raL{\ra}
\def\E0{E_0}
\def\Dik{\Dir^{(k)}}
\def\vi{z}
\def\zi{v}
\def\vt{v}
\def\En{\mathfrak{E}}
\def\TAM{A_{(>)}}
\def\TAm{A_{(<)}}
\def\InB{\mathfrak{B}^\s}
\def\eu{\varepsilon^{\s}}
\def\eren{\varepsilon^{ren}}
\def\TT{\mathscr{T}}
\def\TTs{\TT^{\s}}
\def\an{\lambda}
\def\ET{\mathcal{E}}
\def\Ep{\epsilon}
\def\b0{\textbf{0}}
\def\Ta{\mathsf{T}}
\def\Bp{{B_{\sp,\ns}}}
\def\Bs{{B_{\si,\ns}}}
\def\sp{{\bar{\si}}}
\def\Sp{{\bar{S}}}
\def\ba{{\bf a}}
\def\bap{\mathfrak{a}}
\def\sip{\mathfrak{p}}
\def\bj{\sip(j)}
\def\Op{\mathfrak{O}}
\def\fo{\mathfrak{F}}
\def\HE{H}
\def\uu{v}
\def\AM{A}
\def\Am{a}
\def\DD{U}
\def\blp{\bar{\bf l}}
\def\bkl{b_{\bk\bl}}
\def\bhl{b_{\bh\bl}}
\def\bhlp{\bar{b}_{\bh\blp}}
\def\bxp{\bar{\bx}}
\def\bhp{\bar{\bh}}
\def\CB{C}
\def\aa{\alpha}
\def\bxk{\bx_{\star}}
\def\xk{x_\star}
\def\Co{\lozenge}
\def\NCo{\blacksquare}
\def\Toro{{\bf T}}
\def\atanh{\mbox{arcth}\,}
\def\Iun{\mathfrak{J}}
\def\GaL{\mathfrak{g}}
\def\Mk{M_{\mm,k}}
\def\ri{\mathfrak{r}}
\def\oms{\omega_{*}}
\def\Fock{\mathfrak{F}}
\def\Ee{\mathfrak{E}}
\def\ee{\epsilon}
\def\Ome{\Om_\ee}
\def\bxe{\bx_\ee}
\def\FI{\boma{S}}
\def\dFI{\mathfrak{S}}
\def\ff{\mathcal{F}}
\def\hh{\mathcal{H}}
\def\Mel{\mathfrak{M}}
\def\Tr{\mbox{Tr}\,}
\def\xb{\bar{x}}
\def\rb{\bar{\rho}}
\def\rr{\bar{r}}
\def\uno{\mbox{\textbf{1}}}
\def\tez{\al}
\def\Res{\mbox{Res}}
\def\L2m0{L^2_0}
\def\Tm{\tilde{T}}
\def\II{\mathfrak{I}}
\def\Ns{\mathfrak{N}}
\def\Tti{\tilde{T}}
\def\DF{\mathscr{D}}
\def\F{{\mathcal F}}
\def\mm{\kappa}
\def\oo{\varpi}
\def\0{{\bf 0}}
\def\UU{{\mathcal U}}
\def\Hank{\mathfrak{H}}
\def\t{\mathfrak{t}}
\def\Dir{D}
\def\l{\left}
\def\r{\right}
\def\ha{\widehat{a}}
\def\ak{\ha_k}
\def\ah{\ha_h}
\def\had{\ha^{\dagger}}
\def\akd{\had_k}
\def\ahd{\had_h}
\def\GD{\mathfrak{G}}
\def\Fk{F_k}
\def\Fh{F_h}
\def\Fkc{\overline{\Fk}}
\def\Fhc{\overline{\Fh}}
\def\Fkd{\mathfrak{F}_{k_1}}
\def\fk{f_k}
\def\fh{f_h}
\def\fkc{\overline{\fk}}
\def\fhc{\overline{\fh}}
\def\bl{{\bf l}}
\def\bn{{\bf n}}
\def\bk{{\bf k}}
\def\bh{{\bf h}}
\def\bx{{\bf x}}
\def\by{{\bf y}}
\def\bz{{\bf z}}
\def\bw{{\bf w}}
\def\bq{{\bf q}}
\def\bp{{\bf p}}
\def\Nab{\square}
\def\Fi{\widehat{\phi}}
\def\s{u}
\def\Fis{\Fi^{\s}}
\def\Fiseps{\Fi^{\eps \s}}
\def\Ti{\widehat{T}}
\def\Tis{\Ti^{\s}}
\def\Tiseps{\Ti^{\eps \s}}
\def\Aa{\widehat{A}}
\def\Bb{\widehat{B}}
\def\al{\alpha}
\def\be{\beta}
\def\de{\delta}
\def\eps{\varepsilon}
\def\ga{\gamma}
\def\lam{\lambda}
\def\om{\omega}
\def\si{\sigma}
\def\te{\theta}
\def\Ga{\Gamma}
\def\Om{\Omega}
\def\Si{\Sigma}
\def\dd{\displaystyle}
\def\la{\langle}
\def\ra{\rangle}
\def\leqs{\leqslant}
\def\geqs{\geqslant}
\def\sc{\cdot}
\def\restriction{\upharpoonright}
\def\parn{\par\noindent}
\def\complessi{{\bf C}}
\def\reali{{\bf R}}
\def\razionali{{\bf Q}}
\def\interi{{\bf Z}}
\def\naturali{{\bf N}}
\def\AA{{\mathcal A}}
\def\BB{{\mathcal B}}
\def\FF{{\mathcal F}}
\def\EE{{\mathcal E}}
\def\GG{{\mathcal G} }
\def\HH{{\mathcal H}}
\def\JJ{{\mathcal J}}
\def\KK{{\mathcal K}}
\def\LL{{\mathcal L}}
\def\MM{{\mathcal M}}
\def\OO{{\mathcal O}}
\def\PP{{\mathcal P}}
\def\QQ{{\mathcal Q}}
\def\RR{{\mathcal R}}
\def\SS{{\mathcal S}}
\def\cir{{\scriptscriptstyle \circ}}
\def\parn{\par \noindent}
\def\salto{\vskip 0.2truecm \noindent}
\def\beq{\begin{equation}}
\def\feq{\end{equation}}
\def\barray{\begin{array}}
\def\farray{\end{array}}
\newcommand{\rref}[1]{(\ref{#1})}
\def\vain{\rightarrow}
\def\spazio{\vskip 0.5truecm \noindent}
\def\fine{\hfill $\square$ \vskip 0.2cm \noindent}
\def\ffine{\hfill $\lozenge$ \vskip 0.2cm \noindent}

\setcounter{secnumdepth}{5}

\makeatletter \@addtoreset{equation}{section}
\renewcommand{\theequation}{\thesection.\arabic{equation}}
\makeatother
\begin{titlepage}
{~}
\vspace{-2cm}
\begin{center}
{\Large \textbf{Local zeta regularization}}
\vskip 0.2cm
{~}
\hskip -0.4cm
{\Large \textbf{and the scalar Casimir effect IV.}}
\vskip 0.2cm
{~}
\hskip -0.4cm
{\Large\textbf{The case of a rectangular box}}
\end{center}
\vspace{0.5truecm}
\begin{center}
{\large
Davide Fermi$\,{}^a$, Livio Pizzocchero$\,{}^b$({\footnote{Corresponding author}})} \\
\vspace{0.5truecm}
${}^a$ Dipartimento di Matematica, Universit\`a di Milano\\
Via C. Saldini 50, I-20133 Milano, Italy\\
e--mail: davide.fermi@unimi.it \\
\vspace{0.2truecm}
${}^b$ Dipartimento di Matematica, Universit\`a di Milano\\
Via C. Saldini 50, I-20133 Milano, Italy\\
and Istituto Nazionale di Fisica Nucleare, Sezione di Milano, Italy \\
e--mail: livio.pizzocchero@unimi.it
\end{center}
\begin{abstract}
Applying the general framework for local zeta regularization proposed
in Part I of this series of papers, we compute the renormalized vacuum
expectation value of several observables (in particular, of the stress-energy
tensor and of the total energy) for a massless scalar field confined
within a rectangular box of arbitrary dimension.
\end{abstract}
\vspace{0.2cm} \noindent
\textbf{Keywords:} Local Casimir effect, renormalization, zeta regularization.
\hfill \parn
\par \vspace{0.3truecm} \noindent \textbf{AMS Subject classifications:} 81T55, 83C47.
\par \vspace{0.3truecm} \noindent \textbf{PACS}: 03.70.+k, 11.10.Gh, 41.20.Cv~.
\end{titlepage}
\tableofcontents
\vfill \eject \noindent
\section{Introduction.}
In Part I of this series \cite{PI,PII,PIII} we have considered in general
local (and global) zeta regularization for a quantized neutral scalar
field in a $d$-dimensional spatial domain $\Om$, in the environment of
$(d+1)$-dimensional Minkowski spacetime. \parn
In the present Part IV we consider a massless field confined within
a $d$-dimensional rectangular domain $\Om = (0,a_1) \times ... \times (0, a_d)$,
with Dirichlet boundary conditions; applying the general scheme of
Part I we renormalize several observables, namely: the vacuum
expectation value (VEV) of the stress-energy tensor, the pressure
on boundary points, the total energy VEV and of the total force acting
on any side of the box. All these renormalized observables are
represented as sums of series converging with exponential speed,
for which we give quantitative remainder estimates. Our results
hold for an arbitrary spatial dimension $d$; we subsequently
specialize them to the subcase ($d=1$ and) $d=2$. \parn
Let us make a comparison with the existing literature on the subject.
The total energy and the forces on the sides of a rectangular box have
been discussed in a lot of works, some of them using zeta regularization;
here we only mention some of them.
The foremost computation was performed by Lukosz \cite{EnBox,LukWed}
for the electromagnetic field, by means of exponential regularization
and Abel-Plana formula; the same technique was used by Mamaev and
Trunov \cite{MaTru1,MaTru2} (also see \cite{MaTru3,MaTru4}) to discuss,
amongst other models, the case of a conformal scalar field
({\footnote{These authors also consider the cases of electromagnetic,
(massless) gluon and spinor fields in $(3+1)$-dimensional Minkowski
spacetime; furthermore, they derive the renormalized average over the
spatial domain of the stress-energy VEV.}}).
Alternative derivations of the total energy for a scalar field based
on global zeta regularization were given by Ruggerio, Vilanni and Zimerman
\cite{Zim1,Zim2} in two and three spatial dimensions, and by Ambj{\o}rn
and Wolfram \cite{Wolf1} (see also \cite{Wolf2} for the electromagnetic case)
in the case of a multidimensional rectangular cavity for several boundary
conditions. The same configurations were later re-examined by X. Li, Cheng,
J. Li and Zhai \cite{BoxF} by means of a zeta strategy, and by A. Edery
\cite{Ede1,Ede2} using a so called ``multidimensional cut-off technique''.
Let us also cite the papers by Estrada, Fulling et al. \cite{FulKir0,FulKir},
on which we return later in this Introduction.
Finally, let us mention the monographies of Elizalde et al. \cite{AAP0,10AppZ}
and Bordag et al. \cite{Bord};
these can be taken as standard references
for the study of global aspects. \parn
In the present paper we consider global observables, such as the total energy
and forces, mainly to complete the analysis of local aspects (i.e., the VEV
of the stress-energy tensor), which occupy most of our analysis. Our series
representations for these global observables are different, but equivalent
to the ones of \cite{Bord} (which appear to converge exponentially, like
ours). \parn
Concerning local aspects in the previous literature for a scalar field
in a rectangular box let us first mention the two seminal papers
\cite{ActBox1} and \cite{ActBox2} by Actor. In these papers $d=3$,
the framework is Euclidean and the author renormalizes by analytic
continuation the effective Lagrangian density and the VEV $\la 0|\Fi^{\,2}(x)|0\ra$.
The information contained in these works is equivalent, in the language
of our papers, to the specification of the Dirichlet function $\Dir_s(\bx,\by)
:= \la \de_\bx|(-\Delta)^s \de_\by \ra$ along the diagonal $\by=\bx$
(with $\Delta$ the Laplacian and $\bx,\by \in \Om$),
including its analytic continuation with respect to the complex parameter $s$.
However, the VEV of the stress-energy tensor depends on $\Dir_s(\bx,\bx)$
as well as on the \textsl{derivatives} of $\Dir_s(\bx,\by)$ with respect
to $\bx$ and $\by$, at points of the diagonal (see Part I of the present
series). For this reason the results of \cite{ActBox1,ActBox2} are not
sufficient to determine the stress-energy VEV, which in fact is not
mentioned therein. \parn
The renormalized VEV of the stress-energy tensor is derived in a work
of Svaiter et al. \cite{SvaBox2}; the latter employs, again, analytic
continuation methods in a formulation closely related to the approach
of Actor. However, in \cite{SvaBox2} the authors consider an infinite
rectangular waveguide, rather than a box; more precisely, it is assumed
$d=3$ and the spatial domain is $\Om = (0,a_1) \times (0,a_2) \times \reali$.
Apart from the domain, there are other differences bewteen the approach
of Svaiter et al. and ours; in particular, the methods employed in \cite{SvaBox2}
ultimately yield a representation of the stress-energy VEV via series
converging with polynomial speed.
Even though this approach could be extended to treat a box domain
($\Om = (0,a_1) \times (0,a_2) \times (0,a_3)$), it would most likely
yield similar series representations, whose polynomial convergence would
be slower compared to the exponential convergence of ours. \parn
To go on, let us return to the already mentioned works by Estrada, Fulling
et al. \cite{FulKir0,FulKir}. Therein the configuration of a $2$-dimensional
rectangular box is analysed for both Dirichlet and Neumann boundary conditions
(along with some related models, such as the Casimir piston and the so-called
``Casimir pistol''). The cited works introduce a regularized version of the
stress-energy VEV, based on an exponential cutoff; this can be expressed in
terms of the so-called cylinder kernel, which was also considered for other
reasons in Parts I and II of this series. The regularized stress-energy VEV is
computed by the method of images, allowing to express it as a sum over infinitely
many optical paths. By integration of the above VEV, the authors of \cite{FulKir0,FulKir}
also obtain the regularized total energy and the regularized force acting on
a side of the boundary. Their position of principle is that the theory with a
cutoff is a more realistic description of the physical system under investigation;
nonetheless, they also point out that the VEV of the observables for the system
under analysis can be renormalized retaining only their regular parts with respect
to the cut-off (an idea somehow related to what we call the ``extended zeta approach'').
Renormalization along these lines is carried out for global observables like the
total energy, and hinted at for local observables (namely, for the energy density).
\parn
The present paper is organized as follows. In Section \ref{Gloss},
as in the other papers of this series, we report a summary of results
from Part I to be used in the present work. Section \ref{secBox}
and the related Appendices \ref{AppBox}, \ref{AppTotEBox}, \ref{AppC}
are the core the paper; therein \hfill
\vfill \eject \noindent
{~}
\vskip -2cm \noindent
we treat the box configuration
$\Om = (0, a_1) \times ... \times (0, a_d)$
in arbitrary spatial dimension $d$, for Dirichlet boundary conditions.
Our starting point is the heat kernel $\la \de_\bx|e^{-\t \AA}\de_\by\ra$
for which we derive two different series representations capturing,
respectively, the behavior for small and large $\t$. As emphasized in
Part I, the heat kernel determines an integral representation
of the Dirichlet function $\Dir_s(\bx,\by) := \la \de_\bx|(-\Delta)^s \de_\by \ra$,
which can be used to construct the analytic continuation in $s$ of the
latter. Combining these general facts with the series representation
for the heat kernel cited above, we ultimately produce series expansions
for the analytic continuations of $\Dir_s(\bx,\by)$ and its derivatives,
with the previously mentioned exponential speed of convergence; these
determine all the local or global renormalized observables indicated
before, from the stress-energy VEV to the force on each side of the box. \parn
In Section \ref{Boxd12}, the previous general results are specialized
to the cases with $d \in \{1,2\}$. For $d=1$, we recover from a different
viewpoint the results already obtained in Section 6 of Part I where,
as a first example of our general formalism, we discussed a massless
scalar field on a segment $(0,a)$ (see this section of I
for some references on this case, especially \cite{MeFu}). For $d=2$, working on $\Om = (0,a_1)
\times (0,a_2)$ and using the previously mentioned series expansions,
we produce several graphs; some of them refer to the components of the
stress-energy VEV for some choices of $a_1, a_2$, while the others are
about the further observables in which we are interested.\salto
Before moving on, let us mention that many of the results presented in
this paper have been derived with the aid of the software $\tt{Mathematica}$
for both symbolic and numerical computations.
\section{A summary of results from Part I} \label{Gloss}
\subsection{General setting.}
Throughout the paper we use natural units, so that
\beq c = 1 ~, \qquad \hbar = 1 ~. \feq
Our approach works in $(d+1)$-dimensional Minkowski spacetime, which is
identified with $\reali^{d+1}$ using a set of inertial coordinates
\beq x = (x^\mu)_{\mu=0,1,...,d} \equiv (x^0,\bx) \equiv (t,\bx) ~; \feq
the Minkowski metric is $(\eta_{\mu \nu}) = \mbox{diag} (-1,1,...\,,1)$\,.
We fix a spatial domain $\Om \subset \reali^d$ and a background static
potential $V\!:\!\Om\!\to\!\reali$. We consider a quantized neutral,
scalar field $\Fi : \reali \times \Om \to \LL_{s a}(\Fock)$ ($\Fock$
is the Fock space and $\LL_{s a}(\Fock)$ are the selfadjoint operators on it);
suitable boundary conditions are prescribed on $\partial \Om$. The field
equation reads
\beq 0 = (-\partial_{tt}+\Delta-V(\bx)) \Fi(\bx,t) \label{daquan} \feq
($\Delta := \sum_{i=1}^d \partial_{ii}$ is the $d$-dimensional Laplacian).
We put
\beq \AA := - \Delta + V ~, \label{defaa} \feq
keeping into account the boundary conditions on $\partial\Om$, and consider
the Hilbert space $L^2(\Om)$ with inner product $\la f|g\raL :=
\int_{\Om}d\bx\,\overline{f}(\bx)g(\bx)$. We assume $\AA$ to be selfadjoint
in $L^2(\Om)$ and strictly positive (i.e., with spectrum $\si(\AA) \subset
[\eps^2,+\infty)$ for some $\eps >0$). \parn
We often refer to a complete orthonormal set $(\Fk)_{k \in \KK}$ of
(proper or improper) eigenfunctions of $\AA$ with eigenvalues
$(\om^2_k)_{k \in \KK}$ ($\om_k\!\geqs\!\eps$ for all $k\!\in\!\KK$). Thus
\begin{equation}\begin{split}
& \Fk : \Om \to \complessi; \qquad \AA\Fk= \om^2_k \Fk ~; \\
& \hspace{-0.7cm} \la \Fk | \Fh \raL = \de(k, h) \quad
\mbox{for all $k,h \in \KK$} ~. \label{eigenf}
\end{split}\end{equation}
The labels $k\!\in\!\KK$ can include both discrete and continuous
parameters; $\int_\KK dk$ indicates summation over all labels and
$\de(h,k)$ is the Dirac delta function on $\KK$. \parn
We expand the field $\Fi$ in terms of destruction and creation
operators corresponding to the above eigenfunctions, and assume the
canonical commutation relations; $|0 \ra \in \Fock$ is the vacuum
state and VEV stands for ``vacuum expectation value''. \parn
The quantized stress-energy tensor reads ($\xi\!\in\! \reali$ is a parameter)
\beq \Ti_{\mu \nu} := \l(1 - 2\xi \r) \partial_\mu \Fi \circ \partial_\nu \Fi
- \l({1\over 2} - 2\xi \r)\eta_{\mu\nu}(\partial^\lam\Fi \, \partial_\lam \Fi + V\Fi^2)
- 2 \xi \, \Fi \circ \partial_{\mu \nu} \Fi \,; \label{tiquan} \feq
in the above we put $\Aa \circ \Bb := (1/2) (\Aa \Bb + \Bb \Aa)$
for all $\Aa, \Bb \in \LL_{s a}(\Fock)$, and all the bilinear terms
in the field are evaluated on the diagonal (e.g., $\partial_\mu \Fi
\circ \partial_\nu \Fi$ indicates the map $x \mapsto \partial_\mu \Fi(x)
\circ \partial_\nu \Fi(x)$). The VEV $\la 0|\Ti_{\mu\nu}|0\ra$ is
typically divergent.
\vspace{-0.4cm}
\subsection{Zeta regularization.}
The \textsl{zeta-regularized field operator} is
\beq \Fis := (\mm^{-2} \AA)^{-\s/4} \Fi ~, \label{Fis} \feq
where $\AA$ is the operator \rref{defaa}, $\s \in \complessi$
and $\mm > 0$ is a ``mass scale'' parameter; note that
$\Fis|_{\s = 0} = \Fi$, at least formally. The \textsl{zeta
regularized stress-energy tensor} is
\beq \Tis_{\mu \nu} := (1\!-\!2\xi)\partial_\mu \Fis\!\circ\partial_\nu \Fis\!
- \!\l({1\over 2}\!-\!2\xi\!\r)\!\eta_{\mu\nu}\!
\l(\!\partial^\lam\Fis\partial_\lam \Fis\!+\!V(\Fis)^2\r)
- 2 \xi \, \Fis\!\circ \partial_{\mu \nu} \Fis \,. \label{tiquans} \feq
The VEV $\la 0|\Tis_{\mu\nu}|0\ra$ is well defined for $\Re\s$ large enough
(see the forthcoming subsection \ref{DirT}); moreover, in the region of definition
it is an analytic function of $\s$. The same can be said of many related observables
(including global objects, such as the total energy VEV). \parn
For any one of these observables, let us denote with $\F(\s)$ its
zeta-regularized version and assume this to be analytic for $\s$
in a suitable domain $\UU_0 \subset \complessi$.
The zeta approach to renormalization can be formulated in two versions. \parn
i) \textsl{Restricted version}. Assume the map $\UU_0\!\to\!\complessi$,
$\s\!\mapsto\!\F(\s)$ to admit an analytic continuation (indicated with the
same notation) to an open subset $\UU\!\subset\!\complessi$ with $\UU\!\ni\!0$\,;
then we define the renormalized observables as
\beq \F_{ren} := \F(0) ~. \label{ren} \feq
ii) \textsl{Extended version}. Assume that there exists an open subset
$\UU\!\subset\!\complessi$ with $\UU_0\!\subset\!\UU$, such that
$0\!\in\!\UU$ and the map $\s\!\in\!\UU_0 \mapsto \F(\s)$ has an
analytic continuation to $\UU\!\setminus\!\{0\}$ (still denoted with $\F$).
Starting from the Laurent expansion $\F(\s) = \sum_{k = -\infty}^{+\infty}
\F_k \s^k$, we introduce the \textsl{regular part} $(RP\,\F)(\s) :=
\sum_{k=0}^{+\infty} \F_k \s^k$ and define
\beq \F_{ren} := (RP\,\F)(0)~. \label{renest} \feq
Of course, if $\F$ is regular at $\s = 0$ the defnitions \rref{ren}
\rref{renest} coincide. \parn
Differently from the other papers of this series, the observables
considered in the present Part IV will never exhibit singularities
at $\s = 0$; thus, we will never need to use the prescription (ii). \parn
In the sequel we will often refer to the stress-energy VEV, for which
the prescription (i) gives
\beq \la 0 | \Ti_{\mu \nu}(x) | 0 \ra_{ren} :=
\la 0 | \Tis_{\mu \nu}(x) | 0 \ra \Big|_{\s=0} ~. \label{pri} \feq
\subsection{Conformal and non-conformal parts of the stress-energy VEV.} \label{ConfSubsec}
These are indicated by the superscripts $(\Co)$ and $(\NCo)$, respectively;
they are defined by
\beq \la 0|\Ti_{\mu\nu}|0\ra_{ren} = \la 0|\Ti^{(\Co)}_{\mu\nu}|0\ra_{ren}
+ (\xi\!-\!\xi_d)\,\la 0|\Ti^{(\NCo)}_{\mu\nu}|0\ra_{ren} ~, \label{TRinCo}\feq
where we are considering for the parameter $\xi$ the critical value
\beq \xi_d := {d\!-\!1 \over 4d} ~. \label{xic} \feq
\vspace{-0.9cm}
\subsection{Integral kernels.} If $\BB$ is a linear operator in $L^2(\Om)$,
its integral kernel is the (generalized) function $(\bx,\by) \in
\Om\! \times\!\Om \mapsto \BB(\bx,\by) := \la
\de_{\bx}|\BB\,\de_{\by}\raL$ ($\de_\bx$ is the Dirac delta at
$\bx$). The trace of $\BB$, assuming it exists, fulfills $\Tr \BB
= \int_\Om d\bx\,\BB(\bx,\bx)$\,. \parn In the following
subsections $\AA$ is a strictly positive selfadjoint operator in
$L^2(\Om)$, with a complete orthonormal set of eigenfunctions as
in Eq. \rref{eigenf}. In typical applications, $\AA$
is the operator \rref{defaa}. \vspace{-0.4cm}
\subsection{The Dirichlet kernel and its relations with the stress-energy VEV.}\label{DirT}
For (suitable) $s \in \complessi$, the $s$-th Dirichlet kernel of
$\AA$ is \beq \Dir_s(\bx, \by) := \AA^{-s}(\bx,\by) = \int_\KK {dk
\over \om_k^{2s}}\; \Fk(\bx) \Fkc(\by) ~. \label{eqkerdi} \feq If
$\Om$ is a bounded domain with smooth boundary $\partial\Om$
and $\AA = -\Delta + V$ (with
$V$ a smooth potential) is a strictly positive operator on
$L^2(\Om)$, considering the closure $\ov{\Om} := \Om \cup \partial\Om$
one can show that
the map $\Dir_s(~,~) : \ov{\Om} \times \ov{\Om} \to
\complessi$, $(\bx,\by) \mapsto \Dir_s(\bx,\by)$ is continuous
along with all its partial derivatives up to order $j \in
\naturali$, for all $s \in \complessi$ with $\Re s > d/2+j/2$\,;
moreover, the eigenfunction expansion in Eq. \rref{eqkerdi}
converges absolutely and uniformly on $\ov{\Om}$ with all the
derivatives up order $j$ for $\Re s > d + j/2$. \salto Recalling
Eq. \rref{tiquans}, the regularized stress-energy VEV can be
expressed as follows: \beq {~}\hspace{-0.5cm} \la 0 | \Tis_{0
0}(\bx) | 0 \ra \! = \!\mm^\s
\!\!\l[\!\l(\!\frac{1}{4}\!+\!\xi\!\r)\!\!\Dir_{{\s - 1\over
2}}(\bx,\by) \!+\!\l(\!\frac{1}{4}\!-\!\xi\!\r)\!\!
(\partial^{x^\ell}\!\partial_{y^\ell}\!+\!V(\bx)) \Dir_{{\s + 1
\over 2}}(\bx,\by)\r]_{\by = \bx}\!, \label{Tidir00} \feq \beq \la
0 | \Tis_{0 j}(\bx) | 0 \ra = \la 0 | \Tis_{j 0}(\bx) | 0 \ra = 0
~, \label{Tidiri0} \feq
\begin{equation}\begin{split}
& {~}\hspace{3.8cm} \la 0 | \Tis_{i j}(\bx) | 0 \ra =
\la 0 | \Tis_{j i}(\bx) | 0 \ra = \\
& = \mm^\s \!\l[\!\Big({1\over 4} - \xi\Big) \de_{i j}
\Big(\!\Dir_{{\s - 1 \over 2}}(\bx,\by) -
(\partial^{\,x^\ell}\!\partial_{y^\ell}\!+\!V(\bx))
\Dir_{{\s + 1 \over 2}}(\bx,\by) \Big) \, + \r. \\
& \hspace{5cm} \l. + \l(\!\Big({1\over 2} - \xi\Big)\partial_{x^i y^j}
- \xi\,\partial_{x^i x^j}\!\r)\!
\Dir_{{\s + 1 \over 2}}(\bx,\by) \r]_{\by = \bx}
\label{Tidirij}
\end{split}\end{equation}
($\la 0|\Tis_{\mu \nu}(\bx)|0\ra$ is short for $\la 0|\Tis_{\mu\nu}(t,\bx)|0\ra$;
indeed, the VEV does not depend on $t$). Of course, the map $\Om \to \complessi$,
$\bx \mapsto \la 0|\Tis_{\mu \nu}(\bx)|0\ra$ possesses the same regularity as the
functions $\bx \in \Om \mapsto \Dir_{\s\pm 1 \over 2}(\bx,\bx),
\partial_{zw}\Dir_{\s+1 \over 2}(\bx,\bx)$ ($z,w$ any two spatial variables);
so, due to the previously mentioned results, $\bx \mapsto \la 0|\Tis_{\mu \nu}(\bx)|0\ra$
is continuous on $\ov{\Om}$ for $\Re \s > d+1$. \parn
Assume that $\Dir_{\s - 1 \over 2}$ and $\partial_{zw}\Dir_{\s + 1 \over 2}$
(for $z, w$ any two spatial variables) have analytic continuations regular at
$\s = 0$ (which happens for the configuration considered in this paper); then,
we can define
\beq \Dir_{\pm{1 \over 2}}(\bx,\by) :=
\Dir_{\s \pm 1 \over 2}(\bx,\by)\Big|_{\s= 0} \,, \label{Dik} \feq
\beq \partial_{z w} \Dir_{{1 \over 2}}(\bx,\by) :=
\partial_{z w} \Dir_{\s + 1 \over 2}(\bx,\by)\Big|_{\s= 0}\,. \label{Dikzw} \feq
The renormalized stress-energy VEV $\la 0|\Ti_{\mu \nu}(\bx)|0\ra_{ren} :=
\la 0|\Tis_{\mu \nu}(\bx)|0\ra|_{\s = 0}$ (see Eq. \rref{pri}) can be expressed as
\beq {~}\hspace{-0.3cm} \la 0 | \Ti_{0 0}(\bx)|0\ra_{ren}\! =
\!\l[\!\l(\!\frac{1}{4}\!+\!\xi\!\r)\!\Dir_{\!-{1\over 2}}(\bx,\by)
\!+\!\l(\!\frac{1}{4}\!-\!\xi\!\r)\!
(\partial^{x^\ell}\!\partial_{y^\ell}\!+\!V(\bx))
\Dir_{\!+{1 \over 2}}(\bx,\by)\r]_{\by = \bx} \!,\! \label{Tidir00R} \feq
\beq \la 0 | \Ti_{0 j}(\bx) | 0 \ra_{ren} =
\la 0 | \Ti_{j 0}(\bx) | 0 \ra_{ren} = 0 ~, \label{Tidiri0R} \feq
\begin{equation}\begin{split}
& \hspace{4.1cm} \la 0 | \Ti_{i j}(\bx) | 0 \ra_{ren} =
\la 0 | \Ti_{j i}(\bx) | 0 \ra_{ren} = \\
& = \!\l[\!\Big({1\over 4} - \xi\Big) \de_{i j}
\Big(\!\Dir_{\!-{1 \over 2}}(\bx,\by) -
(\partial^{\,x^\ell}\!\partial_{y^\ell}\!+\!V(\bx))
\Dir_{\!+{1 \over 2}}(\bx,\by) \Big)\, + \r. \\
& \hspace{5.4cm} \l. + \l(\!\Big({1\over 2} - \xi\Big)\partial_{x^i y^j}
- \xi\,\partial_{x^i x^j}\!\r)\!
\Dir_{\!+{1 \over 2}}(\bx,\by) \r]_{\by = \bx} . \label{TidirijR}
\end{split}\end{equation}
(More generally, in presence of a singularity at $\s = 0$ we should
consider the regular parts of the functions $\mm^\s \Dir_{\s - 1 \over 2}$,
$\mm^\s \partial_{zw}\Dir_{\s + 1 \over 2}$).
\subsection{The heat kernel.} For $\t \in [0,+\infty)$, this is given by
\beq K(\t\,;\bx,\by) := e^{-\t\AA}(\bx,\by) =
\int_\KK dk\;e^{- \t\,\om_k^2}\,\Fk(\bx)\Fkc(\by) ~. \label{eqheat} \feq
If $\Om$ is a bounded with smooth boundary $\partial\Om$ and $\AA = -\Delta + V$
($V$ smooth) is strictly positive, the map $K(\t\,;~,~) :\ov{\Om} \times \ov{\Om} \to \complessi$,
$(\bx,\by) \mapsto K(\t\,;\bx,\by)$ is continuous along with all its partial
derivatives of any order for all $\t > 0$\,; moreover, the eigenfunction expansion
in Eq. \rref{eqheat} converges absolutely and uniformly (the same holds for all
the corresponding derivatives).
\vspace{-0.4cm}
\subsection{The Dirichlet kernel as Mellin transform of the heat kernel.}
For suitable values of $s\!\in\!\complessi$ (see Part I), there holds
\beq \Dir_s(\bx,\by) = {1 \over \Ga(s)} \int_0^{+\infty}
\!d\t \; \t^{s-1}\,K(\t\,;\bx,\by) ~. \label{DirHeat} \feq
\vspace{-0.7cm}
\subsection{The case of product domains. Factorization of the heat kernel.} \label{prodDom}
Let $\AA := -\Delta + V$ and consider the case where
\beq \Om = \Om_1 \times \Om_2 \ni \bx = (\bx_1, \bx_2)\,,
\by = (\by_1,\by_2) ~, \label{omfact} \feq
\beq V(\bx) = V_1(\bx_1) + V_2(\bx_2) \feq
($\Om_a\!\subset\!\reali^{d_a}$ is an open subset, for $a \in \{1,2\}$;
$d_1\!+\!d_2=d$); assume the boundary conditions on $\partial \Om$
to arise from suitable boundary conditions prescribed separately
on $\partial \Om_1$ and $\partial \Om_2$ so that,
for $a=1,2$, the operators
\beq \AA_a := - \Delta_a + V(\bx_a) \feq
(with $\Delta_a$ the Laplacian on $\Om_a$) are selfadjoint and strictly
positive in $L^2(\Om_a)$. Then, the Hilbert space $L^2(\Om)$ and the
operator $\AA$ can be represented as
\beq L^2(\Om) = L^2(\Om_1) \otimes L^2(\Om_2) ~, \qquad
\AA = \AA_1 \otimes \uno + \uno \otimes \AA_2 ~. \label{prodOmAA}\feq
This implies, amongst else, that the heat kernels $K(\t\,;\bx,\by) :=
e^{\t \AA}(\bx, \by)$, $K_a(\t\,;\bx_a,\by_a)$ $:= e^{\t \AA_a}(\bx_a,\by_a)$
($a = 1,2$) are related by
\beq K(\t\,;\bx,\by) = K_1(\t\,;\bx_1,\by_1)\,K_2(\t\,;\bx_1,\bx_2) ~. \label{prodK} \feq
Similarly, writing $K(\t)$, $K_a(\t)$ ($a \in\{1,2\}$) for the heat
traces of $\AA$ and $\AA_a$ ($a\in\{1,2\}$), respectively, we have
\beq K(\t) = K_1(\t)\,K_2(\t) ~. \label{prodKTr} \feq
The above considerations have obvious extensions to product domains
with more than two factors.
\subsection{Pressure on the boundary.} \label{pressuretmunu}
This is the force per unit area produced by the quantized field
inside $\Om$ at a point $\bx \in \partial\Om$. We first consider,
for $\Re \s$ large, the \textsl{regularized pressure} $\bp^\s(\bx)$
with components
\beq p^\s_i(\bx) := \la 0|\Tis_{i j}(\bx)|0\ra \, n^j(\bx) ~; \label{press1} \feq
here and in the remainder of this paper, $\bn(\bx) \equiv (n^i(\bx))$
is the unit outher normal at $\bx\in\partial\Om$. For Dirichlet
boundary conditions the above definition implies
\beq p^{\s}_i(\bx) = \mm^\s \l[\l(\!-{1 \over 4}\,\de_{ij}\,
\partial^{x^\ell}\!\partial_{y^\ell} + {1 \over 2}\,\partial_{x^i y^j}\!\r)\!
\Dir_{\s+1 \over 2}(\bx,\by)\r]_{\by=\bx} n^j(\bx) ~. \label{pEeT} \feq
We can define the \textsl{renormalized pressure} by analytic continuation as
\beq p^{ren}_i(\bx) := p^\s_i(\bx)\Big|_{\s = 0} \label{preren} \feq
(more generally, one should take the regular part if $\s = 0$
is a singular point; this will never happen in the present paper).
Alternatively, we could put
\beq p^{ren}_i(\bx) := \l(\lim_{\bx'\in\Om, \bx'\to\bx}
\la 0|\Ti_{ij}(\bx')|0\ra_{ren} \r) n^j(\bx) ~. \label{alt} \feq
Prescriptions \rref{preren} \rref{alt} do not always agree (for a
counterexample, see Section 5 of Part II \cite{PII} of this series
of papers). In Part I we conjectured that the two approaches agree
when both of them give a finite result; this conjecture is true in
the present case of a box as shown in subsection \ref{ACBoxPres}).
\salto
Contrary to the previous works of this series \cite{PI,PII,PIII}, here
we \textsl{first} discuss the pressure and \textsl{next} pass to the
total energy; we make this choice because quite different computation
techniques are employed for the stress-energy VEV and for the pressure,
on the one hand, and for the total energy and integrated force,
on the other hand.
\vspace{-0.4cm}
\subsection{The total energy.}\label{TotEnSub} The \textsl{zeta-regularized
total energy} is
\beq \EE^\s := \int_\Om d\bx\; \la 0|\Tis_{00}(\bx)|0\ra = E^\s + B^\s ~; \label{EEtot}\feq
the second equality is proved after defining the \textsl{regularized bulk}
and \textsl{boundary energies}, which are
\beq E^\s := {\mm^\s\! \over 2} \int_\Om d\bx\;\Dir_{{\s - 1\over 2}}(\bx,\bx)
= {\mm^\s\! \over 2}\; \Tr\,\AA^{{1 -\s \over 2}} ~, \label{defEs}\feq
\beq B^\s := \mm^\s \l(\!{1 \over 4}-\xi\!\r)\! \int_{\partial \Om}\!
d a(\bx)\,\l.{\partial \over \partial n_{\by}}\,
\Dir_{{\s + 1\over 2}}(\bx,\by)\r|_{\by = \bx} ~. \label{defBs} \feq
One has $B^\s = 0$ for $\Om$ bounded and either Dirichlet or
Neummann boundary conditions on $\partial\Om$. \parn
Assuming the functions \rref{defEs} \rref{defBs} to be finite and
analytic for suitable $\s \in \complessi$, we define the renormalized
energies by the restricted (or generalized) zeta approach; for example, we put
\beq E^{ren} := E^\s \Big|_{\s=0} \label{ERenIV} \feq
(again, if a singularity appears at $\s = 0$ we should take the regular part).
\vspace{-0.4cm}
\subsection{Forces on the boundary.} \label{force}
Let $\Op\!\subset\!\partial\Om$; we consider (for $\Re\s$ large)
the \textsl{regularized integrated force} acting on $\Op$, i.e.,
\beq \fo_\Op^\s := \int_{\Op} da(\bx)\;\bp^\s(\bx)
\label{Freg} \feq
($\bp^\s(\bx)$ is the regularized pressure of Eq. \rref{press1}).
We can define the \textsl{renormalized total force} on $\Op$ as
\beq \fo_\Op^{ren} := \fo_\Op^\s\Big|_{\s = 0} \label{Fren} \feq
(once more, the regular part should be taken if $\s = 0$ is a
singular point, a variation which we shall never employ in the
present paper). Alternatively, we could put
\beq \fo_\Op^{ren} := \int_{\Op} da(\bx)\;\bp^{ren}(\bx) ~,
\label{FrenAlt} \feq
($\bp^{ren}(\bx)$ is the renormalized pressure, defined according to
either Eq. \rref{preren} or Eq. \rref{alt}). In general the alternatives
\rref{Fren} \rref{FrenAlt} give different results.
\section{The case of a massless field inside a box} \label{secBox}
\subsection{Introducing the problem.} In the present section we analyse
the model of a massless scalar field confined within a $d$-dimensional
box, with no external potential; more precisely, we assume
\beq \Om = \times_{i =1}^d (0,a_i) \quad
\mbox{with $a_i > 0$ for $i \in \{1,...,d\}$}~, \qquad V = 0~. \label{boxconf} \feq
The boundary $\partial\Om$ of the spatial domain is composed by the sides
\begin{equation} \begin{split}
& \pi_{\nn,\an} := \{\bx \in \reali^d ~|~ \mbox{$x^\nn = \an\,a_p$\,,
$x^i \in [0,a_i]$ for $i \neq \nn$}\} \\
& \hspace{1.7cm} \mbox{for $\nn \in \{1,...,d\}$, $\an \in \{0,1\}$} ~; \label{boxSides}
\end{split}\end{equation}
for the sake of simplicity, we restrict attention to the case where
the field fulfills Dirichlet boundary conditions on each one of these
sides, meaning that
\beq \Fi(t,\bx) = 0 \quad \mbox{for all $t \in \reali$, $\bx \in \pi_{\nn,\an}$
($\nn \in \{1,...,d\}$, $\an \in \{0,1\}$)}~. \feq
As a matter of fact, all the results to be reported in the following
could be generalized to the case of Neumann or periodic boundary conditions,
possibly including cases where different boundary conditions are prescribed
on different sides of the box; moreover, the methods to be presented
could be adapted with little effort to deal with the cases of a massive
scalar field ($V = m^2$) and of a slab configuration (see Part I) where
\beq \Om = \Om_1 \times \reali^{d_2} \,, \quad \Om_1 = \times_{i=1}^{d_1}
(0,a_i) \;\subset\;\reali^{d_1} \qquad (d_1+d_2 = d) ~. \feq
None of these generalizations will be considered in the present paper.
\vspace{-0.4cm}
\subsection{The heat kernel.} \label{heatBox}
Similarly to the model with a harmonic potential considered in our previous
work \cite{PIII}, in the present setting we are dealing with a product domain
configuration, in the sense of subsection \ref{prodDom}. Working in standard
Cartesian coordinates $(x^i)_{i = 1,...,d}$, the Hilbert space and the
fundamental operator $\AA := - \Delta$ can berepresented as
\beq L^2(\Om) = \bigotimes_{i=1}^d L^2(0,a_i)~, \feq
\begin{equation}\begin{split}
& \AA = \AA_1 \otimes \uno \otimes ... \otimes \uno + ... +
\uno \otimes ... \otimes \uno \otimes \AA_d ~, \\
& \hspace{0.35cm} \AA_i := - {d^2 \over d x^2} \quad \mbox{on $L^2(0,a_i)$}
\qquad (i \in \{1,...,d\}) ~;
\end{split}\end{equation}
for each operator $\AA_i$ we assume the induced Dirichlet boundary conditions
in $x^i = 0$ and $x^i = a_i$\,. \parn
According to the general considerations of subsection \ref{prodDom}
(see, in particular, Eq. \rref{prodK}), in this situation the heat kernel
associated to $\AA$ factorizes; more precisely, we have
\beq K(\t\,;\bx,\by) = \prod_{i=1}^d K_i(\t\,; x^i, y^i) ~, \label{Kfact} \feq
where, for $i \in\{1,...,d\}$, $K_i$ indicates the heat kernel of $\AA_i$\,.
Hereafter we compute the heat kernels $K_i$, giving for all of them
two distinct representations; these are suited to describe, respectively,
the behaviour of the kernels $K_i$ for small and large $\t$ (in a sense to
be  made more precise in the following)
({\footnote{For further information about approximate evaluations of the heat
kernel for a box configuration, see, e.g., \cite{KApp3,KApp1,KApp2}.}}).
Clearly, each one of these representations yields, in turn, alternative
expressions for the total heat kernel \rref{Kfact}.
$\phantom{a}$\vspace{-1.4cm}\\
\subsubsection{First representation for the heat kernel (useful for large $\t$).}
Let us begin noting that, for any $i \in \{1,...,d\}$, a complete orthonormal set
of eigenfunctions $(F_{k_i})_{k_i \in \KK_i}$ for $\AA_i$ in $L^2(0,a_i)$,
with corresponding eigenvalues $(\om^2_{k_i})_{k_i \in \KK_i}$, is given by
\beq F_{k_i}(x^i)\!:=\!\sqrt{2 \over a_i}\,\sin(k_i x^i) \,,
\quad\! \om^2_{k_i} := k_i^2 \quad \mbox{for }\,k_i\!\in\!\KK_i\!\equiv\!
\l\{{n_i \pi \over a_i} \,\Big|\, n_i = 1,2,3,...\r\} . \label{Fki} \feq
Using the eigenfunction expansion \rref{eqheat}, we obtain for the
$1$-dimensional heat kernel $K_i$ the following expression
({\footnote{Let us mention that the series in Eq. \rref{KB1Large} could be
explicitly evaluated to yield
$$ K_i(\t\,;x^i,y^i) = {e^{-(x^i\!-y^i)^2 \over 4\t} \over \sqrt{4\pi \t}}\,
\te_3\Big(\!-i\,{a_i(x^i\!-\!y^i) \over 2\t}\,,e^{-{a_i^2 \over \t}}\Big)
- {e^{-(x^i\!+y^i)^2 \over 4\t} \over \sqrt{4\pi \t}}\,
\te_3\Big(\!-i\,{a_i(x^i\!+\!y^i) \over 2\t}\,,e^{-{a_i^2 \over \t}}\Big) ~, $$
where $\te_3(~,~)$ denotes one of the Jacobi elliptic theta functions;
see \cite{Calin} for more details.}}):
\beq K_i(\t\,;x^i,y^i) = {2 \over a_i} \sum_{n_i = 1}^{+\infty}
e^{-{n_i^2 \pi^2\!\over a_i^2}\,\t} \sin\!\Big({n_i \pi\over a_i}\,y^i\Big)
\sin\!\Big({n_i \pi \over a_i}\,x^i\Big) ~. \label{KB1Large} \feq
The above relation, along with Eq. \rref{Kfact}, yields
\beq K(\t\,;\bx,\by) = {2^d \over a_1\,...\,a_d} \sum_{\bn \in \naturali^d}
e^{-\om_\bn^2 \t} \;\CB_\bn(\bx,\by) ~, \label{KBLarge}\feq
where, for the sake of brevity, we put
\beq \naturali := \{1,2,3,...\} ~, \label{defCC} \feq
$$ \om_\bn^2 := \sum_{i = 1}^d {n_i^2 \pi^2 \over a_i^2} ~,
\quad \CB_\bn(\bx,\by) := \prod_{i=1}^d \sin\!\l({n_i \pi \over a_i}\,y^i\r)\!
\sin\!\l({n_i \pi\over a_i}\, x^i\r) \quad \mbox{for $\bn := (n_i)_{i=1,...,d}$}~. $$
Eq. \rref{KBLarge} is easily seen to give a large $\t$ expansion for
the heat kernel $K(\t\,;\bx,\by)$ of $\AA$; with this, we mean that the
series over $\bn \in \naturali^d$ in the cited equation is mainly determined
by the terms corresponding to small values of $n_i$, for $i \in \{1,...,d\}$.
\vspace{-0.4cm}
\subsubsection{Second representation for the heat kernel (useful for small $\t$).}
Let us move on and note that, with little effort (namely, writing the sines
in terms of complex exponentials), we can rephrase Eq. \rref{KB1Large} as
\beq K_i(\t\,;x^i,y^i) = {1 \over 2 a_i} \sum_{n_i = -\infty}^{+\infty}
e^{-{n_i^2 \pi^2\!\over a_i^2}\,\t} \,\Big[e^{i{n_i \pi \over a_i}\,(x^i-y^i)}
- e^{i{n_i \pi \over a_i}\,(x^i+y^i)}\Big] ~. \label{smallK1}\feq
Using the Poisson summation formula
({\footnote{The Poisson summation formula states that, for any sufficiently
regular function $f: \reali \to \complessi$, there holds
$$ \sum_{n = -\infty}^{+\infty} f(n) = \sum_{h = -\infty}^{+\infty} \hat{f}(h)~, \quad
\mbox{where} \quad \hat{f}(h) := \int_{-\infty}^{+\infty}\! dz\;e^{-2i\pi h z}f(z) $$
($\hat{f}$ is, essentially, the Fourier transform of $f$).
Eq. \rref{13} gives $\hat{f}$ in the case we are considering.}})
and noting that
\begin{equation}\begin{split}
& \int_{-\infty}^{+\infty}\!\!\!dz\;e^{-{\pi^2 \over a_i^2}\,\t\,z^2}
e^{i{\pi \over a_i}((x^i\pm y^i) - 2 a_i h_i)z} = {a_i \over \sqrt{\pi\t}}\;
e^{-{(2a_i h_i-(x^i\pm y^i))^2 \over 4\t}} \\
& \hspace{1.7cm} \mbox{for\; $h_i\!\in\!\interi$\; with\;
$\interi := \{0,\pm 1,\pm 2,...\}$} ~, \label{13}
\end{split}\end{equation}
it follows from Eq. \rref{smallK1} that
(\footnote{The very same result of Eq. \rref{KB1Small} could be obtained
via the method of reflections, starting with the heat kernel associated
to the laplacian $-\partial_{x^1 x^1}$ on $\reali$\,, i.e.
$$ K(\t\,;\bx,\by) = {1 \over (4\pi\t)^{d/2}}\; e^{-{|\bx-\by|^2\! \over 4\t}} ~. $$})
\beq K_i(\t\,;x^i,y^i) = {1 \over \sqrt{4\pi \t}} \sum_{h_i = - \infty}^{+\infty}
\Big[e^{-{(2a_i h_i-(x^i - y^i))^2 \over 4\t}}
- e^{-{(2a_i h_i-(x^i+y^i))^2 \over 4\t}} \Big] ~. \label{KB1Small}\feq
The above identity can be rephrased as follows:
\beq K_i(\t\,;x^i,y^i) = {1 \over \sqrt{4\pi \t}}
\sum_{h_i \in \interi,\,l_i \in \{1,2\}}
\de_{l_i}\; e^{-{1 \over \t}\,a_i^2 (h_i - \DD_{l_i}(x^i,y^i))^2} \label{KB1Small2}\feq
where
\beq \de_{l_i} := \l\{\!\!\barray{cc} 1 &
\mbox{for $l_i = 1$} \\ -1 & \mbox{for $l_i = 2$} \farray\r. ~,
\qquad \DD_{l_i}(x^i,y^i) := \l\{\!\!\barray{cc} {x^i - y^i \over 2 a_i} &
\mbox{for $l_i = 1$} \\ {x^i + y^i \over 2 a_i} & \mbox{for $l_i = 2$} \farray\r.\!.
\label{bkldef1} \feq
Eq. \rref{KB1Small2}, along with Eq. \rref{Kfact}, allows us to infer for the
heat kernel the alternative representation
\beq K(\t\,;\bx,\by) = {1 \over (4\pi\t)^{d/2}} \sum_{\bh \in \interi^d,\,
\bl \in \{1,2\}^d} \de_\bl\; e^{-{1 \over \t}\,\bhl(\bx,\by)} ~, \label{KBSmall} \feq
where, for simplicity of notation, we have put (recall Eq. \rref{bkldef1})
\begin{equation}\begin{split}
& \bh := (h_i)_{i=1,...,d} ~, \quad \bl := (l_i)_{i=1,...,d} ~, \quad
\de_\bl := \prod_{i=1}^d \de_{l_i} ~, \\
& \hspace{1.1cm} \bhl(\bx,\by) := \sum_{i=1}^d a_i^2 (h_i\!-\!\DD_{l_i}(x^i,y^i))^2 ~.
\label{bkldef}
\end{split}\end{equation}
Notice that, for small $\t$, the sum of the series appearing in Eq. \rref{KBSmall}
is mainly determined by the terms corresponding to small values of $|h_i|$,
for $i \in \{1,...,d\}$; thus, the mentioned equation yields a small $\t$ expansion
for the heat kernel $K(\t\,;\bx,\by)$\,.
\vspace{-0.4cm}
\subsubsection{Considerations on the sign of some coefficients.}
Before moving on, let us emphasize a number of facts on the expansions
\rref{KBLarge} \rref{KBSmall}; we will resort
to them in the following subsections, when performing the analytic
continuation of the Dirichlet kernel and of its derivatives. \salto
i) On the one hand, we have
\beq \om_\bn^2 > 0 \quad \mbox{for all $\bn \in \naturali^d$} ~. \label{ombn}\feq
ii) On the other hand, notice that
\beq \bhl(\bx,\by) \geqs 0 \quad \mbox{for all $\bh \in \interi^d$,
$\bl \in \{1,2\}^d$, $\bx,\by \in \Om$} ~. \label{bkl1}\feq
In particular, since
\beq \DD_{l_i}(x^i,y^i) \in \l\{\!\barray{cc} \l[-{1 \over 2},{1 \over 2}\r]
& \mbox{for $l_i = 1$} \\ \l[0,1\r] & \mbox{for $l_i = 2$} \farray\r. \feq
(see the definition of $\DD_{l_i}$ in Eq. \rref{bkldef1}), it follows that
\begin{equation}\begin{split}
& \hspace{4.8cm} \bhl(\bx,\by) = 0 \quad \Leftrightarrow \\
& \Leftrightarrow \quad \mbox{for each $i\!\in\!\{1,...,d\}$\,, one has} ~
\l\{\!\!\barray{ll}
h_i = 0, ~ l_i = 1, ~ y^i = x^i \in [0,a_i]  \\
\mbox{or}~ h_i = 0, ~ l_i = 2, ~ y^i = x^i = 0 \\
\mbox{or}~ h_i = 1, ~ l_i = 2, ~ y^i = x^i = a_i \farray\r. ; \label{bkl2}
\end{split}\end{equation}
let us stress that this implies, in particular, $\bhl(\bx,\by) \neq 0$ for $\bx \neq \by$.
\vspace{-0.4cm}
\subsection{The Dirichlet kernel.}\label{ACBox} Consider the integral
representation \rref{DirHeat} for the Dirichlet kernel $\Dir_s$ in terms
of the heat kernel $K$ of $\AA$\,. We are going to construct the analytic
continuation of $\Dir_s$ in the style of Minakshisundaram (see \cite{Minak1});
to this purpose, let us fix arbitrarily
\beq T \in (0,+\infty) \feq
and re-express the cited integral representation as
\beq \Dir_s(\bx,\by) = \Dir^{(>)}_s(\bx,\by) + \Dir^{(<)}_s(\bx,\by)~,
\qquad \mbox{where} \label{DirHeatBox} \feq
\beq \Dir^{(>)}_s(\bx,\by) := {1 \over \Ga(s)} \int_T^{+\infty}\! d\t \;
\t^{s-1}\, K(\t\,;\bx,\by) ~, \label{DirSup}\feq
\beq \Dir^{(<)}_s(\bx,\by) := {1 \over \Ga(s)} \int_0^T d\t \;
\t^{s-1}\, K(\t\,;\bx,\by)  \label{DirInf}\feq
(notice that $\Dir^{(>)}_s$ and $\Dir^{(<)}_s$ depend
on $T$, but their sum $\Dir_s$ does not!).
The idea we are going to pursue in the following is to substitute into
Eq.s \rref{DirSup} \rref{DirInf}, respectively, the large and small $\t$
expansions \rref{KBLarge} \rref{KBSmall} for the heat kernel $K$
({\footnote{\label{footInt}In these manipulations (and in some related computations)
we often take for granted that certain series can be integrated or
differentiated term by term. In all cases under analysis, rigorous
justifications could be given using the Lebesgue dominated convergence
theorem or the Fubini-Tonelli theorem, but we will not go into the
details; the estimates on the series in Appendices \ref{AppBox} and
\ref{AppC} could be connected to such rigorous proofs.}}).
\subsubsection{Series expansion and analytic continuation of $\boma{\Dir^{(>)}_s(\bx,\by)}$.}
Using Eq.s \rref{KBLarge} \rref{DirSup}, we readily infer
\beq \Dir^{(>)}_s(\bx,\by) = {2^d \over a_1...a_d\,\Ga(s)}
\sum_{\bn \in \naturali^d}\CB_\bn(\bx,\by) \int_T^{+\infty}\!\!d\t \;
\t^{s-1}\,e^{-\om_\bn^2 \t} ~. \feq
Concerning the integral over $(T,+\infty)$, via the change of variable
$\tau := \om_\bn^2 \t$ (recall that $\om_\bn^2 > 0$ for all $\bn \in
\naturali^d$; see Eq. \rref{ombn}), we obtain
\beq \int_T^{+\infty}\!\!\!d\t \; \t^{s-1}\,e^{-\om_\bn^2 \t} =
\om_\bn^{-2s}\! \int_{\om_\bn^2 T}^{+\infty}\!\!\!d\tau \; \tau^{s-1}\,e^{-\tau} =
\om_\bn^{-2s}\,\Ga(s,\om_\bn^2\,T) \quad \mbox{for $s\!\in\!\complessi$}\,,
\label{intGa} \feq
where the last passage contains the upper incomplete gamma function
(see \cite{NIST}, p.174, Eq.8.2.2)
\beq \Ga(s,z) := \int_{z}^{+\infty}\!\! dw\; e^{-w}\,w^{s-1} \qquad
(s \in \complessi,\, z \in (0,+\infty)) ~. \label{UpInGa} \feq
Summing up, we have
\beq \Dir^{(>)}_s(\bx,\by) = {2^d \over a_1...a_d\,\Ga(s)}
\sum_{\bn \in \naturali^d}\om_\bn^{-2s}\,\Ga(s,\om_\bn^2\,T)\;
\CB_\bn(\bx,\by) ~. \label{DirSupExp} \feq
The above expression can be used to evaluate derivatives of any order
of the function $\Dir^{(>)}_s(\bx,\by)$; for example, for any pair of
spatial variables $z,w$, we obtain
\beq \partial_{zw}\Dir^{(>)}_s(\bx,\by) = {2^d \over a_1...a_d\,\Ga(s)}
\sum_{\bn \in \naturali^d}\om_\bn^{-2s}\,\Ga(s,\om_\bn^2\,T)\;
\partial_{zw}\CB_\bn(\bx,\by) ~. \label{DzwDirSupExp} \feq
Let us anticipate that the series in the right-hand sides of Eq.s
\rref{DirSupExp} \rref{DzwDirSupExp} converge for all $s \in \complessi$,
even for $\by = \bx$ (see subsection \ref{NumBox} and Appendices
\ref{AppBox}, \ref{AppC} for more details); so, Eq.s \rref{DirSupExp}
\rref{DzwDirSupExp} yield automatically the analytic continuations
of the maps $s \mapsto \Dir^{(>)}_s(\bx,\by)$, $\partial_{zw}\Dir^{(>)}_s(\bx,\by)$
to the whole complex plane.
$\phantom{a}$\vspace{-0.9cm} \\
\subsubsection{Series expansion and analytic continuation of $\boma{\Dir^{(<)}_s(\bx,\by)}$.}\label{subPole}
Proceeding similarly to what we did for the function $\Dir_s^{(>)}(\bx,\by)$,
we can use Eq.s \rref{KBSmall} \rref{DirInf} to deduce
\beq \Dir^{(<)}_s(\bx,\by) = {1 \over (4\pi)^{d/2}\Ga(s)}
\sum_{\bh \in \interi^d,\,\bl \in \{1,2\}^d} \de_\bl \int_0^T d\t\;
\t^{s - {d \over 2} - 1}\, e^{-{1 \over \t}\,\bhl(\bx,\by)} ~.
\label{Dirmin} \feq
To go on, for $\be \geqs 0$ and $s \in \complessi$ satisfying the conditions
in the forthcoming Eq. \rref{PropPP1}, let us introduce the function
\beq \PP_{s}(\be) := \int_0^1 d\tau\;\tau^{s-1}e^{-{\be \over \tau}} ~;
\label{defPP} \feq
it is easy to check that
\beq \PP_s(\be) = \l\{\!\barray{ll}
s^{-1} & \mbox{for $\be = 0$, $\Re s > 0$} \\
\be^s\,\Ga(-s,\be) & \mbox{for $\be > 0$} \farray\r.  \label{PropPP1}\feq
where, again, $\Gamma(~,~)$ denotes the upper incomplete gamma function.
Starting from Eq. \rref{defPP}, we find as well that
\beq \partial_\be^\ell \PP_s(\be) = (-1)^\ell \,\PP_{s-\ell}(\be) \quad\;
\mbox{for all $\ell \in \{0,1,2,...\}$} ~. \label{PropPP2} \feq
Let us return to Eq. \rref{Dirmin} (recalling that $\bhl(\bx,\by) \geqs 0$
for all $\bh \in \interi^d$, $\bl \in \{1,2\}^d$; see Eq. \rref{bkl1}) and
make therein the change of variable $\tau := \t/T$; comparing with the
definition \rref{defPP} of $\PP_\bullet(~)$, we get
\beq \Dir^{(<)}_s(\bx,\by) = {T^{s - {d \over 2}} \over (4\pi)^{d/2}\Ga(s)}
\sum_{\bh \in \interi^d,\,\bl \in \{1,2\}^d} \de_\bl ~
\PP_{s - {d \over 2}}\!\l({\bhl(\bx,\by) \over T}\r) \,. \label{DirInfExp}\feq
The above result can be used, along with Eq. \rref{PropPP2}, to infer
analogous representations for the derivatives of any order of $\Dir^{(<)}_s$;
for example, if $z,w$ are any two spatial variables, derivating term by term
Eq. \rref{DirInfExp} we obtain
\beq \partial_{zw}\Dir^{(<)}_s(\bx,\by) = \label{DzwDirInfExp} \feq
$$ {T^{s-{d\over 2}-2} \over (4\pi)^{d/2}\Ga(s)}
\sum_{\bh \in \interi^d,\,\bl \in \{1,2\}^d}\! \de_\bl\!
\l[\PP_{s - {d \over 2} - 2}\!\l({\bhl\over T}\r)\!
\partial_z\bhl\,\partial_w \bhl - T\,\PP_{s - {d \over 2} - 1}\!\!
\l({\bhl \over T}\r)\! \partial_{zw} \bhl \r]\!(\bx,\by) ~. $$
Let us point out that, due to the results reported in Eq. \rref{bkl2}, we have
\beq \barray{cc} \mbox{$\bhl(\bx,\by) = 0$ only for $\by = \bx$ and for} \\
\mbox{a finite number of terms in the series of Eq.s \rref{DirInfExp}
\rref{DzwDirInfExp}\,.} \farray \label{bkl0} \feq
The above mentioned terms of Eq.s \rref{DirInfExp} \rref{DzwDirInfExp} deserve
special attention, and we must use the first equality in \rref{PropPP1} to
evaluate them; on the contrary, for the infinitely many terms with $\bhl(\bx,\by)>0$\,,
the second line in Eq. \rref{PropPP1} gives expressions in terms of upper
incomplete gamma functions. In this way we obtain
\begin{equation}\begin{split}
& \hspace{0.7cm} \Dir^{(<)}_s(\bx,\by) = {T^{s - {d \over 2}} \over
(4\pi)^{d/2}\Ga(s)(s-{d \over 2})}
\l(\sum_{\substack{\bh \in \interi^d,\,\bl \in \{1,2\}^d \\
\mbox{\scriptsize{s.t.} }\bhl(\bx,\by)=0 }} \de_\bl \r) + \\
& + \;{1 \over (4\pi)^{d/2}\Ga(s)} \sum_{\substack{\bh \in \interi^d,
\,\bl \in \{1,2\}^d \\ \mbox{\scriptsize{s.t.} }\bhl(\bx,\by) > 0}}
\de_\bl\,\Bigg(\bhl^{s - {d \over 2}}\,
\Ga\!\l({d \over 2}\!-\!s\,,\,{\bhl \over T}\r)\!\!\Bigg)(\bx,\by) ~. \label{DirInfExp2}
\end{split}\end{equation}
Let us repeat that the first sum in the above expression contains finitely
many terms. Notice that the term in the first line of Eq. \rref{DirInfExp2}
is related to the first equality in Eq. \rref{PropPP1} which, in principle,
would require $\Re s > d/2$\,; however, this term makes sense for all complex
$s$ except $s = d/2$\,, where a simple pole appears. Moreover, the series
in the second line of Eq.s \rref{DirInfExp2} can be proved to converge for
any complex $s$; we defer a comprehensive discussion of this statement to
subsection \ref{NumBox} (see also Appendix \ref{AppBox}). \parn
In view of the above remarks, Eq. \rref{DirInfExp2} gives automatically the
analytic continuation of $\Dir^{(<)}_s(\bx,\by)$ to a meromorphic function
of $s$ on the whole complex plane, with a simple pole singularity only at
\beq s = d/2 \feq
for $\bx,\by$ such that the first sum in Eq. \rref{DzwDirInfExp2} is
non-empty; this happens only for $\by = \bx$ due to Eq. \rref{bkl2}. \salto
A similar analysis can be made for the derivatives of $\Dir^{(<)}_s$.
For example, if $z,w$ are any two spatial variables, we obtain the
following expression from Eq. \rref{DzwDirInfExp}:
\begin{equation}\begin{split}
& \partial_{zw}\Dir^{(<)}_s(\bx,\by) = -\,{T^{s-{d\over 2}-1} \over
(4\pi)^{d/2}\Ga(s)(s - {d \over 2} - 1)}\,\Bigg(\sum_{
\substack{\bh \in \interi^d,\,\bl \in \{1,2\}^d \\ \mbox{\scriptsize{s.t.} }
\bhl(\bx,\by) = 0}}\!\de_\bl\; \partial_{zw} \bhl(\bx,\by)\!\Bigg)\, +\! \\
& \hspace{0.3cm} +\, {1 \over (4\pi)^{d/2}\Ga(s)}
\sum_{\substack{\bh \in \interi^d,\,\bl \in \{1,2\}^d \\ \mbox{\scriptsize{s.t.} }
\bhl(\bx,\by) > 0}}\! \de_\bl
\l[\bhl^{s - {d \over 2} - 2} \l(\Ga\!\l({d \over 2}\!+\!2\!-\!s\,,{\bhl\over T}\r)\!
\partial_z\bhl\,\partial_w \bhl \; + \r.\r. \\
& \hspace{6cm} \l.\l. - \; \Ga\!\l({d \over 2}\!+\!1\!-\!s\,,{\bhl\over T}\r)\!
\bhl\,\partial_{zw} \bhl \r)\r]\!(\bx,\by) ~. \label{DzwDirInfExp2}
\end{split}\end{equation}
Again, the first of the above two sums is made of finitely many terms;
besides, contrary to what one could expect from Eq. \rref{DzwDirInfExp},
this sum contains no term with the first order derivatives $\partial_w \bhl(\bx,\by)$,
$\partial_z \bhl(\bx,\by)$ because they vanish if $\bhl(\bx,\by)=0$.
By considerations analogous to those made above for Eq. \rref{DirInfExp2}
(based on the convergence of the second sum for all $s \in \complessi$),
we infer that Eq. \rref{DzwDirInfExp2} gives automatically the analytic
continuation of $\partial_{z w}\Dir^{(<)}_s(\bx,\by)$ to a meromorphic
function of $s$ on the whole complex plane, with a simple pole singularity
only for $\by = \bx$ at
\beq s = d/2 + 1 ~. \feq
Let us stress that Eq.s \rref{DirInfExp2} \rref{DzwDirInfExp2} are
just the original Eq.s \rref{DirInfExp} \rref{DzwDirInfExp}, rewritten
separating the terms with $\bhl(\bx,\by)=0$ for a better understanding
of the behaviour with respect to $s$\,. In the sequel, even when considering
meromorphic continuations, we will always refer to the more concise
representations \rref{DirInfExp} \rref{DzwDirInfExp}.
\subsubsection{Conclusions for the Dirichlet kernel.}\label{BoxDirCont}
Using Eq. \rref{DirHeatBox} and expressions \rref{DirSupExp} \rref{DirInfExp}
for the functions $\Dir_s^{(>)}$, $\Dir_s^{(<)}$, respectively, we obtain
the analytic continuation of the full Dirichlet kernel $\Dir_s(\bx,\by)$
to a meromorphic function on the whole complex plane. Similar results hold
for the derivatives of the Dirichlet kernel (see Eq.s \rref{DzwDirSupExp}
\rref{DzwDirInfExp}). \parn
The only singularity of $\Dir_s(\bx,\by)$ is a simple pole for
\beq \by = \bx \quad \mbox{and} \quad s = d/2 ~, \label{exc} \feq
while $\partial_{zw}\Dir_s(\bx,\by)$ (for any pair of spatial
variables $z,w$) has a simple pole for
\beq \by = \bx \quad \mbox{and} \quad s = d/2 + 1 ~. \label{prolBox}\feq
In particular the analytic continuation of $\Dir_{\s-1 \over 2}(\bx,\by)|_{\by = \bx}$
and $\partial_{zw}\Dir_{\s+1 \over 2}(\bx,\by)|_{\by = \bx}$, required for
the evaluation of the regularized stress-energy VEV and pressure, are both
regular at $\s = 0$\,.
\vspace{-0.4cm}
\subsection{Convergence and remainder estimates for the series in Eq.s \rref{DirSupExp} \rref{DzwDirSupExp} and \rref{DirInfExp} \rref{DzwDirInfExp}.} \label{NumBox}
This subject is discussed in more detail in Appendix \ref{AppBox};
here we only report the main results. In the mentioned Appendix we
show that the series cited in the title of this subsection are
absolutely convergent; moreover, we derive fully quantitative remainder
estimates when these series are approximated by finite sums. \parn
To report here these estimates we need some notations, introduced
hereafter. First of all we put
\begin{equation}\begin{split}
& \hspace{1.3cm} \Am := \min_{i \in \{1,...,d\}} \{a_i\} ~, \qquad
\AM := \max_{i \in \{1,...,d\}} \{a_i\} ~; \\
& |\bz| := \l(\sum_{i=1}^d z_i^2\r)^{\!\!1/2} \qquad
\mbox{for $\bz = \bn \in \naturali^d$ or $\bz = \bh \in \interi^d$} ~. \label{deflL}
\end{split}\end{equation}
Besides, for $N\!\in\!(2\sqrt{d},+\infty)$, $\al\!\in\!(0,1)$,
$\be\!\in\!(0,+\infty)$, $\si,\rho \in \reali$, we set
\beq \HE^{(d)}_N(\al,\be;\si,\rho) := \label{defHEN}\feq
$$ {\pi^{d/2} \over (1\!-\!\al)^\si (\al\,\be)^{{d+\rho \over 2}}\,
\Ga({d \over 2})} \l({N\!-\!\sqrt{d} \over N\!-\!2\sqrt{d}}\r)^{\!\!\!d-1}
\!\Ga(\si,(1\!-\!\al)\be N^2)\;\Ga\!\l({d\!+\!\rho \over 2}\,;
\al\,\be (N\!-\!2\sqrt{d})^2\r) \,; $$
it can be shown that there holds the asymptotic expansion
\begin{equation}\begin{split}
& \HE^{(d)}_N(\al,\be;\si,\rho) = {\pi^{d/2} \,\be^{\si-2} e^{-4 \al\,\be d} \over \al(1\!-\!\al)\,
\Ga({d \over 2})}\; \,
e^{-\be N(N-4\al\be\sqrt{d})} N^{2\si+\rho+d-4} \,(1 + O(N^{-1})) \\
& \hspace{5.7cm} \mbox{for $N\to +\infty$} ~. \label{HENAsym}
\end{split}\end{equation}
Finally, we put
\beq C_{\Am,\AM}^{(d)}(\si,N) := \max\l[\l(\!\Am\l(1\!-\!{\sqrt{d}\over N}\r)\!\r)^{\!\!\!2\si}\!,
\l(\!\AM\l(1\!+\!{\sqrt{d}\over N}\r)\!\r)^{\!\!\!2\si}\r] . \label{finaset} \feq
Having introduced the above notations, in the next two subsections we
report the remainder estimates of Appendix \ref{AppBox} for the series
expansions of $\Dir_s^{(>)}$, $\Dir_s^{(<)}$ and of their derivatives.
In all cases the remainder is controlled by the function $H_N^{(d)}$;
due to the exponential decay of this function for large $N$
(see Eq. \rref{HENAsym}), good approximations of all the series under
investigation can be obtained by just summing the first few terms.
\vspace{-0.4cm}
\subsubsection{Estimates for the series \rref{DirSupExp} \rref{DzwDirSupExp}.}
Consider any $s \in \complessi$; keeping in mind Eq. \rref{DirSupExp},
for any $N \in (0,+\infty)$ let us write
\begin{equation}\begin{split}
& \hspace{2.1cm} \Dir^{(>)}_s(\bx,\by) = \Dir^{(>)}_{s,N}(\bx,\by)
+ R^{(>)}_{s,N}(\bx,\by) ~, \\
& \Dir^{(>)}_{s,N}(\bx,\by) := {2^d \over a_1...a_d\,\Ga(s)}
\sum_{\bn \in \naturali^d,\,|\bn| \leqs N}\om_\bn^{-2s}\,\Ga(s,\om_\bn^2\,T)\;\CB_\bn(\bx,\by) ~, \\
& R^{(>)}_{s,N}(\bx,\by) := {2^d \over a_1...a_d\,\Ga(s)}
\sum_{\bn \in \naturali^d,\,|\bn| > N}\om_\bn^{-2s}\,\Ga(s,\om_\bn^2\,T)\;\CB_\bn(\bx,\by) ~.
\label{DirSupEst}
\end{split}\end{equation}
This equation implies a similar representation for the (analytically continued)
derivatives of $\Dir^{(>)}_s$ in terms of the derivatives of $\Dir^{(>)}_{s,N}$
and $R^{(>)}_{s,N}$\,. For the remainder function $R^{(>)}_{s,N}(\bx,\by)$
and for its derivatives with respect to any two spatial variables $z,w$,
we have the following uniform estimates:
\begin{equation}\begin{split}
& \hspace{0.75cm} \l|R^{(>)}_{s,N}(\bx,\by)\r| \leqs {\max(\Am^{2\Re s},\AM^{2\Re s})
\over a_1...a_d\,\pi^{2\Re s} |\Ga(s)|}\;
\HE^{(d)}_N\!\l(\al,{\pi^2\,T \over \AM^2}\,;\Re s,-2\Re s\r) \\
& \mbox{for either \quad $\Re s\!\geqs\!0$, $N\!>\!2\sqrt{d}$} \qquad
\mbox{or} \quad \mbox{$\Re s\!<\!0$, $\dd{N\!>\!2\sqrt{d}\!
+\!{\AM \over \pi}\sqrt{|\Re s| \over \al\,T}}$} ~;
\label{RRSupEst}
\end{split}\end{equation}
\begin{equation}\begin{split}
& \l|\partial_{zw}R^{(>)}_{s,N}(\bx,\by)\r| \leqs {\max(\Am^{2\Re s},\AM^{2\Re s})
\over a_1...a_d\,\Am^2\,\pi^{2(\Re s-1)} |\Ga(s)|}\;
\HE^{(d)}_N\!\l(\al,{\pi^2\,T \over \AM^2}\,;\Re s,2(1\!-\!\Re s)\r) \\
& \hspace{0.15cm}\mbox{for either \qquad $\Re s\!\geqs\!1$, $N\!>\!2\sqrt{d}$}
\quad \mbox{or} \quad \mbox{$\Re s\!<\!1$, $\dd{N\!>\!2\sqrt{d}\!
+\!{\AM \over \pi}\sqrt{(1\!-\!\Re s)\over \al\,T}}$} ~. \label{DzwRRSupEst}
\end{split}\end{equation}
In the above, $\al$ is a parameter that can be freely chosen in $(0,1)$;
of course, the best choice is the one minimizing the right-hand sides
of Eq.s \rref{RRSupEst} and \rref{DzwRRSupEst}, which depends on the
other parameters (e.g., $N,T$) involved in these considerations.
\subsubsection{Estimates for the series \rref{DirInfExp} \rref{DzwDirInfExp}.}
Let $s \in \complessi$, and exclude the case \rref{exc}; keeping in mind
Eq. \rref{DirInfExp}, for any $N \in (0,+\infty)$ we put
\begin{equation}\begin{split}
& \hspace{2.2cm} \Dir^{(<)}_s(\bx,\by) = \Dir^{(<)}_{s,N}(\bx,\by)
+ R^{(<)}_{s,N}(\bx,\by) ~, \\
& \Dir^{(<)}_{s,N}(\bx,\by) := {T^{s - {d \over 2}} \over (4\pi)^{d/2}\Ga(s)}
\sum_{\bh \in \interi^d,\,|\bh| \leqs N,\,\bl \in \{1,2\}^d} \de_\bl\;
\PP_{s - {d \over 2}}\!\l({\bhl(\bx,\by) \over T}\r) ~, \\
& R^{(<)}_{s,N}(\bx,\by) := {T^{s - {d \over 2}} \over (4\pi)^{d/2}\Ga(s)}
\sum_{\bh \in \interi^d,\,|\bh| > N,\,\bl \in \{1,2\}^d} \de_\bl\;
\PP_{s - {d \over 2}}\!\l({\bhl(\bx,\by) \over T}\r) ~.
\label{DirInfEst}
\end{split}\end{equation}
The above equation can be used to derive similar representations for the
(analytically continued) derivatives of $\Dir^{(<)}_s$ in terms of the
derivatives of $\Dir^{(<)}_{s,N}$ and $R^{(<)}_{s,N}$,
with the exclusion of the case \rref{prolBox}\,. For the remainder
function $R^{(<)}_{s,N}$ and for its derivatives with respect to any two
spatial variables $z^i,w^j$ ($i,j \in \{1,...,d\}$), we have the following
uniform estimates:
\begin{equation}\begin{split}
& \l|R^{(<)}_{s,N}(\bx,\by)\r| \leqs {C^{(d)}_{\Am,\AM}(\Re s-{d \over 2},N)
\over \pi^{d/2}|\Ga(s)|}\; \HE^{(d)}_N\!\l(\al,{\Am^2(1\!-\!{\sqrt{d} \over N})^2\over T}\,;
{d \over 2}\!-\!\Re s, 2\Re s\!-\!d\r) \\
& \mbox{for either \quad $\dd{\Re s\!\leqs\!{d \over 2}}$, $N\!>\!2\sqrt{d}$}
\quad \mbox{or} \quad \mbox{$\dd{\Re s\!>\!{d \over 2}}$,
$\dd{N\!>\!3\sqrt{d}\!+\!{1 \over \Am}\sqrt{(\Re s\!-\!{d \over 2})
T \over \al}}$} ~; \label{RRInfEst}
\end{split}\end{equation}
\beq \l|\partial_{z^i w^j}R^{(<)}_{s,N}(\bx,\by)\r| \leqs \label{DzwRRInfEst} \feq
\begin{equation*}\begin{split}
& {1 \over \pi^{d/2}|\Ga(s)|}\!\!\l[\!\l(\!1\!+\!{\sqrt{d} \over N}\!\r)^{\!\!2}\!\!\AM^2
C_{\Am,\AM}^{(d)}\!\!\l(\!\Re s\!-\!{d \over 2}\!-\!2,N\!\r)\!
\HE^{(d)}_N\!\!\l(\!\al,{\Am^2 (1\!-\!{\sqrt{d} \over N})^2 \over T};
{d \over 2}\!+\!2\!-\!\Re s,2\Re s\!-\!d\!-\!2\!\r) \r. \\
& \hspace{0.4cm} \l. \phantom{\l(\!1\!+\!{\sqrt{d} \over N}\!\r)^{\!\!2}}
+{1 \over 2}\;\de_{ij}\, C_{\Am,\AM}^{(d)}\!\!\l(\!\Re s\!-\!{d \over 2}\!-\!1,N\!\r)\!
\HE^{(d)}_N\!\!\l(\!\al,{\Am^2 (1\!-\!{\sqrt{d} \over N})^2 \over T}\,;
{d \over 2}\!+\!1\!-\!\Re s,2\Re s\!-\!d\!-\!2\!\r)\!\r]
\end{split}\end{equation*}
$$ \mbox{for either \quad $\dd{\Re s\!\leqs\!{d \over 2}\!+\!1}$,
$N\!>\!2\sqrt{d}$} \quad \mbox{or} \quad
\mbox{$\dd{\Re s\!>\!{d \over 2}\!+\!1}$,
$\dd{N\!>\!3\sqrt{d}\!+\!{1 \over \Am}
\sqrt{(\Re s\!-\!{d\over 2}\!-\!1) T \over \al}}$} ~. $$
Again, the parameter $\al$ can be freely taken in $(0,1)$
and it is conventient to choose for it the value minimizing
the right-hand sides of Eq.s \rref{RRInfEst} \rref{DzwRRInfEst},
keeping into account the choices made for the other parameters
(in particular, $N$).
\vspace{-0.4cm}
\subsection{The stress-energy tensor.}\label{ACBoxTmn} Consider the
representations deduced in subsection \ref{ACBox} for the analytic
continuations of the Dirichlet kernel and of its derivatives.
Resorting to Eq.s (\ref{Tidir00}-\ref{Tidirij}), we obtain the
following expressions for the components of the regularized VEV of
the stress-energy tensor:
\beq \la 0 | \Tis_{\mu\nu}(\bx) | 0 \ra = T^{\s,(>)}_{\mu\nu}(\bx)
+ T^{\s,(<)}_{\mu\nu}(\bx) ~, \qquad \label{TmnBox1}\feq
where, for $\bullet$ equal to $>$ or $<$, $T^{\s,(\bullet)}_{\mu\nu}(\bx)$
has the espression corresponding to Eq.s (\ref{Tidir00}-\ref{Tidirij}),
with $\Dir_s$ replaced by $\Dir^{(\bullet)}_s$.
Thus, for $i,j \in \{1,...,d\}$, we have
\beq T^{\s,(\bullet)}_{0 0}(\bx) = \mm^\s \!\l[\l(\!\frac{1}{4}+\xi\!\r)\!
\Dir^{(\bullet)}_{{\s - 1\over 2}}(\bx,\by) + \l(\!\frac{1}{4}-\xi\!\r)\!
\partial^{x^\ell}\!\partial_{y^\ell}\Dir^{(\bullet)}_{{\s + 1 \over 2}}(\bx,\by)
\r]_{\by = \bx} , \label{Tidir00box} \feq
\beq T^{\s,(\bullet)}_{0 j}(\bx) = T^{\s,(\bullet)}_{j 0}(\bx) = 0 ~,
\label{Tidiri0box} \feq
\begin{equation}\begin{split}
& T^{\s,(\bullet)}_{i j}(\bx) = \mm^\s \!\l[\!\l(\!{1\over 4}-\xi\!\r)\!
\de_{i j} \Big(\!\Dir^{(\bullet)}_{{\s - 1 \over 2}}(\bx,\by) -
\partial^{\,x^\ell}\!\partial_{y^\ell}
\Dir^{(\bullet)}_{{\s + 1 \over 2}}(\bx,\by)\Big)\, + \r. \\
& \hspace{5cm} \l. + \l(\!\Big({1\over 2} - \xi\Big)\partial_{x^i y^j}
- \xi\,\partial_{x^i x^j}\!\r)\!
\Dir^{(\bullet)}_{{\s + 1 \over 2}}(\bx,\by) \r]_{\by = \bx} .
\label{Tidirijbox}
\end{split}\end{equation}
To proceed notice that, concerning the analyticity of the above functions,
there hold considerations analogous to those presented in subsection
\ref{BoxDirCont} for the Dirichlet kernel and its derivatives; in
particular, it appears that $\s = 0$ is a regular point for each
component of the regularized stress-energy VEV so that, according to
the restricted version of the zeta approach, we can simply put
\beq \la 0|\Ti_{\mu\nu}(\bx)|0\ra_{ren} :=
\la 0|\Tis_{\mu\nu}(\bx)|0\ra\Big|_{\s = 0} ~. \label{TrinBox}\feq
In the forthcoming subsections \ref{NumBox1d} and \ref{NumBox2d},
dealing with the cases $d = 1$ and $d = 2$, we will use approximate
expressions for all the components of $\la 0|\Ti_{\mu\nu}(\bx)|0\ra_{ren}$
obtained replacing each Dirichlet function $\Dir^{(\bullet)}_s$ in
Eq.s (\ref{Tidir00box}-\ref{Tidirijbox}) with the truncations
$\Dir^{(\bullet)}_{s,N}$ of a fixed (sufficiently large) order
$N$, given by Eq.s \rref{DirSupEst} \rref{DirInfEst}. Let us recall
that we have explicit remainder bounds for these truncations
(see Eq.s \rref{RRSupEst} \rref{DzwRRSupEst}
and Eq.s \rref{RRInfEst} \rref{DzwRRInfEst}); these will allow
us to infer error estimates for the approximate expressions of
$\la 0|\Ti_{\mu\nu}(\bx)|0\ra_{ren}$ described above.
\vspace{-0.4cm}
\subsection{The pressure on the boundary.}\label{ACBoxPres}
We refer to the general discussion of subsection \ref{pressuretmunu};
so, we have two alternative definitions for the renormalized pressure
on each of the sides $\pi_{\nn,\an}$ of the box ($\nn\!\in\!\{1,...,d\}$,
$\an\!\in\!\{0,1\}$; see Eq. \rref{boxSides}). \parn
Let $\bx$ be any point interior to one of the sides $\pi_{\nn,\an}$; let
us stress that we exclude $\bx$ to be on an edge of the box (i.e., on
the intersection of two or more sides), where the outer normal is
ill-defined. As an example, let us assume $\bx$ to be an inner point
of the side $\pi_{1,0}$, so that the unit outer normal at $\bx$ is
$\bn(\bx) = (-1,0,...,0)$. We first consider the regularized pressure
\beq p^\s_i(\bx) := \la 0|\Tis_{i j}(\bx)|0\ra\,n^j(\bx)
= - \,\la 0|\Tis_{i1}(\bx)|0\ra ~; \label{pregBox} \feq
this can be expressed using the general rule \rref{pEeT} for the case
of Dirichlet boundary conditions which, in the present case, gives
({\footnote{In the application of Eq. \rref{pEeT} to the present case,
we use the previous expression for $\bn(\bx)$ and the fact that
$$ \partial_{x^i y^j}\Dir_s(\bx,\by)\Big|_{\by = \bx} = 0
\quad \mbox{for all $i,j \in \{1,...,d\}$ such that $i \neq 1$
or $j \neq 1$}~; $$
this follows straightforwardly from the Dirichlet conditions
prescribed on the boundary of $\Om$ and from the eigenfunction
expansion in Eq. \rref{eqkerdi}, taking into account
the factorized structure of the eigenfunctions in our case.}})
\begin{equation}\begin{split}
& \hspace{1.3cm} p^\s_i(\bx) = -\,\de_{i 1}\,{\mm^\s \over 4}\;
\partial_{x^1 y^1}\Dir_{{\s + 1 \over 2}}(\bx,\by)\Big|_{\by = \bx} = \\
& = -\,\de_{i 1}\,{\mm^\s \over 4}\l[\partial_{x^1 y^1}
\Dir^{(>)}_{{\s + 1 \over 2}}(\bx,\by)\Big|_{\by = \bx} +
\Dir^{(<)}_{{\s + 1 \over 2}}(\bx,\by)\Big|_{\by = \bx}\r]\, .
\label{alt1BoxPi1}
\end{split}\end{equation}
The equality in the second line of the above equation, involving the
derivatives of the Dirichlet functions $\Dir^{(\bullet)}_{\s+1 \over 2}$,
follows from Eq. \rref{DirHeatBox} and will be useful for later purposes. \parn
On the one hand, we can put
\begin{equation}\begin{split}
& \hspace{0.95cm} p^{ren}_i(\bx) := p^\s_i(\bx)\Big|_{\s = 0}\! =
- \,\la 0|\Tis_{i1}(\bx)|0\ra \Big|_{\s = 0} = \\
& = -\,{\de_{i 1} \over 4}\l[
\partial_{x^1 y^1}\Dir^{(>)}_{1/2}(\bx,\by)\Big|_{\by = \bx}
+ \partial_{x^1 y^1}\Dir^{(<)}_{1/2}(\bx,\by)\Big|_{\by = \bx} \r]\,.
\label{alt1Box}
\end{split}\end{equation}
On the other hand, we have the alternative definition
\beq \hspace{-0.05cm} p^{ren}_i(\bx)\! := \!\l(\lim_{\bx'\in\Om,\,\bx'\to\bx}
\la 0|\Ti_{i j}(\bx') |0\ra_{ren} \r)\! n^j(\bx) =
-\l(\lim_{\bx'\in\Om,\,\bx'\to\bx} \la 0|\Ti_{i1}(\bx')|0\ra_{ren}\r).\!
\label{alt2Box} \feq
In the next two paragraphs we show the \textsl{equivalence of the
alternative prescriptions \rref{alt1Box} \rref{alt2Box}} and the
\textsl{non-integrable divergence near the edges of the renormalized
pressure} (evaluated, equivalently, according to either of the two cited
prescriptions). \salto
Before proceeding to the proof of the above statements, let us anticipate
that in subsections \ref{NumBox1d} and \ref{NumBox2d} (dealing with the
cases $d = 1$ and $d = 2$, respectively) we will evaluate the pressure
starting from Eq. \rref{alt1Box} and substituting the functions
$\Dir_{1/2}^{(\bullet)}$ therein with the truncations
$\Dir_{1/2,N}^{(\bullet)}$ of a sufficiently large order $N$;
the errors of these approximants will be evaluated using Eq.s
\rref{RRSupEst} \rref{DzwRRSupEst} and \rref{RRInfEst} \rref{DzwRRInfEst}.
\vspace{-0.4cm}
\subsubsection{Equivalence of prescriptions \rref{alt1Box} \rref{alt2Box}.}
The proof of this equivalence, given hereafter, uses arguments similar
to the ones proposed in subsection 4.6 of Part II, where an analogous
statement was derived for a domain $\Om$ bounded by orthogonal hyper-planes. \parn
Let us consider again an inner point $\bx = (0,x^2,...,x^d)$ of the side $\pi_{1,0}$.
When expressing $\la 0|\Tis_{i1}(\bx)|0\ra|_{\s = 0}$ (see Eq. \rref{alt1Box})
or $\lim_{\bx' \to \bx} \la 0|\Ti_{i1}(\bx') |0\ra_{ren}$ (see Eq. \rref{alt2Box})
in terms of the Dirichlet kernel and of its series expansions, there is
only one type of potentially troublesome terms, which could give different
contributions in the two cases. These are terms arising from the summand
$T^{\s,(<)}_{i 1}(\bx)$ in Eq. \rref{Tidirijbox}, when we use for $\Dir^{(<)}_s$
and its derivatives the series expansions \rref{DirInfExp} \rref{DzwDirInfExp};
more precisely, potential troubles could arise from terms in the cited expansions with
\beq \mbox{$\bhl(\bx,\bx) = 0$} \qquad \mbox{and}
\qquad \mbox{$\bhl(\bx',\bx') \neq 0$~ for $\bx' \in \Om$} ~, \feq
corresponding to the choice
\beq \mbox{$h_1 = 0$, $l_1 = 2$} \qquad \mbox{and} \qquad
\mbox{$h_i = 0$, $l_i = 1$~ for $i \neq 1$} ~. \label{bkbl} \feq
Consider the expression for $T^{\s,(<)}_{i 1}(\bx')$ obtained using
Eq. \rref{Tidirijbox} and Eq.s \rref{DirInfExp} \rref{DzwDirInfExp},
where $\bx'$ is either a point in the interior of the spatial domain
$\Om$ or the boundary point $\bx$ under consideration; the contribution
to this expression from problematic terms, with $(\bh,\bl)$ as in
Eq. \rref{bkbl}, is
\beq \!\de_{i 1}\!\!\l(\!{1\over 4}\!-\!\xi\!\r)\!{\mm^\s\, T^{\s - d -1 \over 2}
\over (4\pi)^{d/2} \Ga({\s + 1 \over 2})\!}\! \l[ {\s\!-\!d\!-\!1 \over 2}\,
\PP_{\s-d-1 \over 2}(z)\!+\!z \,\PP_{\s-d-3 \over 2}(z)\r]_{\!z = {({x'}^1)^2 \over T}}
\!\!\! \equiv \!f(\s, {x'}^1)\,.\! \label{ProbBox} \feq
To show the equivalence between prescriptions \rref{alt1Box} \rref{alt2Box}
at the point $\bx$ in consideration, we must show that
\beq f(\s,0) \Big|_{\s=0} = \lim_{{x'}^1 \in (0,a_1),\, {x'}^1 \to 0}
\l( f(\s, {x'}^{1}) \Big|_{\s=0} \r) \label{tocheck} \feq
where $|_{\s=0}$ indicates, as usual, the analytic continuation at $\s=0$\,.
In order to check the equality in Eq. \rref{tocheck} we first notice that
\begin{equation}\begin{split}
& f(\s,0) = \de_{i 1} \l({1\over 4}\!-\!\xi\r) {\mm^\s\, T^{\s - d -1 \over 2}
\over (4\pi)^{d/2}\,\Ga({\s + 1 \over 2})}\l[{\s\!-\!d\!-\!1 \over 2}\;
\PP_{\s-d-1 \over 2}(0) \r] = \\
& \hspace{2.5cm} = \de_{i 1} \l({1\over 4}\!-\!\xi\r) {\mm^\s\, T^{\s- d - 1 \over 2}
\over (4\pi)^{d/2}\,\Ga({\s + 1 \over 2})} \label{ProbBox1a}
\end{split}\end{equation}
(the second equality relies on Eq. \rref{PropPP1} for $\PP_s(0)$).
The last expression in \rref{ProbBox1a} gives the analytic
continuation of $f(\s,0)$ to the whole complex plane; in
particular, when evaluated at $\s = 0$, it yields
\beq f(\s,0) \Big|_{\s=0} = \de_{i 1} \l(\!{1\over 4}\!-\!\xi\!\r)
{2 \over (4 \pi\,T)^{d+1 \over 2}} ~. \label{ProbBox1b} \feq
On the other hand, returning to Eq. \rref{ProbBox} we see that,
for any ${x'}^1\!\in\!(0,a_1)$,
\beq f(\s, {x'}^{1}) \Big|_{\s=0}\!\! = \de_{i 1}\!\l(\!{1\over 4}\!-\!\xi\!\r)\!
{2 \over (4\pi\,T)^{d+1 \over 2}}\!\l[-{d\!+\!1 \over 2}\,\PP_{\!-{d+1 \over 2}}(z)\!
+ \!z\,\PP_{\!-{d+3 \over 2}}(z)\r]_{\! z = ({x'}^1)^2/T}\!.\! \label{ProbBox2} \feq
Resorting to the second equality in Eq. \rref{PropPP1} for the
functions $\PP_\bullet(z)$ and using a recursive relation for the
upper incomplete  gamma function (see \cite{NIST}, p.178, Eq.8.8.2), we obtain
\beq f(\s,{x'}^{1}) \Big|_{\s=0} = \de_{i 1} \l({1\over 4}\!-\!\xi\r)
{2 \over (4 \pi\,T)^{d+1 \over 2}}\; e^{-{({x'}^1)^2 \over T}} ~;
\label{ProbBoxAs} \feq
comparing with Eq. \rref{ProbBox1b}, we immediately obtain the desired
relation \rref{tocheck}, that is $f(\s,0)|_{\s=0} = \lim_{{x'}^1 \in (0,a_1),\,
{x'}^1 \to 0} \left( f(\s, {x'}^{1})|_{\s=0} \right)$\,. \parn
This concludes our analysis of the renormalized pressure at points
in the interior of $\pi_{1,0}$; needless to say, the very same
results also hold for the pressure acting on any other of the
sides delimiting $\Om$\,.
\vspace{-0.4cm}
\subsubsection{Non-integrable behaviour of the renormalized pressure near the edges.}\label{parNI}
Let us first remark that at points on the edges of the box
(i.e., on the corners which appear whenever $d > 1$) the outer
normal and, consequently, the pressure are both ill-defined. \parn
In this paragraph we discuss the behaviour of the pressure
at points in the neighborhood of the edges; more precisely,
we show that the renormalized pressure evaluated at inner
points of one side diverges in a non-integrable manner when
moving towards anyone of the edges. \parn
In order to fix our ideas, let us consider the renormalized
pressure $p_i^{ren}$ on the side $\pi_{1,0}$; in the following
we discuss the behaviour of this quantity in a neighborhood of
the corner placed at $\bx = \b0$, i.e., at the intersection of
all the sides $\pi_{\nn,0}$ ($\nn\in\{1,...,d\}$). To this purpose,
let us consider the expression \rref{alt1Box} for $p_i^{ren}$;
by considerations analogous to those of the previous paragraph,
we see that contributions diverging for $\bx \to \b0$ arise from
$\partial_{x^1 y^1}\Dir^{(<)}_{1/2}$ when we use for it the
series expansion \rref{DzwDirInfExp}. More precisely, these
contributions arise from the terms for which
\beq \mbox{$\bhl(\b0,\b0) = 0$} \qquad \mbox{and} \qquad
\mbox{$\bhl(\bx,\bx) \neq 0$~ for $\bx$ interior to $\pi_{1,0}$,
$\bx \neq \b0$} ~; \feq
by a simple inspection, these terms are seen to correspond
to the following choices
\begin{equation}\begin{split}
& \hspace{4.1cm} h_i = 0 \quad \mbox{for all $i\!\in\!\{1,...,d\}$} ~, \\
& \mbox{$l_i \in \{1,2\}$ for $i \in \{1,...,d\}$} \quad \mbox{and} \quad
\mbox{$l_i = 2$ for at least one $i \in \{2,...,d\}$} ~. \label{bkblpre}
\end{split}\end{equation}
As an example, let us focus on one of the above terms; for any
fixed $\nn \in \{2,...,d\}$, we consider the one with
\beq h_i = 0~~ \mbox{for all $i\!\in\!\{1,...,d\}$} \quad \mbox{and}
\quad l_i = \l\{\!\barray{ll} 2 & \mbox{for $i\in\{2,...,\nn\}$} \\
1 & \mbox{for $i \in\{\nn+1,...,d\}$} \farray \r. . \label{bkln} \feq
With simple but long computations
({\footnote{In particular note that, with the present choices
\rref{bkblpre} \rref{bkln}, we have
$$ \bhl(\bx,\by)\Big|_{\by = \bx,\,x^1 = 0} = \sum_{i = 2}^\nn (x^i)^2 ~, $$
$$ \partial_{x^1}\bhl(\bx,\by)\Big|_{\by = \bx,\,x^1 = 0}\! =
\partial_{y^1}\bhl(\bx,\by)\Big|_{\by = \bx,\,x^1 = 0}\! = 0 ~,
\qquad \partial_{x^1 y^1}\bhl(\bx,\by)\Big|_{\by = \bx,\,x^1 = 0}\! =
- {\de_{l_1} \over 2} ~. $$}}),
the contribution of this term to $\partial_{x^1 y^1}\Dir^{(<)}_{1/2}$
can be expressed as
\beq {2\,(-1)^{d-\nn} \over (4\pi\,T)^{d+1 \over 2}}\;
\PP_{-{d+1\over 2}}\!\l({(x^2)^2\!+\!...\!+\!(x^\nn)^2 \over T}\r)
\equiv g_{\nn,\bl}(\bx) ~. \feq
Using the second relation in Eq. \rref{PropPP1} and resorting
to the asymptotic expansion (see \cite{NIST}, p.178, Eq.8.7.3)
\beq \Ga(s,z) = \Ga(s)(1+O(z^s)) \qquad \mbox{for $\Re s > 0$, $z \to 0$} ~, \feq
we readily infer, for $\bx\to\b0$,
\beq g_{\nn,\bl}(\bx) = {2\,(-1)^{d-\nn} \Ga({d + 1\over 2})
\over (4\pi)^{d+1 \over 2}} \l[{1 \over z^{d+1}}\;(1 + O(z^{d+1}))
\r]_{z = \sqrt{(x^2)^2+...+(x^\nn)^2}} \,. \feq
The above expression shows the non-integrable divergence of $g_{\nn,\bl}$
(for all $\nn\!\in\!\{2,...,d\}$); the same conclusion holds as well for the
terms corresponding to all the other choices of $(\bh,\bl)$ in Eq. \rref{bkblpre}.
Let us stress that no cancellation of the divergent terms can occur,
since all the contributions with the same degree of divergence happen
to have the same sign; besides, due to the remainder estimate \rref{DzwRRInfEst},
no compensation of these terms can either arise from the full series
expansion \rref{DzwDirInfExp}. \parn
The above comments prove the non-integrable behaviour of $p_i^{ren}(\bx)$
for $\bx \to \b0$\,; let us remark that this fact is of utmost importance
when attempting to evaluate the total force acting on any side of the box,
a topic we discuss in detail in subsection \ref{TotForBox}.
\vspace{-0.4cm}
\subsection{The total energy.} \label{EnBox}
First of all, let us recall from subsection \ref{TotEnSub} that the total
energy consists of the sum of both a bulk and a boundary contribution;
since the latter vanishes identically due to the Dirichlet conditions
assumed on the boundary (see the comments below Eq. \rref{defBs}), we
only have to discuss the bulk term. \parn
Consider the representation \rref{defEs} of the bulk energy
({\footnote{Let us mention that, in order to evaluate the bulk energy
$E^\s$, we could proceed in an alternative manner. Namely, we could
consider the representation of $E^\s$ in terms of the trace $\Tr\AA^{1-\s \over 2}$
(see Eq. \rref{defEs}) and determine the analytic continuation of the latter
moving along the same lines we followed for the Dirichlet kernel, using
the heat trace $K(\t)$ in place of the heat kernel $K(\t\,;\bx,\by)$. \\
Nonetheless, since the small $\t$ expansion of the heat trace $K(\t)$
for the present configuration involves quite cumbersome expressions,
we prefer to avoid this approach. Another advantage of the methods proposed
in this subsection is that, after minor variations, they also allow
to evaluate the total force on the boundary (see subsection \ref{TotForBox}).}}):
$$ E^\s := {\mm^\s\! \over 2} \int_\Om d\bx\;
\Dir_{{\s - 1\over 2}}(\bx,\bx) ~. $$
Using the expression \rref{DirHeatBox} for the Dirichlet kernel,
we readily infer
\begin{equation}\begin{split}
& \hspace{3.cm} E^\s = E^{\s,(>)} + E^{\s,(<)} \qquad \mbox{where} \\
& E^{\s,(\bullet)}\! := {\mm^\s \over 2} \int_{(0,a_1)
\times ... \times (0,a_d)} \hspace{-0.8cm} dx^1 ...\,dx^d\;
\Dir^{(\bullet)}_{\s - 1 \over 2}(\bx,\bx) \quad
\mbox{for $\bullet \in \{>,<\}$}\,. \label{EnSupInf}
\end{split}\end{equation}
In the next two paragraphs we discuss the analytic continuation
of $E^\s$ at $\s = 0$, giving the renormalized bulk energy.
To this purpose we introduce series expansions for the functions
$E^{\s,(>)}$ and $E^{\s,(<)}$, which can be used to build their
analytic continuations; we also give fully quantitative remainder
estimates for these series (for more details, see Appendices
\ref{AppTotEBox} and \ref{AppC}).
\vspace{-0.4cm}
\subsubsection{Series expansions and analytic continuations for $\boma{E^{\s,(>)}}$, $\boma{E^{\s,(<)}}$.}
Consider the expression \rref{EnSupInf} for the regularized bulk
energy $E^\s$; hereafter we give series expansions for the two
addenda $E^{\s,(>)}$ and $E^{\s,(<)}$, ultimately yielding the analytic
continuations of these functions to the whole complex plane. \parn
On the one hand, inserting the expansion \rref{DirSupExp} for $\Dir^{(>)}_s$
into Eq. \rref{EnSupInf}, we obtain (see Appendix \ref{AppTotEBox})
\beq E^{\s,(>)} = {\mm^\s \over 2\,\Ga({\s- 1 \over 2})}
\sum_{\bn \in \naturali^d} \om_\bn^{1-\s}\;
\Ga\!\l({\s\!-\!1 \over 2}\,,\om_\bn^2\,T\r) \,. \label{ESupExp} \feq
The right-hand side of above equation can be proved to converge for all
$\s \in \complessi$ (see the next paragraph and Appendix \ref{AppC});
thus, Eq. \rref{ESupExp} gives the analytic continuation of $E^{\s,(>)}$
to the whole complex plane, in particular at $\s = 0$. \parn
On the other hand, inserting the expansion \rref{DirInfExp} for
$\Dir^{(<)}_s$ into Eq. \rref{EnSupInf}, after some effort
we obtain (see again Appendix \ref{AppTotEBox})
\beq E^{\s,(<)}\! = {\mm^\s\,T^{\s-1 \over 2} \over 2^{d+1}\,\Ga({\s-1 \over 2})}\,
\sum_{\ns = 0}^d{(-1)^{d-\ns} \over (d\!-\!\ns)!\ns!} \sum_{\si \in S_d}
{\bap_{\si,\ns} \over (\pi\,T)^{\ns/2}} \sum_{\bh \in \interi^\ns}
\PP_{\s-\ns-1 \over 2}\!\l({\Bs(\bh) \over T}\r)\!.\! \label{EInfExp} \feq
In the above $S_d$ indicates the symmetric group with $d$ elements
and we have put
\begin{equation}\begin{split}
& \hspace{2.7cm} \bap_{\si,0} := 1 ~, \qquad \interi^0 := \{\b0\} ~, \qquad
B_{\si,0}(\b0) := 0 ~, \\
& \bap_{\si,\ns} := \prod_{i = 1}^\ns a_{\si(i)} ~, \quad
\Bs(\bh) := \sum_{i = 1}^{\ns} (a_{\si(i)}h_i)^2
\qquad \mbox{for $\si\!\in\!S_d$, $\ns\!\in\!\{1,...,d\}$} ~; \label{Bsk}
\end{split}\end{equation}
note that the term with $\ns = 0$ in Eq. \rref{EInfExp} is just
$(-1)^d\,\PP_{\s-1 \over 2}(0)$\,. \parn
Let us stress that the functions $\PP_s$ in Eq. \rref{EInfExp} must be
evaluated according to Eq. \rref{defPP}; in particular, recall that
the first relation in Eq. \rref{PropPP1} gives $\PP_s(0) = 1/s$\,.
Thus, for all $\ns \in\{0,...,d\}$, the terms in the series \rref{EInfExp}
with $\Bs(\bh) = 0$, i.e., those with $\bh = \b0$ (see Eq. \rref{Bsk}),
are singular at $\s = \ns+1$ where they have a simple pole.
The series obtained from the right-hand side of Eq. \rref{EInfExp}
removing the finitely many terms with $\bh = \b0$ is (rapidly) convergent
for all $\s \in \complessi$, a fact that we discuss in the next paragraph
and in Appendix \ref{AppC}. \parn
Because of the above considerations, the expression \rref{EInfExp} gives
the analytic continuation of $E^{\s,(<)}$ to a meromorphic function
on the whole complex plane, with simple poles at
\beq \s \in \{1,2,...,d + 1\} ~. \feq
Summing up, $\s = 0$ is a regular point for the analytic continuations
of both $E^{\s,(>)}$ and $E^{\s,(<)}$ so that, according to the
restricted zeta approach (see Eq. \rref{ERenIV}), we can put
\beq E^{ren} = E^{\s,(>)}\Big|_{\s = 0} + E^{\s,(<)}\Big|_{\s = 0} \label{ERenBox} \feq
where the two addenda on the right-hand side simply indicate the
expressions \rref{ESupExp} and \rref{EInfExp} evaluated at $\s = 0$.
\vspace{-0.4cm}
\subsubsection{Estimates for the series \rref{ESupExp} \rref{EInfExp}.} \label{EnRem}
The scheme followed in this paragraph closely resembles the one of
subsection \ref{NumBox}, where we dealt with the series expansions for
the Dirichlet functions $\Dir^{(\bullet)}_s$; we retain here the same
notations introduced therein (see, in particular, Eq.s \rref{deflL} \rref{defHEN}). \parn
Let us consider the series in the right-hand side of Eq. \rref{ESupExp},
and the series in the right-hand side of Eq. \rref{EInfExp} after
removing from it the finitely many singular terms with $\bh = \b0$. In Appendix
\ref{AppC} we prove the convergence of these series for all
$\s \in \complessi$, and derive the remainder estimates reported hereafter.
For $N \in (0,+\infty)$, let us write
\begin{equation}\begin{split}
& \hspace{2cm}E^{\s,(>)} = E_N^{\s,(>)} + R_N^{\s,(>)} ~, \\
& E_N^{\s,(>)} := {\mm^\s \over 2\,\Ga({\s- 1 \over 2})}
\sum_{\bn \in \naturali^d,\,|\bn| \leqs N} \om_\bn^{1-\s}\; \Ga\!\l({\s\!-\!1 \over 2}\,,
\om_\bn^2\,T\r) \,, \\
& R_{N}^{\s,(>)} := {\mm^\s \over 2\,\Ga({\s - 1 \over 2})}
\sum_{\bn \in \naturali^d,\,|\bn| > N} \om_\bn^{1-\s}\,
\Ga\!\l({\s\!-\!1 \over 2}\,,\om_\bn^2\, T\r)\,; \label{ESupApp}
\end{split}\end{equation}
the remainder $R_N^{u,>}$ has the bound (Appendix \ref{AppC})
\begin{equation}\begin{split}
& \hspace{1cm} \l|R_{N}^{\s,(>)}\r| \leqs {\mm^{\Re\s}\,
\max(\Am^{\Re\s - 1},\AM^{\Re\s - 1}) \over 2^{d+1}\,
\pi^{\Re\s - 1}\,|\Ga({\s - 1 \over 2})|}\;
\HE^{(d)}_N\!\l(\al,{\pi^2\,T \over \AM^2}\,;
{\Re\s\!-\!1 \over 2}\,,1\!-\!\Re\s\r) \\
& \mbox{for either \quad $\Re\s \geqs 1$, $N > 2\sqrt{d}$} \quad
\mbox{or} \quad \mbox{$\Re\s < 1$, $\dd{N>2\sqrt{d} + {\AM \over \pi}
\sqrt{(1\!-\!\Re\s) \over 2\al T}}$} ~. \label{ESupEst}
\end{split}\end{equation}
Again for $N \in (0,+\infty)$, let us put
\beq E^{\s,(<)} = E_N^{\s,(<)} + R_N^{\s,(<)} ~, \label{EInfApp} \feq
$$ E_N^{\s,(<)} := {\mm^\s\,T^{\s-1 \over 2} \over 2^{d+1}\,\Ga({\s-1 \over 2})}\,
\sum_{\ns = 0}^d\,{(-1)^{d-\ns} \over (d\!-\!\ns)!\ns!} \sum_{\si \in S_d}
{\bap_{\si,\ns} \over (\pi\,T)^{\ns \over 2}}
\sum_{\bh \in \interi^\ns,\, |\bh| \leqs N}
\PP_{\s-\ns-1 \over 2}\!\l({\Bs(\bh) \over T}\r)\,, $$
$$ R_N^{\s,(<)} := {\mm^\s\,T^{\s-1 \over 2} \over 2^{d+1}\,\Ga({\s-1 \over 2})}\,
\sum_{\ns = 1}^d\,{(-1)^{d-\ns} \over (d\!-\!\ns)!\ns!} \sum_{\si \in S_d}
{\bap_{\si,\ns} \over (\pi\,T)^{\ns \over 2}}
\sum_{\bh \in \interi^\ns,\, |\bh| > N}
\PP_{\s-\ns-1 \over 2}\!\l({\Bs(\bh) \over T}\r) $$
(where $|\bh| := \sqrt{h_1^2+...+h_\ns^2}$); the remainder $R_N^{\s,(<)}$,
which contains no singular term with $\bh = \b0$, fulfills (see, again,
Appendix \ref{AppC})
\begin{equation}\begin{split}
& \l|R_{N}^{\s,(<)}\r| \leqs {\mm^{\Re\s} \over 2^{d+1}\,|\Ga({\s-1 \over 2})|}\,
\sum_{\ns = 1}^d {\max(\Am^{\Re \s - \ns - 1},\AM^{\Re \s - \ns - 1}) \over
\pi^{\ns/2}\,(d\!-\!n)!n!}~ \cdot \\
& \hspace{4.5cm} \cdot \l(\sum_{\si \in S_d} \bap_{\si,\ns}\r)\!
\HE^{(\ns)}_N\!\l(\al,{\Am^2 \over T}\,;{\ns\!+\!1\!-\!\Re \s \over 2},
\Re \s\!-\!\ns\!-\!1\r) \\
& \mbox{for either \,$\dd{\Re \s \leqs 2}$, $N > 2\sqrt{d}$}
\quad\! \mbox{or} \quad\! \mbox{$\dd{\Re \s > d\!+\!1}$, $\dd{N > 2\sqrt{d}\!
+\!{1\over \Am}\,\sqrt{(\Re \s\!-\!2) T \over 2\al}}$} ~. \label{EInfEst}
\end{split}\end{equation}
As sketched in Appendix \ref{AppC}, we could give reminder estimates even
for the case $2 < \Re \s \leqs d + 1$, excluded from \rref{EInfEst}; however,
these would involve rather complicated expressions. Taking into account that,
in the sequel, we will be mainly interested in the case $\s=0$, we prefer
not to report these cumbersome expressions. \parn
In subsections \ref{NumBox1d} and \ref{NumBox2d}, moving along the same
lines as for the VEV of stress-energy tensor, we will use the truncations
$E_N^{\s,(>)}$ and $E_N^{\s,(<)}$ of a fixed sufficiently large order
$N$ to obtain approximate expressions for the functions $E^{\s,(>)}$
and $E^{\s,(<)}$, respectively; these will be used to evaluate the
renormalized bulk energy (see Eq. \rref{EnSupInf}), giving explicit
errors estimates.
\vspace{-0.4cm}
\subsection{The total force on a side of the box.} \label{TotForBox}
Let us consider the framework of subsection \ref{force}; following
the general scheme outlined therein for boundary forces, we can
in principle consider two alternative approaches to define the
total force acting on any side of the box. \parn
As an example, let us focus on the force acting on $\pi_{1,0}$, i.e.,
the side contained in the hyperplane $\{x^1\!= 0\}$; recall that the
unit outer normal at points $\bx$ interior to $\pi_{1,0}$ is $\bn(\bx)
= (-1,0,...,0)$. \parn
We first consider the regularized integrated force
\beq \fo_{1,0}^\s := \int_{\pi_{1,0}}\!\! da(\bx)\;n^i(\bx)\, p^\s_i(\bx)
= - \int_{\pi_{1,0}}\!\! da(\bx)\;p^\s_1(\bx) \label{FregBox} \feq
(compare with Eq. \rref{Freg}, here employed with $\Op = \pi_{1,0}$),
where $p^\s_i(\bx)$ indicates the regularized pressure \rref{pregBox}.
We will prove in the sequel that this can be analytically continued
up to $\s=0$, so that we can define the renormalized integrated force as
\beq \fo_{1,0}^{ren} := \fo_{1,0}^\s \Big|_{\s = 0} ~. \label{FrenBox} \feq
On the other hand, we have the alternative prescription (corresponding
to Eq. \rref{FrenAlt}, with $\Op = \pi_{1,0}$)
\beq \fo_{1,0}^{ren} := \int_{\pi_{1,0}}\!\! da(\bx)\;n^i(\bx)\,
p^{ren}_i(\bx) = - \int_{\pi_{1,0}}\!\! da(\bx)\;p^{ren}_1(\bx) ~,
\label{FrenBoxAlt} \feq
where $p^{ren}_i(\bx)$ is the renormalized pressure, defined equivalently
according to either prescription \rref{alt1Box} or \rref{alt2Box}. \parn
As a matter of fact, we know from the previous subsection that the
renormalized pressure diverges in a non-integrable manner near the
edges of the box; in consequence of this, the prescription \rref{FrenBoxAlt}
gives an infinite value for the total force on $\pi_{1,0}$.
Since this result is patently physically unacceptable, in the following
we only consider the approach \rref{FregBox} \rref{FrenBox}. \salto
Let us therefore consider the regularized expression \rref{FregBox};
using Eq. \rref{alt1BoxPi1} for the regularized pressure on $\pi_{1,0}$
(we are referring, in particular, to the representation in the
second line of the cited equation), we readily infer
\begin{equation}\begin{split}
& \hspace{3.5cm} \fo_{1,0}^\s = \fo_{1,0}^{\s,(>)}
+ \fo_{1,0}^{\s,(<)} \qquad \mbox{where} \\
& \fo_{1,0}^{\s,(\bullet)}\! := {\mm^\s \over 4} \int_{(0,a_2)
\times ... \times (0,a_d)} \hspace{-0.8cm} d x^2 ...\,dx^d\;
\partial_{x^1 y^1}\Dir^{(\bullet)}_{\s + 1 \over 2}(\bx,\by)
\Big|_{\by = \bx,\,x^1 = 0} \quad \mbox{for $\bullet \in \{>,<\}$}\,.
\label{foSupInf}
\end{split}\end{equation}
In the next two paragraphs we introduce convergent series expansions
for the functions $\fo_{1,0}^{\s,(>)}$ and $\fo_{1,0}^{\s,(<)}$; these
expansions ultimately yield the analytic continuations of these functions
and of $\fo_{1,0}^{\s}$ at $\s = 0$ (and so, they determine the renormalized
force on $\pi_{1,0}$ according to Eq. \rref{FrenBox}). We also give
remainder estimates for these series.
\vspace{-0.4cm}
\subsubsection{Series expansions and analytic continuations for $\boma{\fo_{1,0}^{\s,(>)}}$, $\boma{\fo_{1,0}^{\s,(<)}}$.} Inserting into Eq. \rref{foSupInf} the series \rref{DzwDirSupExp}
for $\partial_{x^1 y^1}\Dir_s^{(>)}(\bx,\by)$ and integrating term by term,
we have
\beq \fo_{1,0}^{\s,(>)} = {\mm^\s \over 2a_1\, \Ga({\s + 1 \over 2})}
\sum_{\bn \in \naturali^d} \!\l({n_1 \pi \over a_1}\r)^{\!\!2}
\om_\bn^{-(\s + 1)}\, \Ga\!\l({\s\!+\!1 \over 2}\,,\om_\bn^2\,T\r) \,. \label{foSup1} \feq
The above expression can be proved to converge for all $\s \in \complessi$
(by a simple variation of the proof of the convergence of the expansion
\rref{ESupExp} for $E^{\s,(>)}$; see paragraph \ref{ApSubEM} of
Appendix \ref{AppC}); thus, Eq. \rref{foSup1} gives the analytic
continuation of $\fo_{1,0}^{\s,(>)}$ to the whole complex plane,
in particular at $\s = 0$. \parn
On the other hand, using the series \rref{DzwDirInfExp} for
$\partial_{x^1 y^1}\Dir_s^{(<)}(\bx,\by)$ along with the definition
\rref{foSupInf}, we can show with some effort that
\begin{equation}\begin{split}
& \fo_{1,0}^{\s,(<)} = -\,{\mm^\s\,T^{\s-3 \over 2} \over 2^{d+1}\,
\Ga({\s+1 \over 2})}\,\sum_{\ns = 1}^d\,{(- 1)^{d-\ns} \over
(d\!-\!\ns)!(\ns\!-\!1)!} \sum_{\sp \in \Sp_d} {\bap_{\sp,\ns} \over
(\pi\,T)^{\ns \over 2}} \; \cdot \\
& \hspace{1cm} \cdot\,\sum_{\bh \in \interi^\ns} \l[(a_1 h_1)^2\,
\PP_{\s-\ns-3 \over 2}\!\l({\Bp(\bh) \over T}\r) - {T \over 2}\,
\PP_{\s-\ns-1 \over 2}\!\l({\Bp(\bh) \over T}\r)\r] \,; \label{foInf1}
\end{split}\end{equation}
in the above we have put, for brevity, \beq \Sp_d := \{\sp
\!\in\! S_d ~|~ \sp(1) = 1 \} ~, \qquad \bap_{\sp,1} := 1 ~,
\qquad B_{\sp,1}(\bh) := (a_1 h_1)^2 ~, \label{defSp} \feq
$$ \bap_{\sp,\ns} := \prod_{i = 2}^\ns a_{\sp(i)} ~, \quad
\Bp(\bh) := (a_1 h_1)^2 + \sum_{i = 2}^{\ns} (a_{\sp(i)}h_i)^2
\qquad \mbox{for $\sp\!\in\!\Sp_d$, $\ns\!\in\!\{2,...,d\}$} ~. $$
The last result is derived in a manner similar to expansion \rref{EInfExp}
for $E^{\s,(<)}$ (see Appendix \ref{AppTotEBox}); besides, there hold
considerations analogous to those below the cited equation. More precisely:
the terms in the right-hand side of Eq. \rref{foInf1} with $\ns \in \{1,...,d\}$
and $\bh = \b0$ have a simple pole at $\s = \ns + 1$; after removing
these finitely many terms, the series in the right-hand side of Eq. \rref{foInf1}
converges for all $\s \in \complessi$. Summing up, Eq. \rref{foInf1}
gives the analytic continuation of $\fo_{1,0}^{\s,(<)}$ to a meromorphic
function on the whole complex plane, with simple poles at
\beq \s \in\{2,3,...,d,d + 1\} ~. \feq
Summing up, $\s = 0$ is a regular point for the analytic continuations
of both $\fo_{1,0}^{\s,(>)}$ and $\fo_{1,0}^{\s,(<)}$; now, recalling
the definition \rref{FrenBox} for the renormalized total force on
$\pi_{1,0}$ and comparing with Eq. \rref{foSupInf}, we get
\beq \fo_{1,0}^{ren} = \fo_{1,0}^{\s,(>)}\Big|_{\s = 0} +
\fo_{1,0}^{\s,(<)}\Big|_{\s = 0} \label{foren} \feq
where the two addenda on the right-hand side simply indicate the
expressions \rref{foSup1} and \rref{foInf1} evaluated at $\s = 0$.
\vspace{-0.4cm}
\subsubsection{Estimates for the series \rref{foSup1} \rref{foInf1}.} \label{FoRem}
The arguments of paragraph \ref{EnRem} and Appendix \ref{AppC}
for the expansions \rref{ESupExp} \rref{EInfExp} of $E^{\s,(>)}$ and
$E^{\s,(<)}$ can also be adapted to deduce remainder estimates on
the series \rref{foSup1} \rref{foInf1} for $\fo_{1,0}^{\s,(>)}$ and
$\fo_{1,0}^{\s,(<)}$. Here we only report the final results, concerning
truncation at a suitable order $N$\,; in our presentation we adopt
the notations introduced in subsection \ref{NumBox}
(see, in particular, Eq.s \rref{deflL} \rref{defHEN}). \parn
For $N \in (0,+\infty)$, let us write
\begin{equation}\begin{split}
& \hspace{3.7cm}\fo_{1,0}^{\s,(>)} = \fo_{1,0,N}^{\s,(>)} + \Ro_{1,0,N}^{\s,(>)} ~, \\
& \fo_{1,0,N}^{\s,(>)} := {\mm^\s \over 2a_1\, \Ga({\s + 1 \over 2})}
\sum_{\bn \in \naturali^d,\,|\bn| \leqs N} \!\l({n_1 \pi \over a_1}\r)^{\!\!2}
\om_\bn^{-(\s+1)}\,\Ga\!\l({\s\!+\!1 \over 2}\,,\om_\bn^2\,T\r) \,, \\
& \Ro_{1,0,N}^{\s,(>)} := {\mm^\s \over 2a_1\, \Ga({\s + 1 \over 2})}
\sum_{\bn \in \naturali^d,\,|\bn| > N} \!\l({n_1 \pi \over a_1}\r)^{\!\!2}
\om_\bn^{-(\s+1)}\,\Ga\!\l({\s\!+\!1 \over 2}\,,\om_\bn^2\,T\r) \,; \label{foSupApp}
\end{split}\end{equation}
the remainder $\Ro_{1,0,N}^{\s,(>)}$ has the bound
\begin{equation}\begin{split}
& \hspace{0.15cm} \l|\Ro_{1,0,N}^{\s,(>)}\r| \leqs {\mm^{\Re\s}\pi^{1-\Re\s}
\max(\Am^{\Re\s + 1}\!,\AM^{\Re\s + 1}) \over
2^{d+1}\,a_1^3\,|\Ga({\s + 1 \over 2})|}\,
\HE^{(d)}_N\!\l(\!\al,{\pi^2\,T \over \AM^2}\,;{\Re\s\!+\!1 \over 2}\,,1\!-\!\Re\s\!\r) \\
& \mbox{for either \quad $\Re\s \geqs 1$, $N>2\sqrt{d}$} \quad \mbox{or}
\quad \mbox{$\Re \s < 1$, $\dd{N>2\sqrt{d}+{\AM \over \pi}
\sqrt{(1\!-\!\Re\s) \over 2\al T}}$} ~. \label{FoSupEst}
\end{split}\end{equation}
Moreover, let
\begin{equation}\begin{split}
& \hspace{4.7cm}\fo_{1,0}^{\s,(<)} = \fo_{1,0,N}^{\s,(<)} + \Ro_{1,0,N}^{\s,(<)} ~, \\
& \fo_{1,0,N}^{\s,(<)} := -\,{\mm^\s\,T^{\s-3 \over 2} \over 2^{d+1}\,
\Ga({\s+1 \over 2})}\,\sum_{\ns = 1}^d\,{(- 1)^{d-\ns} \over
(d\!-\!\ns)!(\ns\!-\!1)!} \sum_{\sp \in \Sp_d} {\bap_{\sp,\ns} \over
(\pi\,T)^{\ns \over 2}} ~ \cdot \\
& \hspace{1.5cm} \cdot \sum_{\bh \in \interi^\ns,\, |\bh| \leqs N}
\l[(a_1 h_1)^2\, \PP_{\s-\ns-3 \over 2}\!\l({\Bp(\bh) \over T}\r)\!
- {T \over 2}\,\PP_{\s-\ns-1 \over 2}\!\l({\Bp(\bh) \over T}\r)\r]\,, \\
& \Ro_{1,0,N}^{\s,(<)} := -\,{\mm^\s\,T^{\s-3 \over 2} \over 2^{d+1}\,
\Ga({\s+1 \over 2})}\,\sum_{\ns = 1}^d\,{(- 1)^{d-\ns} \over
(d\!-\!\ns)!(\ns\!-\!1)!} \sum_{\sp \in \Sp_d} {\bap_{\sp,\ns} \over
(\pi\,T)^{\ns \over 2}} ~ \cdot \\
& \hspace{1.5cm} \cdot \sum_{\bh \in \interi^\ns,\, |\bh| > N}
\l[(a_1 h_1)^2\, \PP_{\s-\ns-3 \over 2}\!\l({\Bp(\bh) \over T}\r)\!
- {T \over 2}\,\PP_{\s-\ns-1 \over 2}\!\l({\Bp(\bh) \over T}\r)\r] \label{foInfApp}
\end{split}\end{equation}
(recall that $|\bh| := \sqrt{h_1^2+...+h_\ns^2}$\,); as for the remainder
$\Ro_{1,0,(N)}^{\s,(<)}$ (containing no singular term with $\bh = \b0$),
there holds the estimate
\pagebreak
\begin{equation}\begin{split}
& \hspace{2.2cm} \l|\Ro_{1,0,N}^{\s,(<)}\r| \leqs {\mm^{\Re\s} \over 2^{d+1}\,
|\Ga({\s+1 \over 2})|}\sum_{\ns = 1}^d{1 \over (d\!-\!\ns)!(\ns\!-\!1)!}
\sum_{\sp \in \Sp_d}\!{\bap_{\sp,\ns} \over \pi^{\ns/2}}\;\cdot \\
& \cdot \l[a_1^2 \max(\Am^{\Re\s-\ns-3},\AM^{\Re\s-\ns-3})\,
\HE^{(\ns)}_N\!\l(\!\al,{a^2 \over T};{\ns\!+\!3\!-\!\Re\s \over 2}\,,
\Re\s\!-\!\ns\!-\!1\r) + \r. \\
& \hspace{1.5cm}\l. +\,{1 \over 2}\,\max(\Am^{\Re\s-\ns-1},\AM^{\Re\s-\ns-1})\,
\HE^{(\ns)}_N\!\l(\!\al,{a^2 \over T};{\ns\!+\!1\!-\!\Re\s \over 2}\,,
\Re\s\!-\!\ns\!-\!1\r)\r] \label{FoInfEst}
\end{split}\end{equation}
$$ \mbox{for either ~ $\Re\s \leqs 2$, $N > 2\sqrt{d}$} \quad
\mbox{or} \quad \mbox{$\Re\s > d\!+\!1$, $\dd{N > 2\sqrt{d}
+ {1 \over \Am}\sqrt{(\Re\s\!-\!2)T \over 2\al}}$} ~. $$
Concerning the case $2 < \Re \s \leqs d + 1$, not taken into account in
Eq. \rref{FoInfEst}, there hold considerations analogous to the ones
below Eq. \rref{EInfEst}; also in this case the corresponding reminder
estimates would involve rather cumbersome expressions which we choose
not to discuss here in view of the fact that we will be interested only
in the case $\s = 0$. \salto
The evaluation of the force on $\pi_{1,0}$ presented in the subsequent
subsections \ref{NumBox1d} \ref{NumBox2d} for $d=1,2$ will be based
on the truncated expansions and on the related remainder bounds discussed
in this paragraph.
\vspace{-0.4cm}
\subsection{Scaling considerations.} \label{Scal}
\label{notescaling} From Eq.s (\ref{TmnBox1}-\ref{Tidirijbox}) and from
the expressions for the Dirichlet functions given in subsection \ref{ACBox},
we easily infer the following relation for each component of the
stress-energy VEV ($\mu,\nu \in \{0,...,d\}$):
\beq \la 0 | \Tis_{\mu\nu}(\bx) | 0 \ra = a_1^{\s-d-1}\,
\mathrm{T}_{\mu\nu}^{\s}(\bxk; \mbox{$\boma{\rho}$})~, \label{TmnSc}\feq
where $\mathrm{T}_{\mu\nu}^{\s}$ is a suitable function and $\bxk$,
$\boma{\rho}$ are, respectively, the $d$-tuple and the $(d-1)$-tuple
with compoments
\beq \xk^i := {x^i \over a_i} \in (0,1)~\mbox{for $i \in \{1,...,d\}$} ~, \qquad
\rho_i := {a_i \over a_1} \quad \mbox{for $i \in \{2,...,d\}$} ~. \label{defxk} \feq
For $d=1$, the variables $\rho_i$ are not defined and $\mathrm{T}_{\mu\nu}^{\s}$
only depends on $\xk^1 = x^1/a^1$
({\footnote{It is apparent from Eq.s \rref{Fki} \rref{defCC} that the
eigenfunctions $F_{\bk}(\bx) := F_{k_1}(x^1)...F_{k_d}(x^d)$ and the
corresponding eigenvalues $\om_\bk^2$ can be written in the form
$$ F_\bk(\bx) = a_1^{-{d \over 2}}\; \varphi_{\bn,\boma{\rho}}(\bx_\star) ~,
\quad \om_\bk^2 = a_1^{-2} \lam_{\bn,\boma{\rho}}^2 \qquad
\mbox{for $\dd{k_i = {n_i \pi \over a_i}}$\, ($i \in \{1,...,d\}$)} ~, $$
for some suitable functions $\varphi_{\bn,\boma{\rho}}$, and some coefficients
$\lam_{\bn,\boma{\rho}}$. Using the eigenfunction expansion
for the Dirichlet kernel (see Eq. (3.19) in Part I), we obtain
$$ \Dir_s(\bx,\by) = a_1^{-(d-2)s} \mathfrak{D}_{s,\boma{\rho}}(\bx_\star,\by_\star)~, $$
for some suitable function $\mathfrak{D}_{s,\boma{\rho}}$; from the above
relation we can easily infer that, for any pair $z,w$ of spatial variables,
$$ \partial_{zw}\Dir_s(\bx,\by) = a_1^{-(d-2)s-2} \partial_{z_\star w_\star}
\mathfrak{D}_{s,\boma{\rho}}(\bx_\star,\by_\star) ~. $$
Using these results, one can easily deduce Eq. \rref{TmnSc} and the subsequent
statements of this subsection.}}).
Similarly, for the regularized pressure acting on any point $\bx$ in the interior
of the side $\pi_{1,0}$, we deduce from Eq.s \rref{pregBox} \rref{TmnSc} that
\beq p^\s_i(\bx) = a_1^{\s-d-1} \,\mathrm{p}^\s_i(\bxk;\mbox{$\boma{\rho}$})
\qquad \mbox{for $i \in \{1,...,d\}$} \label{pregSc} \feq
where $\mathrm{p}^\s_i$ are suitable functions and $\bxk$ is defined as
in Eq. \rref{defxk} at points on the boundary. Clearly, the same conclusions
can be drawn for the pressure on any other side $\pi_{\nn,\an}$
($\nn \in \{1,...,d\}$, $\an \in \{0,1\}$). \parn
Analogous considerations hold for the total energy and integrated force
on the boundary of the spatial domain. On the one hand, concerning the
bulk energy, from the expansions derived in subsection \ref{EnBox} we
easily infer (indicating with $\mathrm{E}^\s$ a suitable function)
\beq E^\s = a_1^{\s - 1}\,\mathrm{E}^\s(\mbox{$\boma{\rho}$}) ~. \feq
On the other hand, as for the total force on $\pi_{1,0}$ (for example),
from Eq.s  \rref{foSup1} \rref{foInf1} it follows that
\beq \fo_{1,0}^\s = a_1^{\s-2} \,\mathrm{F}_{1,0}^\s(\mbox{$\boma{\rho}$})
\label{ftotSc} \feq
for some suitable function $\mathrm{F}_{1,0}^\s$. Again, similar results
hold for the total force on any other side $\pi_{\nn,\an}$. \parn
By analytic continuation at $\s = 0$, we obtain the renormalized
counterparts of the above relations:
more precisely, we have
\begin{equation}\begin{split}
& \la 0 | \Ti_{\mu\nu}(\bx) | 0 \ra_{ren} = a_1^{-(d+1)}\,
\mathrm{T}_{\mu\nu}(\bxk;\mbox{$\boma{\rho}$}) ~, \qquad
p^{ren}_i(\bx) = a_1^{-(d+1)} \,\mathrm{p}_i(\bxk;\mbox{$\boma{\rho}$}) ~, \\
& \hspace{3.2cm} E^{ren} = a_1^{-1}\,\mathrm{E}(\mbox{$\boma{\rho}$}) ~,
\qquad \fo_{1,0}^{ren} = a_1^{-2}\,\mathrm{F}_{1,0}(\mbox{$\boma{\rho}$})
\label{rescEq}
\end{split}\end{equation}
(where the right-hand sides of the above relations are obtained evaluating at
$\s = 0$ the functions in the right-hand sides of Eq.s (\ref{TmnSc}-\ref{ftotSc}). \salto
Due to the remarks of this subsection, for any spatial dimension $d$
the analysis of the renormalized stress-energy VEV, total energy, pressure
and of the integrated force can always be reduced to the case $a_1 = 1$;
we will use this fact in the next section on the cases $d=1$ and $d=2$.
\section{The previous results in spatial dimension $\boma{d\!\in\!\{1,2\}}$}\label{Boxd12}
\subsection{Case $\boma{d=1}$ (the segment).} \label{NumBox1d}
\begin{figure}[h!]
    \centering
        \begin{subfigure}[b]{0.49\textwidth}
                \includegraphics[width=\textwidth]{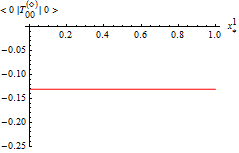}
        \end{subfigure}
        \begin{subfigure}[b]{0.49\textwidth}
                \includegraphics[width=\textwidth]{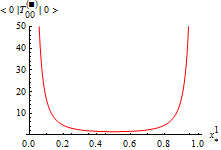}
        \end{subfigure}
       \caption{$d = 1$: graphs of $\la 0|\Ti^{(\Co)}_{00}|0\ra_{ren}$
       and $\la 0|\Ti^{(\NCo)}_{00}|0\ra_{ren}$\,.} \label{fig:T00Box1}
\end{figure}

\begin{figure}[h!]
    \centering
        \begin{subfigure}[b]{0.49\textwidth}
                \includegraphics[width=\textwidth]{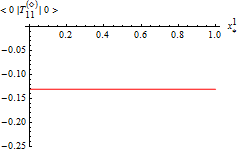}
        \end{subfigure}
        \begin{subfigure}[b]{0.49\textwidth}
                \includegraphics[width=\textwidth]{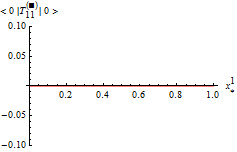}
        \end{subfigure}
       \caption{$d = 1$: graphs of $\la 0|\Ti^{(\Co)}_{11}|0\ra_{ren}$
       and $\la 0|\Ti^{(\NCo)}_{11}|0\ra_{ren}$\,.}
       \label{fig:T11Box1}
\end{figure}

As a first application of our framework, let us consider the case
\beq d = 1 ~, \qquad \Om = (0,a_1) \quad (a_1 > 0) ~. \label{presc1} \feq
As a matter of fact, we already analysed this configuration in Section
6 of Part I; the present Eq. \rref{presc1} corresponds to Eq. (6.1)
of Part I (with $a = a_1$). \parn
In Part I we performed the exact computation for the renormalized VEV
of the stress-energy tensor and of the pressure for various types of
boundary conditions.
Here we carry out an approximate evaluation of $\la 0|\Ti_{\mu\nu}(\bx)|0\ra_{ren}$,
$p^{ren}_1(\bx)|_{x^1 = 0}$ and of $E^{ren}$ for Dirichlet boundary
conditions, truncating the series expansions for these quantities derived
in the present work for a box in arbitrary spatial dimension. Our aim
is just to check the validity of the general methods developed here;
to this purpose, we compare the results obtained for the renormalized
stress-energy VEV, pressure and total energy, respectively,
with those reported in Eq.s (6.24), (6.26) and (6.27) of Part I. \parn
In our computations we only consider the case with
\beq a_1 = 1 ~, \label{a1} \feq
which cause no loss of generality due to the scaling considerations
discussed in subsection \ref{Scal} (of course, due to Eq. \rref{a1}, the
rescaled variable $x_\star^1 := x^1 / a_1$ in fact coincides with $x^1$)
({\footnote{\label{footd1} Let us stress that, for $a_1 = 1$, the
quantities $\la 0|\Ti_{\mu\nu}|0\ra_{ren}$ and $p^{ren}_1$ are,
respectively, equal to the rescaled functions $\mathrm{T}_{\mu\nu}^{ren}$
and $\mathrm{p}^{ren}_i$ (see Eq. \rref{rescEq}).}}). \parn
Before proceeding to the evaluation of the renormalized VEVs
$\la 0|\Ti_{\mu\nu}|0\ra_{ren}$ and $p^{ren}_1$, let us recall
the representation (\ref{Tidir00box}-\ref{Tidirijbox}) of the regularized
stress-energy VEV in terms of the $(>)$ and $(<)$ parts of the Dirichlet
kernel; these parts depend on the choice of a parameter $T>0$, which
however has no effect on the sum $\Dir_s = \Dir^{(>)}_s + \Dir^{(<)}_s$.
After fixing $T$, we can approximate $\Dir^{(>)}_s$ and $\Dir^{(<)}_s$
truncating their series expansions at some sufficiently large order $N$,
giving estimates on the remainders as well. Analogous considerations
hold for the spatial integral of the diagonal Dirichlet kernel,
ultimately giving the renormalized bulk energy $E^{ren}$ (see
Eq.s \rref{defEs} \rref{EnSupInf} and \rref{ERenBox}). \parn
Here we choose
\beq T = 1 ~, \qquad N = 5 ~;  \label{N6} \feq
the truncated sums are evaluated numerically, and the remainder estimates
(\ref{DirSupEst}-\ref{DzwRRInfEst}) and \rref{ESupEst} \rref{EInfEst}
are also taken into account, fixing
({\footnote{\label{footal}We make the choice \rref{al3} for $\al$ because
it is close to the value minimizing the error estimates cited above; a more
precise evaluation of this optimal value for $\al$ would require a laborious
numerical analysis which we prefer to avoid here.}})
\beq \al = 0.03 ~. \label{al3} \feq
Of course, the bounds obtained for the remainders associated to $\Dir^{(>)}_s$
and $\Dir^{(<)}_s$ allow us, in turn, to infer error estimates for the
approximate expressions of both $\la 0|\Ti_{\mu\nu}|0\ra_{ren}$ and
$p^{ren}_1$. \salto
Let us first consider the \textsl{renormalized stress-energy VEV}
$\la 0|\Ti_{\mu\nu}|0\ra_{ren}$; according to the analysis of
subsection \ref{ACBoxTmn}, this is obtained setting $\s = 0$ in
Eq.s (\ref{TmnBox1}-\ref{Tidirijbox}). In reporting our results,
we distinguish between the conformal and non-conformal parts of
each component, which are respectively denoted as usual with
$\la 0|\Ti^{(\Co)}_{\mu\nu}|0\ra_{ren}$ and
$\la 0|\Ti^{(\NCo)}_{\mu\nu}|0\ra_{ren}$ (see Eq. \rref{TRinCo});
notice that (since $d = 1$) Eq. \rref{xic} gives
\beq \xi_1 = 0 ~. \feq
The graphs of the conformal and noncomformal parts of each
stress-energy component are shown in Fig.s \ref{fig:T00Box1}
and \ref{fig:T11Box1} (recall that we refer to the rescaled
variable $x_\star^1 := x^1 / a_1 \equiv x^1$).
Let us comment briefly on the above graphs. Apart from
$\la 0|\Ti^{(\NCo)}_{00}|0\ra_{ren}$, it appears that all the
other components of the renormalized stress-energy VEV are constants
and $\la 0|\Ti^{(\NCo)}_{11}|0\ra_{ren}$ is very small;
indeed, our computations with $N = 5$ ($T=1$, $\al = 0.03$) ensure,
for all $\bx \equiv \xk^1 \in (0,1)$,
\begin{equation}\begin{split}
& \la 0|\Ti^{(\Co)}_{00}(\bx)|0\ra_{ren} = \la 0|\Ti^{(\Co)}_{11}(\bx)|0\ra_{ren}
= -\,0.1308997 \pm 8 \cdot 10^{-7} ~, \\
& \hspace{3cm} |\la 0|\Ti^{(\NCo)}_{11}(\bx)|0\ra_{ren}| \leqs 4 \cdot 10^{-6} ~.
\label{T1N1}
\end{split}\end{equation}
Concerning $\la 0|\Ti^{(\NCo)}_{00}(\bx)|0\ra_{ren}$ we have, for
example,
\begin{equation}\begin{split}
& \la 0|\Ti^{(\NCo)}_{00}(\bx)|0\ra_{ren} = \l\{\! \barray{ll}
3.141593 \pm 10^{-6} & \mbox{for $\xk^1 = 1/4$} \\
1.570796 \pm 10^{-6} & \mbox{for $\xk^1 = 1/2$} \farray\r. ~.
\label{T1N2}
\end{split}\end{equation}
These results are in agreement with the exact calculations
of subsection 6.6 of Part I, which gave the following outcomes
(see Eq. (6.24) of the cited subsection with $a \equiv a_1 = 1$)
({\footnote{To make a comparison with Eq.s \rref{T1N1} \rref{T1N2},
note that
$$ -{\pi \over 24} = -0.13089969... ~, \qquad
{\pi \over 2 \sin^2(\pi x^1)} = \l\{\! \barray{ll}
3.14159265... & \mbox{for $x^1 = 1/4$} \\
1.57079633... & \mbox{for $x^1 = 1/2$} \farray\r. \,. $$}})
\begin{equation}\begin{split}
& \hspace{1.5cm} \la 0|\Ti^{(\Co)}_{0 0}(\bx)|0\ra_{ren} =
\la 0|\Ti^{(\Co)}_{1 1}(\bx)|0\ra_{ren} = - {\pi \over 24} ~, \\
& \la 0|\Ti^{(\NCo)}_{0 0}(\bx)|0\ra_{ren} = {\pi \over 2 \sin^2(\pi x^1)} ~,
\qquad \la 0|\Ti^{(\NCo)}_{1 1}(\bx)|0\ra_{ren} = 0 ~.
\end{split}\end{equation}
Next, let us pass to the evaluation of
\beq F_{ren}(0) := p^{ren}_1(\bx)\Big|_{x^1 = 0} ~; \feq
this is nominally the ``pressure'' on the boundary point $x^1 = 0$
but in fact coincides with the \textsl{force} on this point, due to the
zero dimensionality of the boundary. \parn
Let us consider the prescriptions \rref{alt1Box}; computing the derivative
of the functions $\Dir^{(>)}_{1/2}$, $\Dir^{(<)}_{1/2}$ appearing therein
with the choices $a_1 = 1$, $T = 1$ and $N = 5$, we obtain
\beq F_{ren}(0) = 0.1308997 \pm 3 \cdot 10^{-7} \feq
(again, the error is obtained using the remainder estimates of subsection
\ref{FoRem} with $\al = 0.03$). The above result is in agreement with the
exact expression $F_{ren}(0) = {\pi \over 24}$ derived in subsection 6.6
of Part I (see Eq.s (6.24) (6.27), and set $a \equiv a_1 = 1$ therein). \salto
Finally, we consider the \textsl{renormalized bulk energy} $E^{ren}$;
the series expansions \rref{ESupExp} \rref{EInfExp} derived in subsection
\ref{EnBox} (with the previous choices of $a_1,T,N$) allow us to infer
\beq E^{ren} = -0.1308996938996 \pm 3 \cdot 10^{-13} \label{E1App} \feq
(where the error is obtained using the remainder estimates of subsection
\ref{EnRem}, again with $\al = 0.03$). The result \rref{E1App} agrees
with the exact computation $E^{ren} = - {\pi \over 24}$ obtained in
subsection 6.6 of Part I (see Eq. (6.26), and set $a \equiv a_1 = 1$ therein). \salto
Let us stress that our approximants by truncation, converge
quite rapidly to the exact results; in order to exemplify this
statement, we notice that, by slightly increasing the value of
the truncation order $N$, we obtain a remarkable improvement of
the error estimates. For example (using again the estimates of
subsection \ref{EnRem} with $T = 1$ and $\al = 0.03$) the error
$\pm 3 \cdot 10^{-13}$ in Eq. \rref{E1App} (for $N = 5$) becomes
$\pm2 \cdot 10^{-46}$ for $N = 10$, $6 \cdot 10^{-101}$ for $N = 15$
and $4 \cdot 10^{-177}$ for $N = 20$.
\vspace{-0.4cm}
\subsection{Case $\boma{d=2}$.} \label{NumBox2d}
\begin{figure}[h!]
    \centering
        \begin{subfigure}[b]{0.49\textwidth}
                \includegraphics[width=\textwidth]{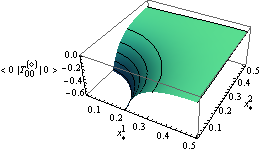}
        \end{subfigure}
        \begin{subfigure}[b]{0.49\textwidth}
                \includegraphics[width=\textwidth]{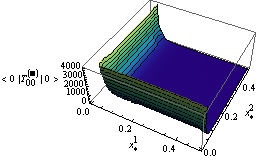}
        \end{subfigure}
       \caption{$d = 2$: graphs of $\la 0|\Ti^{(\Co)}_{00}(\bx)|0\ra_{ren}$
       and $\la 0|\Ti^{(\NCo)}_{00}(\bx)|0\ra_{ren}$ for $a_2 = 1$\,.} \label{fig:T00Box2}
\end{figure}

\begin{figure}[h!]
    \centering
        \begin{subfigure}[b]{0.49\textwidth}
                \includegraphics[width=\textwidth]{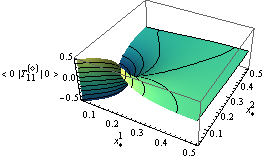}
        \end{subfigure}
        \begin{subfigure}[b]{0.49\textwidth}
                \includegraphics[width=\textwidth]{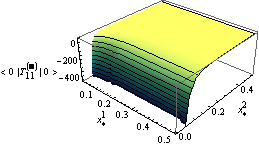}
        \end{subfigure}
       \caption{$d = 2$: graphs of $\la 0|\Ti^{(\Co)}_{11}(\bx)|0\ra_{ren}$
       and $\la 0|\Ti^{(\NCo)}_{11}(\bx)|0\ra_{ren}$ for $a_2 = 1$\,.} \label{fig:T11Box2}
\end{figure}

\begin{figure}[h!]
    \centering
        \begin{subfigure}[b]{0.49\textwidth}
                \includegraphics[width=\textwidth]{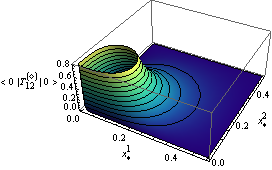}
        \end{subfigure}
        \begin{subfigure}[b]{0.49\textwidth}
                \includegraphics[width=\textwidth]{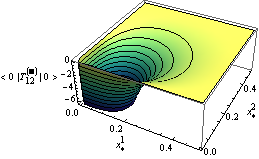}
        \end{subfigure}
       \caption{$d = 2$: graphs of $\la 0|\Ti^{(\Co)}_{12}(\bx)|0\ra_{ren}$
       and $\la 0|\Ti^{(\NCo)}_{12}(\bx)|0\ra_{ren}$ for $a_2 = 1$\,.} \label{fig:T12Box2}
\end{figure}

\begin{figure}[h!]
    \centering
        \begin{subfigure}[b]{0.49\textwidth}
                \includegraphics[width=\textwidth]{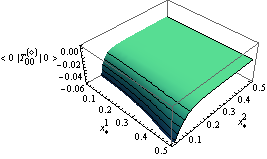}
        \end{subfigure}
        \begin{subfigure}[b]{0.49\textwidth}
                \includegraphics[width=\textwidth]{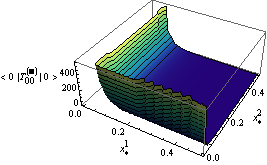}
        \end{subfigure}
       \caption{$d = 2$: graphs of $\la 0|\Ti^{(\Co)}_{00}(\bx)|0\ra_{ren}$
       and $\la 0|\Ti^{(\NCo)}_{00}(\bx)|0\ra_{ren}$ for $a_2 = 5$\,.} \label{fig:T00Box2b}
\end{figure}

\begin{figure}[h!]
    \centering
        \begin{subfigure}[b]{0.49\textwidth}
                \includegraphics[width=\textwidth]{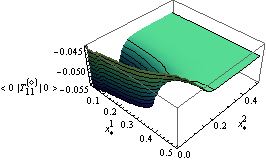}
        \end{subfigure}
        \begin{subfigure}[b]{0.49\textwidth}
                \includegraphics[width=\textwidth]{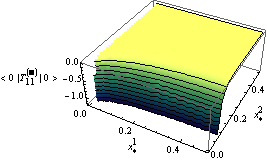}
        \end{subfigure}
       \caption{$d = 2$: graphs of $\la 0|\Ti^{(\Co)}_{11}(\bx)|0\ra_{ren}$
       and $\la 0|\Ti^{(\NCo)}_{11}(\bx)|0\ra_{ren}$ for $a_2 = 5$\,.} \label{fig:T11Box2b}
\end{figure}

\begin{figure}[h!]
    \centering
        \begin{subfigure}[b]{0.49\textwidth}
                \includegraphics[width=\textwidth]{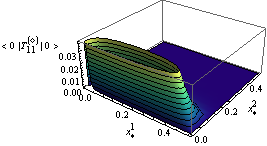}
        \end{subfigure}
        \begin{subfigure}[b]{0.49\textwidth}
                \includegraphics[width=\textwidth]{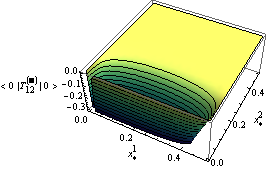}
        \end{subfigure}
       \caption{$d = 2$: graphs of $\la 0|\Ti^{(\Co)}_{12}(\bx)|0\ra_{ren}$
       and $\la 0|\Ti^{(\NCo)}_{12}(\bx)|0\ra_{ren}$ for $a_2 = 5$\,.} \label{fig:T12Box2b}
\end{figure}

\begin{figure}[h!]
    \centering
        \begin{subfigure}[b]{0.49\textwidth}
                \includegraphics[width=\textwidth]{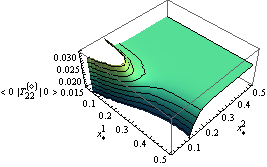}
        \end{subfigure}
        \begin{subfigure}[b]{0.49\textwidth}
                \includegraphics[width=\textwidth]{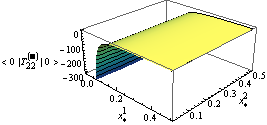}
        \end{subfigure}
       \caption{$d = 2$: graphs of $\la 0|\Ti^{(\Co)}_{22}(\bx)|0\ra_{ren}$
       and $\la 0|\Ti^{(\NCo)}_{22}(\bx)|0\ra_{ren}$ for $a_2 = 5$\,.} \label{fig:T22Box2b}
\end{figure}

\begin{figure}[h!]
     \centering
         \begin{subfigure}[b]{0.49\textwidth}
                 \includegraphics[width=\textwidth]{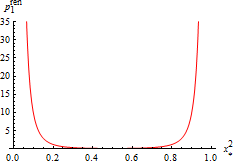}
         \end{subfigure}
         \begin{subfigure}[b]{0.49\textwidth}
                 \includegraphics[width=\textwidth]{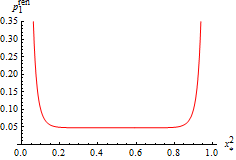}
         \end{subfigure}
        \caption{$d = 2$: graphs of $p_1^{ren}$ for $a_2 = 1$ (left)
        and $a_2 = 5$ (right).} \label{fig:PBox2}
\end{figure}

\begin{figure}[h!]
     \centering
         \begin{subfigure}[b]{0.49\textwidth}
                 \includegraphics[width=\textwidth]{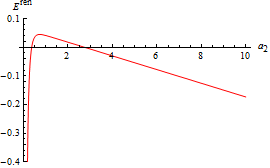}
         \end{subfigure}
     \caption{$d = 2$: graph of $E^{ren}$ as function of $a_2$\,.}
     \label{fig:EnBox}
\end{figure}

\begin{figure}[h!]
     \centering
         \begin{subfigure}[b]{0.49\textwidth}
                 \includegraphics[width=\textwidth]{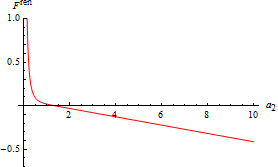}
         \end{subfigure}
     \caption{$d = 2$: graph of $\fo_{1,0}^{ren}$ as function of $a_2$\,.}
     \label{fig:FBox}
\end{figure}

Let us now pass to the $2$-dimensional case:
\beq d = 2 ~, \qquad \Om = (0,a_1)\times(0,a_2) \quad (a_1,a_2 > 0) ~. \feq
As in the previous subsection, we fix
\beq a_1 = 1 \label{a12} \feq
and consider different values of $a_2$; let us repeat that
the above choice does not imply a loss of generality, due
to the scaling properties of subsection \ref{notescaling}.
Moreover, we present the final results in terms of the rescaled
coordinates $x_\star^1 := x^1 /a_1 \equiv x^1$, $x_\star^2 :=
x^2/a_2 \in (0,1)$, defined in Eq. \rref{defxk}
({\footnote{\label{footd2} Similarly to what we said in the
footnote \ref{footd1} on page \pageref{footd1}, for $a_1 = 1$,
the quantities $\la 0|\Ti_{\mu\nu}|0\ra_{ren}$, $p^{ren}_1$
and $\fo_{1,0}^{ren}$ (to be discussed hereafter) do in fact
coincide with the rescaled analogues $\mathrm{T}_{\mu\nu}^{ren}$,
$\mathrm{p}^{ren}_i$, $\mathrm{F}_{1,0}^{ren}$ introduced in
Eq.s \rref{TmnSc} \rref{pregSc} \rref{ftotSc}). Besides, the
lenght $a_2$ of the second side is identified with the ratio
$\rho_2$ (see Eq. \rref{defxk}).}}). \salto
Again, the basic elements to compute the renormalized stress-energy
VEV and the pressure are the Dirichlet functions $\Dir_s^{(>)}$,
$\Dir_s^{(<)}$, along with their spatial derivatives, for which we
use the truncated expansions (\ref{DirSupEst}-\ref{DzwRRInfEst})
and the remainder bounds of Eq.s (\ref{DirSupEst}-\ref{DzwRRInfEst}). \parn
Needless to say, analogous considerations also hold for the renormalized
bulk energy and for the integrated boundary forces (see subsections \ref{EnBox}
and \ref{TotForBox}, respectively). \salto
Let us first consider the stress-energy VEV and the pressure; as examples,
we compute these observables for the two configurations with
\beq a_2 = 1 \qquad \mbox{and} \qquad a_2 = 5 ~. \feq
In these cases, for the parameter $T$ of the decomposition into $(>)$ and
$(<)$ parts and for the truncation order $N$, we make the following choices:
\begin{equation}\begin{split}
& T = 1 ~, \quad N = 7 \qquad \mbox{for $a_2 = 1$} ~; \\
& T = 1 ~, \quad N = 9 \qquad \mbox{for $a_2 = 5$} ~. \label{N10}
\end{split}\end{equation}
The truncation errors in Eq.s (\ref{DirSupEst}-\ref{DzwRRInfEst}) are
evaluated making for the parameter $\al$ therein the choice
({\footnote{This choice can be justified by considerations similar to
the ones in the footnote \ref{footal} of page \pageref{footal}.}})
\beq \al = 0.04 ~. \feq
The \textsl{renormalized stress-energy VEV} $\la 0|\Ti_{\mu\nu}|0\ra_{ren}$
is obtained setting $\s = 0$ in Eq.s (\ref{TmnBox1}-\ref{Tidirijbox}).
Again, we separate the conformal and nonconformal parts, respectively
indicated by the superscripts $(\Co)$ and $(\NCo)$; recall that
Eq. \rref{xic} gives, in the two-dimensional case,
\beq \xi_2 = 1/8 ~. \feq
In the following we present the graphs for $\la 0|\Ti^{(\Co)}_{\mu\nu}|0\ra_{ren}$
and $\la 0|\Ti^{(\NCo)}_{\mu\nu}|0\ra_{ren}$ obtained from the previous
truncated expansions; more precisely, Fig.s \ref{fig:T00Box2}-\ref{fig:T12Box2}
and Fig.s \ref{fig:T00Box2b}-\ref{fig:T22Box2b} show, respectively,
the results obtained for the configurations with $a_2 = 1$ and
$a_2 = 5$\,. In the cited figures we refer to the variables
$x_\star^i := x^i /a_i \in (0,1)$ and, keeping into account
some obvious symmetry considerations
({\footnote{Indeed, every component of the stress-energy VEV can be
shown to be symmetric under the exchange $x^i \leftrightarrow a_i - x^i$
(or $x^i_\star \leftrightarrow 1 - x^i_\star$) for $i\in\{1,2\}$\,.}}),
we only show the graphs for
\beq x_\star^i \in (0,1/2) \qquad \mbox{for $i \in \{1,2\}$} ~; \feq
moreover, in the case of a square box with $a_1 = a_2 = 1$ we do
not report the graphs for the conformal and non-conformal parts
of $\la 0|\Ti_{22}(\bx)|0\ra_{ren}$, since these are equal to the
corresponding parts of $\la 0|\Ti_{11}(\bx)|0\ra_{ren}$.
\eject \vfill \noindent \\
\eject \vfill \noindent \\
\eject \vfill \noindent \\
Concerning the error estimates, for $\mu,\nu \in \{0,1,2\}$ and
$\bullet \in \{\Co,\NCo\}$, let us introduce the following notation:
\beq \ET_{\mu\nu}^{(\bullet)} := \barray{c}\mbox{remainder
corresponding to our approximation} \\ \mbox{by truncation of
$\la 0|\Ti_{\mu\nu}^{(\bullet)}|0\ra_{ren}$\,.} \farray \feq
For $a_2 = 1$, our choice $N = 7$ ($T = 1$, $\al = 0.04$) yields
the uniform bounds
\begin{equation}\begin{split}
& \hspace{1.7cm} |\ET_{00}^{(\Co)}| \leqs 2 \cdot 10^{-12} ~,
\qquad |\ET_{00}^{(\NCo)}| \leqs 2 \cdot 10^{-11} ~; \\
& |\ET_{ij}^{(\Co)}| \leqs 6 \cdot 10^{-12} ~, \qquad
|\ET_{ij}^{(\NCo)}| \leqs 3 \cdot 10^{-11} \qquad \mbox{for $i,j \in \{1,2\}$} ~.
\end{split}\end{equation}
For $a_2 = 5$, our choice $N = 9$ ($T = 1$, $\al = 0.04$) ensures
\begin{equation}\begin{split}
& \hspace{1.8cm} |\ET_{00}^{(\Co)}| \leqs 5 \cdot 10^{-12} ~,
\qquad |\ET_{00}^{(\NCo)}| \leqs 4 \cdot 10^{-11} ~; \\
& |\ET_{ij}^{(\Co)}| \leqs 2 \cdot 10^{-11} ~, \qquad
|\ET_{ij}^{(\NCo)}| \leqs 8 \cdot 10^{-11} \qquad \mbox{for $i,j \in \{1,2\}$} ~.
\end{split}\end{equation}
Now, let us evaluate the \textsl{pressure} $p_i^{ren}(\bx)$ at points
$\bx$ of one side; as in the construction of the general theory we
consider, as an example, the points $\bx \equiv (0,x^2)$ in the
interior of the side $\pi_{1,0}$, making reference to the prescription
\rref{alt1Box}. \parn
Fig. \ref{fig:PBox2} shows the graphs obtained for $p_1^{ren}(\bx)$ as
a function of $x^2_\star := x^2/a_2$ (again, choosing $T = 1$ and
truncating the related expansions to order $N = 7$, for $a_2 = 1$, and
$N = 9$, for $a_2 = 5$).
\parn
As for the error, indicating with $\Ep_1$ the remainder associated
to our approximation by truncation of $p_1^{ren}$, we obtain the
following uniform estimates (setting $\al = 0.04$):
\begin{equation}\begin{split}
& |\Ep_1| \leqs 2 \cdot 10^{-12} \qquad \mbox{for $a_2 = 1$} ~; \\
& |\Ep_1| \leqs 5 \cdot 10^{-12} \qquad \mbox{for $a_2 = 5$} ~.
\end{split}\end{equation}
Before moving on, let us briefly comment on the behaviour of the
renormalized pressure $p_1^{ren}$ near the edge $\bx = \b0$. Indeed,
specializing to the present $d = 2$ case the considerations of
paragraph \ref{parNI} one can prove that, for all $a_2 > 0$,
\beq p_1^{ren}(\bx) = {1 \over 32\pi (x^2)^3} + O((x^2)^2) \qquad
\mbox{for $\bx = (0,x^2)$ and $x^2 \to 0^+$} ~. \feq
Now, let us pass to the computation of the bulk energy and of the
integrated force. For each one of these two observables we consider
several configurations, corresponding to different values of $a_2$;
for any one of these values, we consider the decomposition into
$(>)$ and $(<)$ parts, and choose the truncation order $N$ of the
related expansions so as to obtain error estimates all of approximatively
the same order of magnitude, fixing again
\beq \al = 0.04 ~. \feq
In order to obtain the \textsl{renormalized bulk energy} $E^{ren}$,
we first consider its regularized version \rref{EnSupInf}, along
with the series expansions \rref{ESupExp} \rref{EInfExp}, giving
the analytic continuations of the functions $E^{\s,(>)}$ and
$E^{\s,(<)}$, respectively. Due to the considerations of subsection
\ref{EnBox}, we can simply put $\s = 0$ in these expansions,
since no singularity arises there; next, we truncate the corresponding
series at a suitable order $N$ (see the comments above) and use the
remainder estimates \rref{ESupEst} \rref{EInfEst}. \parn
In this way we obtain, for example, the following results: for $a_1 = 1$
(and with the choice $T = 1$, as usual)
\beq \barray{lll}
E^{ren} = -\,1.73691776 \pm 4 \cdot 10^{-8} \quad & \mbox{for $a_2 = 0.1$} & (N = 40) ~; \\
E^{ren} = +\,0.03524178 \pm 3 \cdot 10^{-8} \quad & \mbox{for $a_2 = 0.5$} & (N = 8) ~; \\
E^{ren} = +\,0.04104060 \pm 9 \cdot 10^{-8} \quad & \mbox{for $a_2 = 1$} & (N = 4) ~; \\
E^{ren} = -\,0.05412096 \pm 2 \cdot 10^{-8} \quad & \mbox{for $a_2 = 5$} & (N = 7) ~; \\
E^{ren} = -\,0.17369178 \pm 1 \cdot 10^{-8} \quad & \mbox{for $a_2 = 10$} & (N = 15) ~.
\farray \feq
Again for $a_1 = 1$, one can plot the renormalized bulk energy $E^{ren}$
as a function of $a_2$; the graph in Fig. \ref{fig:EnBox} has been obtained
using the truncation order $N = 50$ (with $T = 1$; with the choices made for
the parameters $N,T,a_1$, we know that the remainder is smaller than
$2 \cdot 10^{-3}$ for $a_2 \in [0.05,10)$).
\parn
Let us discuss some facts regarding the function $a_2 \mapsto E^{ren}(a_2)$,
which can be read from the graph in Fig. \ref{fig:EnBox} (the results reported
hereafter are obtained using standard numerical methods, implemented in
$\tt{Mathematica}$)
({\footnote{The derivation of the results in items i)-iv), including
the error estimates, should be accounted for; as an example, let us
give some details on Eq. \rref{a2EnMax}. To obtain this result we have
used for $E^{ren}$ an approximant by truncation at order $N=50$ that,
according to the reminder estimates \rref{ESupEst} and \rref{EInfEst},
gives $E^{ren}$ up to an error $\leqs 10^{-272}$ for $a_1 =1$ and
$0.5 \leqs a_2 \leqs 1$.
This large order approximant of $E^{ren}$ has been maximized numerically
with respect to $a_2$ via Mathematica, asking for a precision of order
$10^{-15}$ on the maximum point. The final results have been prudentially
truncated to $8$ digits, yielding Eq. \rref{a2EnMax}; in view of the
previous considerations, they are very likely even though not certified.}}). \parn
i) There is only one point of maximum $a_2^{\max}$; our approximation
by truncation at order $N = 50$ gives
\begin{equation}\begin{split}
& \hspace{0.7cm} a_2^{\max} = 0.72719110 \pm 10^{-8} ~, \\
& E^{ren}(a_2^{\max}) = 0.04472675 \pm 10^{-8} ~. \label{a2EnMax}
\end{split}\end{equation}
ii) $E^{ren}$ vanishes for two values $\bar{a}_2^{(1)} < \bar{a}_2^{(2)}$
of $a_2$; these are found to be
\begin{equation}\begin{split}
& \hspace{0.1cm} \bar{a}_2^{(1)} = 0.36538151 \pm 10^{-8} ~, \\
& \bar{a}_2^{(2)} = 2.73686534 \pm 10^{-8} ~. \label{a2EnZer}
\end{split}\end{equation}
$E^{ren}$ is positive for $\bar{a}_2^{(1)}\!< a_2 < \bar{a}_2^{(2)}$
and negative elsewhere. This feature was also pointed out in \cite{MaTru1};
therein it is stated that $\bar{a}_2^{(2)} = (\bar{a}_2^{(1)})^{-1}$, a
relation (approximately) verified by the numerical values in Eq. \rref{a2EnZer}.\parn
iii) For $a_2 \to 0^+$, $E^{ren}$ has the asymptotic behaviour
\beq E^{ren}(a_2) = {e_0 \over (a_2)^2}\; \Big(1 + O(a_2)\Big) \quad \mbox{with}
\quad e_0 = -0.02391 \pm 10^{-5} ~. \feq
iv) There are indications that $E^{ren}$ approaches an asymptote for $a_2 \to +\infty$;
taking into account values of the abscissa up to $a_2 = 100$, we find that
this asymptote is the straight line
\begin{equation}\begin{split}
y = m_E\,a_2 + q_E \qquad \mbox{with}\quad \l\{\!\barray{l}
m_E = -\,0.02391416 \pm 10^{-8} \\
q_E = +\,0.06544985 \pm 10^{-8} \farray\r. \,.
\end{split}\end{equation}
Finally, let discuss the \textsl{renormalized total force} $\fo_{1,0}^{ren}$
acting on the side $\pi_{1,0}$; following the analysis of subsection
\ref{TotForBox}, we consider the expression \rref{foren} and represent
the functions $\fo_{1,0}^{\s,(>)}$, $\fo_{1,0}^{\s,(<)}$ appearing
therein using the series expansions \rref{foSup1} \rref{foInf1}.
$\fo_{1,0}^{ren}$ is obtained setting $\s = 0$ in these series
expansions; for the actual calculation, these can be truncated
at a fixed, sufficiently large order $N$ using the remainder
estimates \rref{FoSupEst} \rref{FoInfEst}. \parn
In this way we obtain, for example, the following results: for $a_1 = 1$
(and with the choice $T = 1$, as usual)
\beq \barray{lll}
\fo_{1,0}^{ren} = +\,2.3914162 \pm 4 \cdot 10^{-7} \quad & \mbox{for $a_2 = 0.1$} & (N = 50) ~; \\
\fo_{1,0}^{ren} = +\,0.0956400 \pm 2 \cdot 10^{-7} \quad & \mbox{for $a_2 = 0.5$} & (N = 9) ~; \\
\fo_{1,0}^{ren} = +\,0.020520  \pm 6 \cdot 10^{-6} \quad & \mbox{for $a_2 = 1$} & (N = 4) ~; \\
\fo_{1,0}^{ren} = -\,0.173692  \pm 4 \cdot 10^{-6} \quad & \mbox{for $a_2 = 5$} & (N = 7) ~; \\
\fo_{1,0}^{ren} = -\,0.412833  \pm 3 \cdot 10^{-6} \quad & \mbox{for $a_2 = 10$} & (N = 15) ~.
\farray \feq
Again for $a_1 = 1$, one can plot the renormalized force $\fo_{1,0}^{ren}$
as a function of $a_2$; the graph in Fig. \ref{fig:FBox} has been obtained
using the truncation order $N = 60$ (with $T = 1$; with the choices made for
the parameters $N,T,a_1$, we know that the remainder is smaller than
$2 \cdot 10^{-3}$ for $a_2 \in [0.07,10)$).
\parn
In the following we briefly discuss a number of facts concerning
the function $a_2 \mapsto \fo_{1,0}^{ren}(a_2)$, which appear
in the above graph. \parn
i) The function  under analysis is strictly decreasing for $a_2 \in (0,+\infty)$\,. \parn
ii) There is only one value of $a_2$, which we indicate with $\bar{a}_2$,
where $\fo_{1,0}^{ren}$ vanishes; our approximation by truncation at
order $N = 60$ allows us to infer
\beq \bar{a}_2 = 1.3751543 \pm 10^{-7} ~. \label{a2ZerFo} \feq
iii) For $a_2 \to 0^+$, $\fo_{1,0}^{ren}$ has the asymptotic behaviour
\beq \fo_{1,0}^{ren}(a_2) = {f_0 \over (a_2)^2}\; \Big(1 + O(a_2)\Big) \quad \mbox{with}
\quad f_0 = 0.02391 \pm 10^{-5} ~. \feq
iv) It appears that $\fo_{1,0}^{ren}$ approaches an asymptote for
$a_2 \to +\infty$; considering again values of the abscissa up
to $a_2 = 100$, the equation of this asymptote is found to be
\begin{equation}\begin{split}
y = m\,a_2 + q \qquad \mbox{with}\quad \l\{\!\barray{l}
m = -\,0.0478283 \pm 10^{-7} \\
q = +\,0.0654498 \pm 10^{-7} \farray\r. \,.
\end{split}\end{equation}
Let us mention that, in agreement with the general considerations of
subsection 4.4 in Part I, the above results for the renormalized total
force on the side $\pi_{1,0}$ of the box could be equivalently derived by differentiating
the renormalized total energy $E^{ren}$ with respect to the lenght
$a_2$ of the edge of the box perpendicular to $\pi_{1,0}$.
\salto
To conclude, let us compare the previous results about $E^{ren}, \fo_{1,0}^{ren}$
with the calculations of Bordag et al. \cite{Bord}. The authors of \cite{Bord} derive
(both by Abel-Plana formula and by zeta regularization) series
expansions different from ours for the renormalized bulk energy
and for the force on one side; besides, they give no remainder estimates
for these expansions ({\footnote{In fact, a not so trivial analysis
(which we do not report here for brevity) allows to conclude that
the series expansions given in \cite{Bord} also converge with
exponential speed.}}). \parn
The numerical values of $E^{ren}, \fo_{1,0}^{ren}$ given by our
previous analysis are in good agreement with those arising from the
expansions in \cite{Bord}, a fact strongly indicating the equivalence
between our approach and \cite{Bord}. Let us also mention that the results
of \cite{Bord} about $E^{ren}$ or $\fo_{1,0}^{ren}$ are equivalent
to the ones of \cite{Wolf1,10AppZ,FulKir}.
\vskip 0.2cm \noindent
\textbf{Acknowledgments.}
This work was partly supported by INdAM, INFN and by MIUR, PRIN 2010
Research Project  ``Geometric and analytic theory of Hamiltonian systems in finite and infinite dimensions''.
\vfill \eject \noindent
\appendix
\section{Appendix. Absolute convergence and remainder bounds for the series of subsections \ref{ACBox}.}\label{AppBox}
Let us consider the general framework of the subsections mentioned in the
title. Note that the representation \rref{DirSupExp} for the function
$\Dir_s^{(>)}(\bx,\by)$ contains the series
\beq \sum_{\bn \in \naturali^d}\om_\bn^{-2s}\,\Ga(s,\om_\bn^2\,T)\;\CB_\bn(\bx,\by) ~.
\label{AX} \feq
On the other hand, in the expansion \rref{DirInfExp} for $\Dir_s^{(<)}(\bx,\by)$
there appears a series over $\bh\!\in\!\interi^d$ and $\bl\!\in\!\{1,2\}^d$\,,
that we have reexpressed in the equivalent form \rref{DirInfExp2}.
We pointed out in paragraph \ref{subPole} that the terms in this series with
$\bhl(\bx,\by) = 0$ become singular at $s = {d/2}$\,; these have
either $|\bh| = 0$ or $|\bh| = 1$ (see Eq. \rref{bkl2}) and, after removing all terms
with $|\bh| = 0,1$ we are left with the series
\beq \sum_{\substack{\bh \in \interi^d,\,|\bh| > 1 \\ \bl \in \{1,2\}^d}} \de_\bl\;
\PP_{s - {d \over 2}}\!\l({\bhl(\bx,\by) \over T}\r) \,. \label{AY} \feq
Note that the absolute convergence of the series \rref{AY} is equivalent to
the absolute convergence of the series in the second line of Eq. \rref{DirInfExp2}. \parn
In the next subsections we prove the convergence of the series \rref{AX}
\rref{AY} for all $s \in \complessi$\,, deriving as well remainder estimates
for both of them; the statements in Eq.s \rref{DirSupExp} \rref{RRSupEst}
(related to $\Dir_s^{(>)}(\bx,\by)$) and in Eq.s
\rref{DirInfExp} \rref{RRInfEst} (related to
$\Dir_s^{(<)}(\bx,\by)$) follow easily from these results. \parn
The arguments presented in this appendix can be easily generalized to
derive the analogous conclusions for the series in Eq.s \rref{DzwDirSupExp}
\rref{DzwRRSupEst} and \rref{DzwDirInfExp} \rref{DzwRRInfEst} related to the
derivatives $\partial_{zw}\Dir_s^{(>)}(\bx,\by)$ and $\partial_{zw}\Dir_s^{(<)}(\bx,\by)$;
we will briefly return on this topic in the sequel (see subsections
\ref{DirSupSub} and \ref{DirInfSub}). \parn
In the final subsection of this appendix we also derive the asymptotic
\rref{HENAsym} for the function $\HE^{(d)}_N$ in our remainder estimates.
\salto
Before proceeding, let us recall the following definitions (see Eq. \rref{deflL}):
$$ \Am := \min_{i\in\{1,...,d\}} \{a_i\}\,,\;\; \AM := \max_{i\in\{1,...,d\}} \{a_i\}\,,
\quad |\bz|^2 := \sum_{i=1}^d z_i^2 \;\;\mbox{for $\bz = \bn\!\in\!\naturali^d$
or $\bz = \bh\!\in\!\interi^d$}\,. $$
\vspace{-0.8cm}
\subsection{Preliminary estimates.}\label{preEst} First of all, let us
consider the upper incomplete gamma function (see Eq. \rref{UpInGa})
$$ \Ga(s,z) := \int_{z}^{+\infty}\!\! dw\; e^{-w}\,w^{s-1} \qquad
(s \in \complessi,\,z \in (0,+\infty)) $$
and note that the above definition implies the following:
\beq |\Ga(s,z_2)| \leqs \Ga(\Re s,z_1) \qquad \mbox{for all $s\!\in\!\complessi$,
$z_1,z_2\!\in\!(0,+\infty)$ with $0\!<\!z_1\!\leqs\!z_2$} ~; \label{boundGa1} \feq
\beq |\Ga(s,z)| \leqs (1\!-\!\al)^{-\Re s} e^{-\al z}\,\Ga(\Re s,(1\!-\!\al)z) \quad
\mbox{for all $s\!\in\!\complessi$, $z\!\in\!(0,+\infty)$, $\al\!\in\!(0,1)$}\,.\!\label{boundGa2}\feq
Hereafter we are going to prove that
\begin{equation}\begin{split}
& \hspace{2.1cm} \sum_{\bh \in \interi^d,\,|\bh| > N}\hspace{-0.2cm}
|\bh|^\rho\, |\Ga(s,\be |\bh|^2)| \leqs \HE^{(d)}_N(\al,\be;\Re s,\rho) \\
& \barray{c} \mbox{for all $s\in\complessi$, $N,\rho \in \reali$,
$\be \in (0,+\infty)$, $\al\in(0,1)$ such that} \\
\mbox{either $\rho \leqs 0$ and $N > 2\sqrt{d}$} \quad \mbox{or} \quad
\mbox{$\rho > 0$ and $\dd{N > 2\sqrt{d}+\sqrt{\rho \over 2\al \be}}$} ~,
\farray \label{boundNGA}
\end{split}\end{equation}
where $\HE^{(d)}_N$ is defined according to Eq. \rref{defHEN}, i.e.
(for $\si \in \reali$),
$$ \HE^{(d)}_N(\al,\be;\si,\rho) := $$
$$ {\pi^{d/2} \over (1\!-\!\al)^\si (\al\,\be)^{{d+\rho \over 2}}\,
\Ga({d \over 2})} \l({N\!-\!\sqrt{d} \over N\!-\!2\sqrt{d}}\r)^{\!\!\!d-1}
\!\Ga(\si,(1\!-\!\al)\be N^2)\;\Ga\!\l({d\!+\!\rho \over 2}\,;
\al\,\be (N\!-\!2\sqrt{d})^2\r)\,. $$
In the sequel we always assume $s\!\in\!\complessi$, $\rho\!\in\!\reali$,
$N\!\in\!(2\sqrt{d},+\infty)$, $\be\!\in\!(0,+\infty)$, $\al\!\in\!(0,1)$. \parn
In order to derive the estimate \rref{boundNGA} we first point out that,
due to Eq.s \rref{boundGa1} \rref{boundGa2}, there holds
\begin{equation}\begin{split}
& |\Ga(s,\be|\bh|^2)| \leqs (1\!-\!\al)^{-\Re s}\, e^{-\al \be|\bh|^2}\,
\Ga(\Re s,(1\!-\!\al)\be N^2) \\
& \hspace{2.2cm} \mbox{for all $\bh\!\in\!\interi^d$ with $|\bh|\!>\!N$} ~;
\end{split}\end{equation}
this in turn implies
\beq \sum_{\bh \in \interi^d,\,|\bh| > N}\hspace{-0.2cm}|\bh|^\rho\,
|\Ga(s,\be |\bh|^2)|\! \leqs (1\!-\!\al)^{-\Re s}\, \Ga(\Re s,(1\!-\!\al)\be N^2)\!\!
\sum_{\bh \in \interi^d,\,|\bh| > N}\hspace{-0.2cm}|\bh|^\rho\,e^{-\al \be |\bh|^2} ~.
\label{bb2} \feq
To proceed, let us note that, for any non-increasing function
$F : [N\!-\!2\sqrt{d},+\infty)\!\to\![0,+\infty)$, there holds
the following estimate (see, e.g., \cite{MPiz}, p. 691):
\beq \sum_{\bh \in \interi^d,\, |\bh| > N}\hspace{-0.2cm} F(|\bh|) \leqs
{2 \pi^{d/2} \over \Ga({d \over 2})} \int_{N-2\sqrt{d}}^{+\infty}\hspace{-0.1cm}
d\tau\;(\tau\!+\!\sqrt{d})^{d-1} F(\tau) ~; \feq
noting that $\tau\!+\!\sqrt{d} = \Big(1\!+\!{\sqrt{d} \over \tau}\Big) \tau
\leqs \Big({N\!-\!\sqrt{d} \over N\!-\!2\sqrt{d}}\Big) \tau$
(for $\tau\!>\!N\!-\!2\sqrt{d} > 0$), the above inequality yields
\beq \sum_{\bh \in \interi^d,\, |\bh| > N}\hspace{-0.2cm} F(|\bh|) \leqs
{2 \pi^{d/2} \over \Ga({d \over 2})} \l({N\!-\!\sqrt{d} \over N\!-\!2\sqrt{d}}\r)^{\!\!\!d-1}
\!\int_{N-2\sqrt{d}}^{+\infty}\hspace{-0.1cm}d\tau\;\tau^{d-1} F(\tau) ~.
\label{boundF} \feq
We want to employ the above bound to estimate the sum in the right-hand
side of Eq. \rref{bb2}; this case involves the function
\beq F(\tau) = \tau^\rho\,e^{-\al \be\,\tau^2} ~, \label{Fdef}\feq
which is decreasing on $[N\!-\!2\sqrt{d},+\infty)$ for
\beq \mbox{either $\rho \leqs 0$ and $N\!> 2\sqrt{d}$ (as before)} \quad \mbox{or} \quad
\mbox{$\rho > 0$ and $\dd{N\!> 2\sqrt{d}+\!\sqrt{\rho \over 2\al \be}}$}\,. \feq
In each one of the above two situations, making the change of variable
$v := \al\,\be\,\tau^2 \in [\al\,\be(N\!-\!2\sqrt{d})^2,+\infty)$
and recalling the definition \rref{UpInGa} of the upper incomplete
gamma function, we easily obtain
\beq \int_{N-2\sqrt{d}}^{+\infty}\hspace{-0.1cm}d\tau\;\tau^{d+\rho-1}\,
e^{-\al \be\,\tau^2} = {1 \over 2(\al\,\be)^{{d+\rho \over 2}}}\;
\Ga\!\l({d\!+\!\rho \over 2}\,; \al\,\be (N\!-\!2\sqrt{d})^2\r)\,, \feq
which, along with Eq.s \rref{bb2} \rref{boundF} and \rref{Fdef},
proves Eq. \rref{boundNGA}. \salto
In passing we point out that Eq. \rref{boundNGA} implies the analogous estimate
({\footnote{Indeed, it sufficies to observe that for any function
$F: (0,+\infty) \to [0,+\infty)$ there holds
$$ \sum_{\bh \in \interi^d,\,|\bh| > N} F(|\bh|) \geqs
2^d \sum_{\bn \in \naturali^d,\,|\bn| > N} F(|\bn|) ~. $$}})
\begin{equation}\begin{split}
& \hspace{1.8cm} \sum_{\bn \in \naturali^d,\,|\bn| > N}\hspace{-0.2cm}
|\bn|^\rho\, |\Ga(s,\be |\bn|^2)| \leqs {1 \over 2^d}\; \HE^{(d)}_N(\al,\be;\Re s,\rho) \\
& \barray{c} \mbox{for all $s\in\complessi$, $N,\rho \in \reali$, $\be \in (0,+\infty)$,
$\al\in(0,1)$ such that} \\
\mbox{either $\rho \leqs 0$ and $N > 2\sqrt{d}$} \quad \mbox{or} \quad
\mbox{$\rho > 0$ and $\dd{N > 2\sqrt{d} + \sqrt{\rho \over 2\al \be}}$} ~.
\farray \label{boundNGAnat}
\end{split}\end{equation}
\vspace{-0.8cm}
\subsection{The series \rref{AX}; connections with the expansions \rref{DirSupExp} \rref{DzwDirSupExp} for $\boma{\Dir^{(>)}_s}$ and its derivatives.}\label{DirSupSub}
For $s\!\in\!\complessi$ and $N\!\in\!(0,+\infty)$, let us define
\beq \RR^{(>)}_{s,N}(\bx,\by) := \sum_{\bn \in \naturali^d,\,|\bn| > N}
\Big|\om_\bn^{-2s}\,\Ga(s,\om_\bn^2\,T)\; \CB_\bn(\bx,\by)\Big| ~; \label{RRSup} \feq
of course, the series \rref{AX} is absolutely convergent if and only if
$\RR^{(>)}_{s,N}(\bx,\by)<+\infty$ for some $N$\,. The rest of this
subsection will be mostly dedicated to evaluating $\RR^{(>)}_{s,N}(\bx,\by)$\,. \parn
First of all, note that from the definitions \rref{defCC} of $C_\bn,\om_\bn^2$
and \rref{deflL} of $\Am,\AM$ it easily follows, for all $\bn \in \naturali^d$,
\beq |\CB_\bn(\bx,\by)| \leqs 1 ~, \label{CnEst}\feq
\beq {\pi^2 \over \AM^2}\,|\bn|^2 \leqs \om_\bn^2 \leqs
{\pi^2 \over \Am^2}\,|\bn|^2 ~. \label{omnEst}\feq
Using Eq. \rref{omnEst} we infer, for all $s \in \complessi$,
$\bn \in \naturali^d$,
\beq |\om_\bn^{-2s}| \leqs {\max(\Am^{2\Re s},\AM^{2\Re s}) \over \pi^{2\Re s}}\;|\bn|^{-2\Re s} ~,
\qquad |\Ga(s,\om_\bn^2\,T)| \leqs \Ga\!\l(\Re s, {\pi^2\,T \over \AM^2}\,|\bn|^2\r)
\label{bound1} \feq
(to deduce the second inequality we also employ Eq. \rref{boundGa1}).
Returning to the definition \rref{RRSup}, we obtain
\beq \RR^{(>)}_{s,N}(\bx,\by) \leqs {{\max(\Am^{2\Re s},\AM^{2\Re s})}\,
\over \,\pi^{2\Re s}} \sum_{\bn \in \naturali^d,\,|\bn| > N} \hspace{-0.2cm}
|\bn|^{-2\Re s}\, \Ga\!\l(\Re s, {\pi^2\,T \over \AM^2}\,|\bn|^2\r) \,, \feq
and using the estimate \rref{boundNGAnat} we conclude the following,
for any $\al \in (0,1)$:
\begin{equation}\begin{split}
& \hspace{1.2cm} \RR^{(>)}_{s,N}(\bx,\by) \leqs {\max(\Am^{2\Re s},\AM^{2\Re s}) \over
2^d \,\pi^{2\Re s}}\;\HE^{(d)}_N\!\l(\!\al,{\pi^2\,T \over \AM^2}\,;\Re s,-2\Re s\!\r) \\
& \mbox{for either \quad $\Re s \geqs 0$, $N>2\sqrt{d}$} \quad \mbox{or} \quad
\mbox{$\Re s<0$, $\dd{N>2\sqrt{d}+{\AM \over \pi}\sqrt{|\Re s| \over \al T}}$} ~.
\label{RRRSupEst}
\end{split}\end{equation}
As noted before, the finiteness of $\RR^{(>)}_{s,N}(\bx,\by)$ proves
the absolute convergence for all $s \in \complessi$ of the series
\rref{AX}, contained in Eq. \rref{DirSupExp} for $\Dir^{(>)}_s(\bx,\by)$;
the bound \rref{RRSupEst} for the remainder $R^{(>)}_{s,N}(\bx,\by)$
of Eq. \rref{DirSupEst} follows straightforwardly from Eq. \rref{RRRSupEst}
since we have
\beq |R^{(>)}_{s,N}(\bx,\by)| \leqs {2^d \over a_1...a_d\,|\Ga(s)|}\;
\RR^{(>)}_{s,N}(\bx,\by) ~. \label{admb} \feq
Let us mention that the series in Eq. \rref{DzwDirSupExp} for $\partial_{z w}\Dir^{(>)}_s$
can be discussed similarly; in place of $\RR^{(>)}_{s,N}(\bx,\by)$,
we use the quantity
\beq \phantom{\RR^{>}}_{(z,w)}\RR^{(>)}_{s,N}(\bx,\by) :=
\sum_{\bn \in \naturali^d,\,|\bn| > N} \Big|\om_\bn^{-2s}\,\Ga(s,\om_\bn^2\,T)\;
\partial_{z w} \CB_\bn(\bx,\by)\Big| \feq
and we estimate it similarly to what we did with $\RR^{(>)}_{s,N}(\bx,\by)$,
noting that
\beq |\partial_{zw}\CB_\bn(\bx,\by)| \leqs {\pi^2 \over \Am^2}\, |\bn|^2 ~. \feq
The conclusions of this analysis are the absolute convergence for all $s \in \complessi$
of the series \rref{DzwDirSupExp} and the remainder bound \rref{DzwRRSupEst}.
\vspace{-0.4cm}
\subsection{The series \rref{AY}: connections with the expansions \rref{DirInfExp} \rref{DzwDirInfExp} for $\boma{\Dir^{(<)}_s}$ and its derivatives.}\label{DirInfSub}
Keeping in mind the definitions \rref{bkldef1}, \rref{bkldef} and \rref{defPP}
of $\bhl,\de_l$ and $\PP_s$ (and noting that $|\de_l| = 1$), let us put
\beq \RR^{(<)}_{s,N}(\bx,\by) :=
\sum_{\substack{\bh \in \interi^d,\, |\bh| > N \\ \bl \in \{1,2\}^d}}
\l|\PP_{s - {d \over 2}}\!\l({\bhl(\bx,\by) \over T}\r)\r| ~,
\label{RRRInf} \feq
for $s\!\in\!\complessi$ and $N\!\in\!(0,+\infty)$\,. Let us consider
the series \rref{AY}: clearly, this converges absolutely if and only if
$\RR^{(<)}_{s,N}(\bx,\by) < + \infty$ for some $N$\,. \parn
Most of the sequel will be dedicated to evaluating $\RR^{(<)}_{s,N}(\bx,\by)$\,.
To this purpose we first note that, for $\bl \in \{1,2\}^d$, $\bh \in \interi^d$
and $|\bh| > N > \sqrt{d}$,
({\footnote{For example, in order to prove the first inequality in Eq.
\rref{bklBo}, recall the definition \rref{bkldef} of $\bhl$; moreover,
let us put $\boma{\DD}_\bl(\bx,\by) := (\DD_{l_i}(x^i,y^i))_{i = 1,...,d}$
and note that
$$ |\boma{\DD}_\bl(\bx,\by)|^2 := \sum_{i = 1}^d \l(\DD_{l_i}(x^i,y^i)\r)^2
\leqs \sum_{i = 1}^d 1 = d ~. $$
Then, for all $i\in\{1,...,d\}$, we have
\begin{equation*}\begin{split}
& \bkl(\bx,\by) = \sum_{i=1}^d a_i^2 (h_i - \DD_{l_i}(x^i,y^i))^2 \geqs
a^2 |\bh - \boma{\DD}_{\bl}(\bx,\by)|^2 \geqs \\
& \hspace{1cm} \geqs \Am^2 \Big(|\bh| - |\boma{\DD}_{\bl}(\bx,\by)| \Big)^2
\geqs \Am^2 \Big(|\bh|\!-\!\sqrt{d}\,\Big)^{\!2}\!
\geqs \Am^2\!\l(1 - {\sqrt{d} \over N}\r)^{\!\!2}\!|\bh|^2 ~.
\end{split}\end{equation*}}})
\beq \Am^2 \l(1 - {\sqrt{d}\over N}\r)^{\!\!2} |\bh|^2 \leqs \bhl(\bx,\by)
\leqs \AM^2 \l(1 + {\sqrt{d}\over N}\r)^{\!\!2} |\bh|^2 ~. \label{bklBo}\feq
Using these bounds we easily infer, for all $s \in \complessi$,
\beq |\bhl(\bx,\by)^s| \leqs C_{\Am,\AM}^{(d)}(\Re s,N)\; |\bh|^{2\Re s} \label{bklsBo}\feq
where, as in Eq. \rref{finaset}, we have put (for $\si \in \reali$)
$$ C_{\Am,\AM}^{(d)}(\si,N) := \max\l[\l(\!\Am\l(\!1\!-\!{\sqrt{d}\over N}\r)\!\r)^{\!\!2\si}\!,
\l(\!\AM\l(\!1\!+\!{\sqrt{d}\over N}\r)\!\r)^{\!\!2\si}\r] . $$
Next, notice that the identity \rref{PropPP1} implies
\beq |\PP_s(\be)| \leqs |\be|^{\Re s}\,\Ga(-\Re s,\be) \qquad
\mbox{for all $s\in\complessi$, $\be \in (0,+\infty)$} ~; \label{PPBo}\feq
in particular, from the above relation and Eq.s \rref{boundGa1}
\rref{bklBo} \rref{bklsBo}, we deduce
\beq \l|\PP_s\Big({\bhl \over T}\Big)\r| \leqs {C_{\Am,\AM}^{(d)}(\Re s,N) \over T^{\Re s}}\;
|\bh|^{2\Re s}\,\Ga\!\l(\!-\Re s,{\Am^2(1\!-\!{\sqrt{d} \over N})^2 \over T}\; |\bh|^2\!\r) .
\label{PPsbo} \feq
Inserting the previous results into the definition \rref{RRRInf} of
$\RR^{(<)}_{s,N}(\bx,\by)$ we get
({\footnote{Note that, since the estimate obtained no longer depends on $\bl$,
the sum over $\bl \in \{1,2\}^d$ just yields a multiplicative factor $2^d$\,.}})
\beq {~}\hspace{-0.4cm} \RR^{(<)}_{s,N}(\bx,\by)\! \leqs
\!{2^d C^{(d)}_{\Am,\AM}(\Re s-{d \over 2},N) \over
T^{\Re s - d/2}}\!\! \sum_{\bh \in \interi^d,\,|\bh| > N}\hspace{-0.35cm}
|\bh|^{2\Re s - d}\,\Ga\!\l({d \over 2}\!-\!\Re s,{\Am^2(1\!-\!{\sqrt{d} \over N})^2\over T}\;
|\bh|^2\!\r);\! \feq
now, using Eq. \rref{boundNGAnat} we conclude the following, for any $s \in \complessi$
and any $\al \in (0,1)$:
\begin{equation}\begin{split}
& \hspace{0.2cm} \RR^{(<)}_{s,N}(\bx,\by) \leqs {2^d C^{(d)}_{\Am,\AM}(\Re s-{d \over 2},N)
\over T^{\Re s - d/2}}\; \HE^{(d)}_N\!\l(\al,{\Am^2(1\!-\!{\sqrt{d} \over N})^2\over T}\,;
{d \over 2}\!-\!\Re s, 2\Re s\!-\!d\r) \label{RRRInfEst} \\
& \mbox{for either ~ $\dd{\Re s \leqs {d \over 2}}$, $N > 2\sqrt{d}$}
\quad\! \mbox{or} \quad\! \mbox{$\dd{\Re s > {d \over 2}}$,
$\dd{N > 3\sqrt{d} + {1 \over \Am} \sqrt{(\Re s\!-\!{d \over 2})T \over \al}}$} ~.
\end{split}\end{equation}
As anticipated above, the finiteness of $\RR^{(<)}_{s,N}(\bx,\by)$
implies the absolute convergence of the series \rref{AY}, which appears
in the expansion \rref{DirInfExp} for $\Dir_s^{(<)}(\bx,\by)$; besides,
the bound \rref{RRInfEst} for the remainder $R^{(<)}_{s,N}(\bx,\by)$ in
Eq. \rref{DirInfEst} follows easily from Eq. \rref{RRRInfEst} noting that
\beq |R^{(<)}_{s,N}(\bx,\by)| \leqs {T^{\Re s - {d \over 2}} \over (4\pi)^{d/2} |\Ga(s)|}
\; \RR^{(<)}_{s,N}(\bx,\by) ~. \label{admb1} \feq
A similar analysis can be developed for the series in the right-hand side
of Eq. \rref{DzwDirInfExp} for $\partial_{z w} \Dir^{(<)}_s$; namely, in
place of $\RR^{(<)}_{s,N}(\bx,\by)$ we consider
\begin{equation}\begin{split}
& \hspace{4.5cm} \phantom{\RR^{<}}_{(z,w)}\RR^{(<)}_{s,N}(\bx,\by) := \\
& \sum_{\substack{\bh \in \interi^d,\, |\bh| > N \\ \bl \in \{1,2\}^d}}
\l|\l[\PP_{s - {d \over 2} - 2}\!\l({\bhl\over T}\r)\!
{\partial_z\bhl\,\partial_w \bhl \over T^2}
- \PP_{s - {d \over 2} - 1}\!\l({\bhl \over T}\r)\!
{\partial_{zw} \bhl \over T} \r]\!(\bx,\by)\r|
\end{split}\end{equation}
and estimate this quantity using Eq. \rref{PPsbo}, along with the following
relations:
\begin{equation}\begin{split}
& |\partial_{z^i} \bhl(\bx,\by)| \leqs \AM \l(\!1 + {\sqrt{d}\over N}\r) |\bh| ~,
\qquad |\partial_{z^i w^j} \bhl(\bx,\by)| = {1 \over 2}\;\de_{ij} \\
& \hspace{2.5cm} \mbox{for $\bz,\bw = \bx$ or $\by$, $i,j \in \{1,...,d\}$} ~.
\end{split}\end{equation}
The final result proves, in this case, the absolute convergence of the series
in the right-hand side of Eq. \rref{DzwDirInfExp} as well as the remainder
estimate \rref{DzwRRInfEst}.
\subsection{Asymptotics for $\boma{\HE^{(d)}_N}$.}
The upper incomplete gamma function is known to possess the following asymptotic expansion
(see \cite{NIST}, p.179, Eq.8.11.2):
\beq \Ga(s,z) = e^{-z} z^{s-1} (1\!+\!O(z^{-1})) \qquad
\mbox{for $z \in \reali$, $z \to +\infty$, and all $s \in \complessi$} ~. \feq
Comparing with the definition \rref{defHEN} of $\HE^{(d)}_N$
(for any $\si \in \reali$), we readily infer
\begin{equation*}\begin{split}
& \HE^{(d)}_N(\al,\be;\si,\rho) = {\pi^{d/2} \,\be^{\si-2}
e^{-4 \al\,\be d} \over \al(1\!-\!\al)\, \Ga({d \over 2})}\;\,
e^{-\be N(N-4\al\be\sqrt{d})} N^{2\si+\rho+d-4} \,(1 + O(N^{-1})) \\
& \hspace{5.1cm} \mbox{for $N\to +\infty$} ~,
\end{split}\end{equation*}
which is the result stated in Eq. \rref{HENAsym}.
\vspace{-0.4cm}
\section{Appendix. The expansions of subsection \ref{EnBox} for the bulk energy}\label{AppTotEBox}
Consider the general framework of subsection \ref{EnBox}, where we
discuss the regularized total energy that, due to the vanishing of
the boundary contributions, coincides with the regularized bulk
energy $E^{\s}$. For the latter we have (see Eq. \rref{EnSupInf})
\begin{equation*}\begin{split}
& \hspace{2.8cm} E^\s = E^{\s,(>)} + E^{\s,(<)} \qquad \mbox{where} \\
& E^{\s,(\bullet)}\! := {\mm^\s \over 2} \int_{(0,a_1)
\times ... \times (0,a_d)} \hspace{-0.8cm} dx^1 ...\,dx^d\;
\Dir^{(\bullet)}_{\s - 1 \over 2}(\bx,\bx) \quad
\mbox{for $\bullet \in \{>,<\}$} ~.
\end{split}\end{equation*}
Hereafter we show in several steps how to obtain the series expansions
\rref{ESupExp} and \rref{EInfExp} for $E^{\s,(>)}$ and $E^{\s,(<)}$,
respectively
({\footnote{Concerning the interchange of certain sums with integrals
or derivatives, recall the comments of footnote \ref{footInt} on page \pageref{footInt}.}}).
\vspace{-0.4cm}
\subsection{Derivation of the expansion \rref{ESupExp}.} Let us first
consider the function $E^{\s,(>)}$; using the expression \rref{DirSupExp}
for the diagonal Dirichlet kernel $\Dir^{(>)}_{\s - 1 \over 2}(\bx,\bx)$
appearing in the definition \rref{EnSupInf}, we readily obtain
\begin{equation}\begin{split}
& \hspace{5.7cm} E^{\s,(>)} = \\
& {2^{d-1}\,\mm^\s \over a_1...a_d\,\Ga({\s - 1 \over 2})}
\sum_{\bn \in \naturali^d} \om_\bn^{1-\s}\,\Ga\!\l({\s\!-\!1 \over 2}\,,
\om_\bn^2\,T\r) \int_{(0,a_1) \times ... \times (0,a_d)}
\hspace{-0.8cm} dx^1 ...\,dx^d\; \CB_\bn(\bx,\bx) ~. \label{EnSup0}
\end{split}\end{equation}
Recalling the definition \rref{defCC} of $\CB_\bn(\bx,\by)$, the integral
in the above equation can be straightforwardly evaluated to yield
\beq \int_{(0,a_1) \times ... \times (0,a_d)} \hspace{-0.8cm}
d x^1 ...\,dx^d\; \CB_\bn(\bx,\bx) = \prod_{i = 1}^d \int_{0}^{a_i}\!
dx^i \sin^2\!\l({n_i \pi \over a_i}\,x^i\r) = {a_1\,...\,a_d \over 2^d} ~, \feq
which, along with Eq. \rref{EnSup0}, allows us to infer Eq. \rref{ESupExp}.
\subsection{Derivation of the expansion \rref{EInfExp}.} Let us now
discuss the term $E^{\s,(<)}$; to this purpose, we consider the expression
\rref{DirInfExp} for the Dirichlet kernel $\Dir^{(<)}_{\s - 1 \over 2}(\bx,\by)$.
When evaluated along the diagonal $\by = \bx$, this expression reduces to
\beq \Dir^{(<)}_{\s-1 \over 2}(\bx,\by)\Big|_{\by = \bx} =
{T^{\s-d-1 \over 2} \over (4\pi)^{d/2}\,\Ga({\s-1 \over 2})}
\sum_{\bh \in \interi^d,\,\bl \in \{1,2\}^d} \de_\bl ~
\PP_{\s-d-1 \over 2}\!\l({\bhl(\bx) \over T}\r) \,;
\label{DirDia} \feq
here and in the remainder of this appendix, for all $\bh\!\in\!\interi^d$,
$\bl\!\in\!\{1,2\}^d$, we use the short-hand notation (compare with Eq. \rref{bkldef})
\beq \bhl(\bx) := \bhl(\bx,\bx) = \sum_{i = 1}^d a_i^2
(h_i - \DD_{l_i}(x^i,x^i))^2 ~. \label{bkldefEn} \feq
Eq.s \rref{EnSupInf} \rref{DirDia} allow us to infer for $E^{\s,(<)}$
the representation
\beq E^{\s,(<)} = {\mm^\s\,T^{\s-d-1 \over 2} \over 2\,(4\pi)^{d/2}\,
\Ga({\s-1 \over 2})}\; \QQ^{(d,T)}_{\s-d-1 \over 2} \label{EnInf0} \feq
where, for suitable $s\!\in\!\complessi$ and for all $T\!\in\!(0,+\infty)$,
we have introduced the function
\beq \QQ^{(d,T)}_{s} := \sum_{\bh \in \interi^d,\,\bl \in \{1,2\}^d}
\de_\bl \int_{(0,a_1)\times ... \times (0,a_d)} \hspace{-0.5cm}
dx^1 ...\,dx^d\; \PP_{s}\!\l({\bhl(\bx) \over T}\r) \,. \label{defQQS} \feq
In the next subsection we show that this function can be expressed as
\beq \QQ^{(d,T)}_s = \sum_{\ns = 0}^d {(- \sqrt{\pi\,T}\,)^{d-\ns} \over
(d\!-\!\ns)!\ns!} \sum_{\si \in S_d} \bap_{\si,\ns} \sum_{\bh \in \interi^\ns}
\PP_{s+{d-\ns \over 2}}\l({\Bs(\bh) \over T}\r)\,\label{QQExpEn} \feq
where the coefficients $\bap_{\si,\ns},\Bs(\bh)$ are as in Eq. \rref{Bsk}.
Once \rref{QQExpEn} is proved, this equation and \rref{EnInf0} give
the thesis \rref{EInfExp}.
\vspace{-0.4cm}
\subsection{Concluding the previous argument: derivation of Eq. \rref{QQExpEn}.}
First of all, notice that the definitions \rref{defQQS} and \rref{defPP}
of $\QQ_s^{(d,T)}$ and $\PP_s$ give
\beq \QQ^{(d,T)}_s = \int_0^1 d\tau\;\tau^{s-1}\!
\sum_{\bh \in \interi^d,\,\bl \in \{1,2\}^d} \de_{\bl} \int_{(0,a_1)
\times ... \times (0,a_d)} \hspace{-0.8cm} d x^1 ...\,dx^d\;
e^{-{1 \over \tau}{\bhl(\bx) \over T}} ~; \feq
this implies, recalling the definition \rref{bkldefEn} of $\bhl(\bx)$,
\beq \QQ^{(d,T)}_s = \int_0^1\! d\tau\;\tau^{s-1} \prod_{i = 1}^d
\sum_{h_i \in \interi,\,l_i \in \{1,2\}} \!\de_{l_i}\! \int_0^{a_i}\!\!d x^i\;
e^{-{a_i^2 \over \tau\,T}\, (h_i - \DD_{l_i}(x^i,x^i))^2} ~. \label{QQdT} \feq
Let us now focus on the general term in the product over $i \in \{1,...,d\}$;
explicitating the sum over $l_i \in \{1,2\}$ and noting that
$\DD_1(x^i,x^i) = 0$, $\DD_2(x^i,x^i) = x^i/a_i$
(see Eq. \rref{bkldef1}), we get
\begin{equation}\begin{split}
& \sum_{h_i \in \interi,\,l_i \in \{1,2\}} \de_{l_i} \int_0^{a_i}\! d x^i\;
e^{-{a_i^2 \over \tau\,T}\,(h_i - \DD_{l_i}(x^i,x^i))^2} = \\
& \hspace{3cm} =  \sum_{h_i \in \interi} \l[\int_0^{a_i}\!\! d x^i\;
e^{-{(a_i h_i)^2 \over \tau\,T}} - \int_0^{a_i}\!\! d x^i\;
e^{-{1 \over \tau\,T}\,(a_i h_i - x^i)^2}\r] ~. \label{sumkE}
\end{split}\end{equation}
The first integral in the square brackets above is trivial,
since the integrand function is constant; moreover
\begin{equation}\begin{split}
& \hspace{2.5cm} \sum_{h_i \in \interi}\,\int_0^{a_i} d x^i\;
e^{-{1 \over \tau\,T}\,(a_i h_i - x^i)^2} = \\
& = \sum_{h_i \in \interi}\,\int_{a_i (h_i-1)}^{a_i h_i} dz^i\;
e^{-{(z^i)^2 \over \tau\,T}} = \int_{-\infty}^{+\infty} dz^i\;
e^{-{(z^i)^2 \over \tau\,T}} = \sqrt{\pi\,\tau\,T} ~,
\end{split}\end{equation}
where in the second passage we have performed the change of variable
$z^i := a_i h_i - x^i$\,. Summing up, Eq. \rref{sumkE} yields
\beq \sum_{h_i \in \interi,\,l_i \in \{1,2\}} \!\de_{l_i}\!
\int_0^{a_i}\!\! d x^i\; e^{-{a_i^2 \over \tau\,T}\,(h_i - \DD_{l_i}(x^i,x^i))^2}
= a_i \!\l(\,\sum_{h_i \in \interi} e^{-{(a_i h_i)^2 \over \tau\,T}}\r)\!
- \sqrt{\pi\,\tau\,T} ~. \label{prodkiE} \feq
In the following we use the notations introduced in Eq. \rref{Bsk}.
Let us return to Eq. \rref{QQdT} and consider the product over
$i \in \{1,...,d\}$ therein; Eq. \rref{prodkiE} allow us to infer
({\footnote{\label{footProdSum} This result depends on the identity
(already pointed out in footnote 10 of Part II of this series)
$$ \prod_{\ns = 1}^d (a_\ns + b_\ns) = \sum_{\ns = 0}^d {1 \over (d\!-\!\ns)!\ns!}
\sum_{\si \in S_d}\! \l(\prod_{i=1}^\ns\, a_{\si(i)}\r)\!
\l(\prod_{j = \ns+1}^d b_{\si(j)}\r) $$
holding for any $d\!\in\!\{1,2,3,...\}$, $a_\ns,b_\ns\!\in\!\reali$
($\ns\!\in\!\{1,2,...,d\}$), where by convention we intend
$\prod_{i=1}^0 a_{\si(i)} := \prod_{j=d+1}^d b_{\si(j)} := 1$\,.}})
$$ \prod_{i = 1}^d \sum_{h_i \in \interi,\,l_i \in \{1,2\}} \!\de_{l_i}\!
\int_0^{a_i}\!\! d x^i\, e^{-{a_i^2 \over \tau\,T}\,(h_i - \DD_{l_i}(x^i,x^i))^2} =
\prod_{i = 1}^d \l[a_i\!\l(\,\sum_{h_i \in \interi}
e^{-{(a_i h_i)^2 \over \tau\,T}}\r)\! - \sqrt{\pi\,\tau\,T} \r]\! = $$
\beq = \sum_{\ns = 0}^d {(- \sqrt{\pi\,\tau\,T}\,)^{d-\ns} \over (d\!-\!\ns)!\ns!}
\sum_{\si \in S_d} \bap_{\si,\ns} \sum_{\bh \in \interi^\ns}
e^{-{1 \over \tau\,T}\,\Bs(\bh)}\,. \label{fina} \feq
Summing up, Eq.s (\ref{QQdT}-\ref{fina}) imply
\beq \QQ^{(d,T)}_s = \sum_{\ns = 0}^d {(- \sqrt{\pi\,T}\,)^{d-\ns}
\over (d\!-\!\ns)! \ns!} \!\sum_{\si \in S_d} \bap_{\si,\ns}\!
\sum_{\bh \in \interi^\ns} \int_0^1\! d\tau\;\tau^{s+{d-\ns \over 2}-1}
e^{-{1 \over \tau}\,{\Bs(\bh) \over T}} \,. \label{QQtemp}\feq
To conclude and obtain Eq. \rref{QQExpEn}, just note that all the integrals
over $\tau \in (0,1)$ in Eq. \rref{QQtemp} can be evaluated according to
Eq. \rref{defPP} to give the functions $\PP_{s+{d-\ns \over 2}}({\Bs(\bh)\over T})$\,.
\section{Appendix. Absolute convergence and remainder bounds for the series of subsection \ref{EnBox}.}\label{AppC}
Let us recall that the expansion \rref{ESupExp} for $E^{\s,(>)}$
contains the series
\beq \sum_{\bn \in \naturali^d}\, \om_\bn^{1-\s}\;
\Ga\!\l({\s - 1 \over 2}\,,\,\om_\bn^2\,T\r)\,. \label{BX} \feq
On the other hand, the right-hand side of Eq. \rref{EInfExp} for $E^{\s,(<)}$
contains series in $\bh \in \interi^\ns$, for $\ns \in \{0,...,d\}$.
Let us consider one of these series; after removing the term with $\bh = \b0$,
which becomes singular at $\s = \ns + 1$, we are left with
\beq \sum_{\bh \in \interi^\ns,\,\bh \neq \b0}
\PP_{\s - \ns - 1 \over 2}\!\l({\Bs(\bh) \over T}\r)\,. \label{BY} \feq
In the following subsections we prove the absolute convergence of the series
\rref{BX} \rref{BY}, for all $\s \in \complessi$\,; moreover, we derive remainder
estimates for both series, which justify statements \rref{ESupEst} \rref{EInfEst}.
Throughout this appendix we adopt systematically the notations
introduced in Appendix \ref{AppBox} and the results obtained therein
(see, in particular, subsection \ref{preEst}). \parn
$\phantom{a}$ \vspace{-1.4cm}\\
\subsection{The series \rref{BX}; connections with the expansion \rref{ESupExp} of $\boma{E^{\s,(>)}}$.} \label{ApSubEM}
Let us define, for $\s \in \complessi$ and $N \in (0,+\infty)$,
\beq \RR_{N}^{\s,(>)} := \sum_{\bn \in \naturali^d,\,|\bn| > N}\l|\,\om_\bn^{1-\s}\;
\Ga\!\l({\s - 1 \over 2}\,,\,\om_\bn^2\,T\r)\r| \,. \label{RREnSup} \feq
If we can show that $\RR_{N}^{\s,(>)} < +\infty$ for some $N$,
of course the series \rref{BX} converges absolutely. \parn
Hereafter we derive quantitative estimates for $\RR_{N}^{\s,(>)}$.
Using the inequalities \rref{bound1}, we infer from definition
\rref{RREnSup} that
\beq \RR_{N}^{\s,(>)} \leqs {\max(\Am^{\Re \s-1},\AM^{\Re \s-1}) \over \pi^{\Re \s-1}}
\sum_{\bn \in \naturali^d,\,|\bn| > N} |\bn|^{1-\Re \s}\;
\Ga\!\l({\Re\s\!-\!1 \over 2}\,,\,{\pi^2\,T \over \AM^2}\,|\bn|^2\r) \,; \feq
now, using the result \rref{boundNGAnat} we obtain the following,
for any $\al \in (0,1)$\,:
\begin{equation}\begin{split}
& \hspace{1.2cm} \RR_{N}^{\s,(>)} \leqs {\max(\Am^{\Re \s-1},\AM^{\Re \s-1}) \over
2^d\,\pi^{\Re \s-1}}\; \HE^{(d)}_N\!\l(\al,{\pi^2\,T \over \AM^2}\,;
{\Re\s\!-\!1 \over 2}\,,\,1\!-\!\Re\s\r) \\
& \mbox{for either \quad $\Re \s \geqs 1$, $N > 2\sqrt{d}$}
\quad \mbox{or} \quad \mbox{$\Re \s<1$, $\dd{N>2\sqrt{d}
+ {\AM \over \pi}\sqrt{|\Re\s - 1| \over 2\al T}}$} ~. \label{REnSupEst}
\end{split}\end{equation}
Finally, let us point out that the remainder bound \rref{ESupEst}
for $\RR_{N}^{\s,(>)}$ follows easily from the above inequality noting that
\beq \l|R_{N}^{\s,(>)}\r| \leqs {\mm^{\Re\s} \over 2\,|\Ga({\s - 1 \over 2})|}\;
\RR_{N}^{\s,(>)} ~. \label{admbEn1} \feq
\vspace{-0.8cm}
\subsection{The series \rref{BY}; connections with the expansion \rref{EInfExp} of $\boma{E^{\s,(<)}}$.}
Let us define the following functions, for $\s \in \complessi$,
$N \in (0,+\infty)$, $\ns \in \{1,...,d\}$ and $\si \in S_d$:
\beq \RR_{N}^{\s,(<)}(\si,\ns) := \sum_{\bh \in \interi^\ns,\, |\bh| > N}
\l|\PP_{{\s-\ns-1 \over 2}}\!\l(\!{\Bs(\bh) \over T}\!\r)\r| \label{RRInfnsp} \feq
where $\PP_s$ and $\Bs$ are as in Eq.s \rref{defPP} and \rref{Bsk}, respectively. \parn
Hereafter, we show that $\RR_{N}^{\s,(<)}(\si,\ns) < +\infty$ for some suitable $N$;
this implies the absolute convergence of the series \rref{BY}. \parn
In the rest of this paragraph we derive explicit estimates for
the functions $\RR_{N}^{\s,(<)}(\si,\ns)$, for all $\s\!\in\!\complessi$,
$\ns\!\in\!\{1,...,d\}$, $\si\!\in\!S_d$ and some suitable $N$\,. First of all,
we note that, for all $\bh\!\in\!\interi^d$, from the definition of
$\Bs(\bh)$ in Eq. \rref{Bsk} it follows (compare with Eq. \rref{bklBo})
\beq \Am^2 |\bh|^2 \leqs |\Bs(\bh)| \leqs \AM^2 |\bh|^2 ~;
\label{BpBo} \feq
using these inequalitites we easily infer (for all $\s \in \complessi$)
\beq \Big|\Bs(\bh)^{\s-1 \over 2}\Big| \leqs
\max(\Am^{\Re\s-1},\AM^{\Re\s-1})\,|\bh|^{\Re\s-1} ~. \label{BpsBo} \feq
Similarly to the derivation of Eq. \rref{PPsbo}, using the above
bound along with Eq.s \rref{boundGa1} \rref{PPBo} we obtain
\begin{equation}\begin{split}
& \hspace{3.2cm} \l|\PP_{{\s-\ns-1 \over 2}}\!\l(\!{\Bs(\bh) \over T}\!\r)\r| \leqs \\
& {\max(\Am^{\Re\s-\ns-1},\AM^{\Re\s-\ns-1}) \over T^{{\Re\s-\ns-1 \over 2}}}\;
|\bh|^{\Re\s-\ns-1}\,\Ga\!\l({\ns\!+\!1\!-\!\Re\s \over 2}\,,\,
{a^2 \over T}\,|\bh|^2 \r) \,. \label{PPsbo2}
\end{split}\end{equation}
Inserting the previous results into the definition \rref{RRInfnsp}
of $\RR_{\s,N}^{(<)}(\si,n)$, we get
\begin{equation}\begin{split}
& \hspace{5cm} \RR_{N}^{\s,(<)}(\si,\ns) \leqs \\
& {\max(\Am^{\Re\s-\ns-1},\AM^{\Re\s-\ns-1}) \over T^{{\Re\s - \ns -1 \over 2}}}
\sum_{\bh \in \interi^\ns,\, |\bh| > N}\! |\bh|^{{\Re\s-\ns-1}}\,
\Ga\!\l({\ns\!+\!1\!-\!\Re\s \over 2}\,,{a^2 \over T}\,|\bh|^2 \r);
\end{split}\end{equation}
using Eq. \rref{boundNGA} (with $d$ replaced by $n$), we conclude
the following, for any $\al \in (0,1)$:
\beq \RR_{N}^{\s,(<)}(\si,\ns) \leqs
{\max(\Am^{\Re\s-\ns-1},\AM^{\Re\s-\ns-1}) \over T^{{\Re\s - \ns - 1 \over 2}}}\;
\HE^{(\ns)}_N\!\l(\al,{\Am^2 \over T}\,;{\ns\!+\!1\!-\!\Re\s \over 2},\Re\s\!-\!\ns\!-\!1\r)
\label{REnInfEst} \feq
$$ \mbox{for either \,$\dd{\Re \s \leqs \ns\!+\!1}$, $N > 2\sqrt{\ns}$}
\quad \mbox{or} \quad \mbox{$\dd{\Re \s > \ns\!+\!1}$,
$\dd{N > 2\sqrt{\ns}+{1\over \Am}\,\sqrt{|\Re \s\!-\!\ns\!-\!1| T \over 2\al}}$}~. $$
The above relation proves the finiteness of $\RR_{N}^{\s,(<)}(\si,\ns)$
for all $\s \in \complessi$, $\ns \in \{1,...,d\}$ and all $\si \in S_d$,
which implies the absolute convergence of the series \rref{BY}; moreover,
the remainder bound \rref{EInfEst} for $R_{N}^{\s,(<)}$ follows
straightforwardly from Eq. \rref{REnInfEst}, noting that
\beq \l|R_{N}^{\s,(<)}\r| \leqs {\mm^{\Re\s}\,T^{\Re\s-1 \over 2} \over
2^{d+1}\,|\Ga({\s-1 \over 2})|}\,\sum_{\ns = 1}^d{1 \over (d\!-\!\ns)!\ns!}
\sum_{\si \in S_d} {\bap_{\si,\ns} \over (\pi\,T)^{\ns/2}}\;
\RR_{N}^{\s,(<)}(\si,\ns) ~. \label{admbEn2} \feq
\vfill \eject \noindent


\end{document}